\documentclass[aps,prd,nofootinbib,showpacs,superscriptaddress, preprint]{revtex4-1}
\pdfoutput=1
\usepackage[utf8]{inputenc}
\usepackage{amsmath,amssymb}
\usepackage{epsfig}
\usepackage{comment}
\usepackage{graphicx}
\usepackage[usenames,dvipsnames]{color}
\usepackage{float}
\usepackage{bbold}
\usepackage[colorlinks,citecolor=blue]{hyperref}
\usepackage{color}
\usepackage{subfigure}

\begin{document}
\title{WIMPy Leptogenesis in Non-Standard Cosmologies}


\author{Devabrat Mahanta}
\email{devab176121007@iitg.ac.in}
\affiliation{Department of Physics, Indian Institute of Technology Guwahati, Assam 781039, India}

\author{Debasish Borah}
\email{dborah@iitg.ac.in}
\affiliation{Department of Physics, Indian Institute of Technology Guwahati, Assam 781039, India}

\begin{abstract}
We study the possibility of generating baryon asymmetry of the universe from dark matter (DM) annihilations during non-standard cosmological epochs. Considering the DM to be of weakly interacting massive particle (WIMP) type, the generation of baryon asymmetry via leptogenesis route is studied where WIMP DM annihilation produces a non-zero lepton asymmetry. Adopting a minimal particle physics model to realise this along with non-zero light neutrino masses, we consider three different types of non-standard cosmic history namely, (i) fast expanding universe, (ii) early matter domination and (iii) scalar-tensor theory of gravity. By solving the appropriate Boltzmann equations incorporating such non-standard history, we find that the allowed parameter space consistent with DM relic and observed baryon asymmetry gets enlarged with the possibility of lower DM mass in some scenarios. While such lighter DM can face further scrutiny at direct search experiments, the non-standard epochs offer complementary probes on their own.
\end{abstract}

\maketitle

\section{Introduction}
As suggested by cosmological and astrophysical evidences \cite{Aghanim:2018eyx, Zyla:2020zbs}, the present universe contains a large proportion of nun-luminous, non-baryonic form of matter, known as dark matter (DM), amounting to nearly five times the density of ordinary matter or baryons. While relative abundance of DM is approximately $27\%$, it is conventionally reported in terms of density parameter $\Omega_{\rm DM}$ and reduced Hubble constant $h = \text{Hubble Parameter}/(100 \;\text{km} ~\text{s}^{-1} 
\text{Mpc}^{-1})$ as \cite{Aghanim:2018eyx}
\begin{equation}
\Omega_{\text{DM}} h^2 = 0.120\pm 0.001
\label{dm_relic}
\end{equation}
\noindent at 68\% CL. In spite of these irrefutable evidences, the particle nature of DM is not yet known. On the other hand, the baryonic matter content is highly asymmetric leading to another longstanding puzzle of baryon asymmetry of the universe (BAU). This observed excess of baryons over anti-baryons is quantified in terms of the baryon to photon ratio as \cite{Aghanim:2018eyx} 
\begin{equation}
\eta_B = \frac{n_{B}-n_{\overline{B}}}{n_{\gamma}} \simeq 6.2 \times 10^{-10}, 
\label{etaBobs}
\end{equation} 
based on the cosmic microwave background (CMB) measurements which also agrees well with the big bang nucleosynthesis (BBN) estimates \cite{Zyla:2020zbs}. While none of the standard model (SM) particles can be a viable DM candidate, the SM also fails to satisfy the Sakharov's conditions \cite{Sakharov:1967dj} required to generate the observed BAU dynamically. This has led to several beyond standard model (BSM) proposals in the literature to account for the DM and BAU. Among different particle DM scenarios, the weakly interacting massive particle (WIMP) has been the most widely studied one \cite{Kolb:1990vq, Jungman:1995df, Bertone:2004pz, Feng:2010gw, Arcadi:2017kky, Roszkowski:2017nbc}. On the other hand, the mechanism of baryogenesis \cite{Weinberg:1979bt, Kolb:1979qa} which invokes out-of-equilibrium decay of heavy new particles, has been the frontrunner in explaining the BAU. One appealing way to achieve baryogenesis is the leptogenesis \cite{Fukugita:1986hr} route where a non-zero lepton asymmetry is first generated which later gets converted into the BAU via electroweak sphalerons \cite{Kuzmin:1985mm}.

While the above-mentioned scenarios, among others, can explain the origin of DM and BAU independently, the very similarity between their abundances namely, $\Omega_{\rm DM} \approx 5\,\Omega_{\rm Baryon}$ might deserve an explanation. Ignoring the possibility of any numerical coincidence or anthropic origin behind this similarity, one can provide a dynamical origin of it by uniting their production mechanisms. A brief review of such cogenesis mechanisms for DM and BAU can be found in \cite{Boucenna:2013wba}. Such cogenesis mechanisms can be classified into two broad categories. In the first one, the DM sector is also assumed to be asymmetric like the visible one, known as the asymmetric dark matter (ADM) scenario \cite{Nussinov:1985xr, Davoudiasl:2012uw, Petraki:2013wwa, Zurek:2013wia, Barman:2021ost, Cui:2020dly}, where out-of-equilibrium decay of the same heavy particle is responsible for generating similar asymmetries $n_B-n_{\overline{B}} \sim \lvert n_{\rm DM}-n_{\overline{ \rm DM}} \rvert$ in the two sectors. In the second class of this cogenesis scenario, the asymmetry is produced from annihilations \cite{Yoshimura:1978ex, Barr:1979wb, Baldes:2014gca}, where one or more particles involved in the process eventually go out of thermal equilibrium to generate a net asymmetry\footnote{See \cite{Chu:2021qwk} for a hybrid scenario where dark matter annihilates into metastable dark partners whose late decay produces baryon asymmetry.}. The so-called WIMPy baryogenesis \cite{Cui:2011ab, Bernal:2012gv, Bernal:2013bga} belongs to this category, where a DM particle freezes out to generate its own relic abundance while simultaneously producing an asymmetry in the baryon sector. This idea has also been extended to leptogenesis, known as the WIMPy leptogenesis scenario \cite{Kumar:2013uca, Racker:2014uga, Dasgupta:2016odo, Borah:2018uci, Borah:2019epq, Dasgupta:2019lha}.

Irrespective of the specific mechanism of cogenesis, it is usually a high scale phenomena considered to be taking place in the radiation dominated era of standard $\Lambda {\rm CDM}$ cosmology, prior to the BBN epoch. However, as the lower bound on the reheat temperature can be a few MeV \cite{Kawasaki:1999na, Kawasaki:2000en, Hasegawa:2019jsa}, there is no experimental evidence to support radiation domination prior to the BBN era. While such non-standard cosmological history prior to the BBN epoch is allowed experimentally, it can change the dynamics of cogenesis and hence the corresponding constraints on relevant model parameters which can be probed experimentally. In this present work, we consider three such non-standard histories namely, (a) a fast expanding universe (FEU) scenario, (b) an early matter dominated (EMD) phase and (c) scalar-tensor theory of gravity (STG) and study the impact on the WIMPy leptogenesis scenario. While WIMPy leptogenesis has not been studied in non-standard cosmology before, generation of DM abundance in such non-standard history have received lots of attention  \cite{McDonald:1989jd, Kamionkowski:1990ni, Chung:1998rq, Moroi:1999zb, Giudice:2000ex, Allahverdi:2002nb, Allahverdi:2002pu, Acharya:2009zt, Davoudiasl:2015vba, Drees:2018dsj, Bernal:2018ins, Bernal:2018kcw, Arias:2019uol, Delos:2019dyh, Chanda:2019xyl, Bernal:2019mhf, Poulin:2019omz, Maldonado:2019qmp, Betancur:2018xtj, DEramo:2017gpl, DEramo:2017ecx, Biswas:2018iny, Visinelli:2015eka, Visinelli:2017qga, Barman:2021ifu, Arcadi:2020aot, Bernal:2020bfj, Ahmed:2020fhc, Borah:2021inn, Borah:2022byb, Catena:2004ba, Catena:2009tm, Meehan:2015cna, Dutta:2016htz, Dutta:2017fcn, Ghosh:2022fws}. On the other hand, impact of non-standard cosmology on leptogenesis has been studied in \cite{Chen:2019etb,Abdallah:2012nm,Dutta:2018zkg, Chakraborty:2022gob} whereas its effects on scenarios which include both DM as well as leptogenesis were studied in \cite{Mahanta:2019sfo, Chang:2021ose, Konar:2020vuu, Barman:2022gjo, JyotiDas:2021shi, Barman:2021ost}. Here we study the impact of such non-standard cosmological history on cogenesis of DM and BAU by adopting a TeV scale WIMPy leptogenesis framework. Such non-standard cosmology not only leads to different model parameters but can also lower the scale of WIMPy leptogenesis, increasing the discovery prospects at different experiments.

This paper is organised as follows. In section \ref{sec:model}, we briefly discuss the particle physics model followed by the discussion of WIMPy leptogenesis in standard cosmology in section \ref{sec:wimpy}. In section \ref{sec:wimpy2} we discuss the details of WIMPy leptogenesis with three different non-standard cosmological histories and finally conclude in section \ref{sec:conclude}.

\section{The Model}
\label{sec:model}
As pointed out in earlier works on WIMPy baryogenesis or WIMPy leptogenesis, one can satisfy all the Sakharov's conditions with DM annihilations such that some of the processes responsible for WIMP freeze-out can also create a baryon or lepton asymmetry. In order to keep the washout scatterings under control, one has to ensure that the washout scatterings freeze out before WIMP freeze-out \cite{Cui:2011ab}. Based on this central criteria, several WIMPy baryogenesis and leptogenesis models have been constructed. In order to illustrate the effects of non-standard cosmological histories, we consider the model considered in \cite{Dasgupta:2019lha} for simplicity although choosing a more complicated model will not change the generic conclusions reached in this work.

Similar to usual leptogenesis scenarios, WIMPy leptogenesis models are also constructed in a way which explains non-zero neutrino mass too, another observed phenomena which the SM fails to explain. The model proposed in \cite{Dasgupta:2019lha} is an extension of the minimal scotogenic model \cite{Ma:2006km} by a scalar triplet. The minimal scotogenic model extends the SM by three gauge singlet right handed neutrinos (RHN) $N_i, (i=1-3)$ and a scalar doublet $\eta$, all of which are odd under an unbroken $Z_2$ symmetry. While this minimal field content can account for neutrino mass, DM as well as leptogenesis from RHN decay, the extension by a scalar triplet is necessary to realise a WIMPy leptogenesis setup with the neutral real component of inert scalar doublet $\eta$ playing the role of WIMP DM. The scalar triplet is kept $Z_2$ even like SM particles are, in order to realise the necessary interactions. The relevant leptonic Lagrangian can be written as follows. 
\begin{equation}
-\mathcal{L} \supset Y_{i \alpha}^{N} \overline{\ell_{\alpha}} \tilde{\eta} N_{i}+Y_{\alpha \beta}^{\Delta}\overline{\ell^{c}_{\alpha}}  \Delta \ell_{\beta}+ {\rm h.c.}, 
\end{equation}
with $\ell= (\nu,l)^T$ being the SM lepton doublet, C in superscript denotes the charge conjugation, $\tilde{\eta}=i \sigma_{2}\eta^{*}$. The scalar potential of the model is given in Appendix \ref{sec:appen1}. The term $\lambda^{''}_{H\eta}(H^{\dagger}\eta)^2$ in the scalar potential given in Eq. \eqref{Potential} plays a crucial role in generating radiative neutrino mass as well as DM phenomenology, by generating the mass splitting of scalar and pseudoscalar components of $\eta$, the details of which is given in Appendix \ref{sec:appen1}. On the other hand, the trilinear term $\mu_{H \Delta} \tilde{H}^{\dagger} \Delta H$ leads to an induced vacuum expectation value (VEV) of neutral component of $\Delta$ (denoted as $v_{\Delta}$) after electroweak symmetry breaking generating the well known type II seesaw contribution to neutrino mass \cite{Magg:1980ut, Lazarides:1980nt, Mohapatra:1980yp, Cheng:1980qt, Schechter:1980gr, Schechter:1981cv}. The other trilinear term in the scalar potential namely, $\mu_{\eta \Delta} \eta^{\dagger} \Delta^{\dagger} \tilde{\eta}$ plays an important role in WIMPy leptogenesis as it opens up new WIMP annihilation channels into leptons which violate lepton number. The trilinear coupling $\mu_{\eta \Delta}$ is a free parameter of the model and can be complex in general. To generate a net leptonic asymmetry, the coupling $\mu_{\eta \Delta}$ is assumed to be purely imaginary while all other parameters in Eq. \eqref{Potential} are considered real.

\section{WIMPy leptogenesis in standard cosmology}
\label{sec:wimpy}

\begin{figure}[h]
\includegraphics[scale=.35]{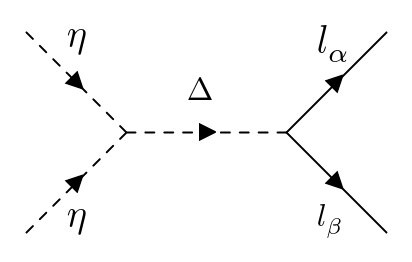}
\includegraphics[scale=.40]{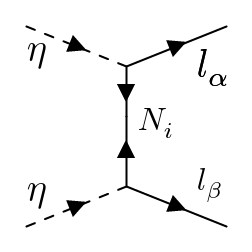}
\caption{Feynmann diagrams for the scattering process $\eta \eta \longrightarrow ll$.}
\label{fig:annihilation}
\end{figure}

We consider the lightest neutral component of the scalar doublet $\eta$ as the WIMP DM candidate. While there exist several gauge and scalar portal diagrams for inert scalar DM annihilations, as discussed in earlier works \cite{Dasgupta:2014hha, LopezHonorez:2006gr,  Hambye:2009pw, Dolle:2009fn, Honorez:2010re, LopezHonorez:2010tb, Gustafsson:2012aj, Goudelis:2013uca, Arhrib:2013ela, Diaz:2015pyv, Ahriche:2017iar}, there are fewer diagrams which violate lepton number in our setup. Such CP and lepton number violating DM annihilation processes are responsible for generating a non-zero lepton asymmetry, as shown in Fig. \ref{fig:annihilation}. In this model the lepton number violating DM annihilation is $\eta \eta \longrightarrow \ell \ell$, which violates lepton number by two units ($\Delta L=2$). Due to the existence of multiple Feynman diagrams for this process as shown in Fig. \ref{fig:annihilation}, it is possible to generate a non-zero CP asymmetry from the interference. Although we show the tree level diagrams only, the propagators are resummed taking the radiative corrections into account, similar to \cite{Borah:2020ivi} where three-body decay involving DM in final state was considered instead of DM annihilation to source lepton asymmetry. In order to enhance the CP asymmetry, we consider the resonance regime $m_{\eta} \simeq m_{\Delta}/2$ such that the scale of WIMPy leptogenesis can be as low as possible with the washout scatterings under control.

In order to implement the constraints from neutrino data, we use the Casas-Ibarra parametrisation \cite{Casas:2001sr} to write the Yukawa couplings $Y^N, Y^{\Delta}$ in terms of neutrino parameters as well as physical masses of different BSM particles. As can be seen from Eq. \eqref{eq:mass_diff} in Appendix \ref{sec:appen1}, the mass difference between $\eta_{R}$ and $\eta_{I}$ depends on $\mu_{\eta \Delta}$, $\lambda_{H\eta}^{''}$ and $v_{\Delta}$. On the other hand, from Eq. \eqref{eq:lambda} and Eq. \eqref{eq:YN} it is clear that increasing mass difference between $\eta_{R}$ and $\eta_{I}$ decreases the Yukawa coupling $Y^{N}$. For leptogenesis one need large a enough $Y^{N}$ to generate the correct asymmetry. Therefore we need to choose $\lambda_{H\eta}^{''}$, $v_{\Delta}$ and $\mu_{ \eta \Delta}$ carefully such that the mass difference between $\eta_{R}$ and $\eta_{I}$ remains small. Also one can not make the mass difference $\Delta m_{\eta^{0}}=m_{\eta_{R}}-m_{\eta_{I}}$ arbitrarily small as it will make the direct detection cross-section for the DM very large \cite{Arina:2007tm}, ruled out by stringent direct detection constraints \cite{LZ:2022ufs}. In this work, we choose the all mentioned parameters such a way that $\Delta m_{\eta^{0}}> 200$ keV, large enough to forbid tree level Z mediated inelastic scattering of DM off nucleons.

The Boltzmann equations (BEs) for comoving number densities of DM and lepton number respectively, can be identified to be
\begin{eqnarray}
\dfrac{dY_{\eta}}{dz} & = & -\dfrac{s}{{\bf H}(z)z}\left[ (Y_{\eta}^{2}-(Y_{\eta}^{\rm eq})^{2}) \langle \sigma v\rangle_{\eta \eta \longrightarrow {\rm SM \, SM}} \right],  \\
\dfrac{dY_{\Delta L}}{dz} & = & \dfrac{s}{{\bf H}(z)z} \left[ (Y_{\eta}^{2}-(Y_{\eta}^{\rm eq})^{2})  \langle \sigma v \rangle^{\delta}_{\eta \eta \longrightarrow \ell \ell} \right] -2 Y_{\Delta L}Y_{l}^{\rm eq}r_{\eta}^{2}\langle \sigma v \rangle_{\eta \eta \longrightarrow \ell \ell} \nonumber \\ & & -2 Y_{\Delta L}Y_{\eta}^{\rm eq} \langle \sigma v \rangle_{\eta \bar{\ell} \longrightarrow \eta \ell},
\label{BE1}
\end{eqnarray}
where $z=m_{\eta}/T \equiv m_{\rm DM}/T$ and $Y_{i}^{\rm eq}=n_{i}^{\rm eq}/s$ are the normalised number densities (in equilibrium) for the particle species $i$ ($s$ being the entropy density of the universe). The $Y_{\Delta L}$ is the comoving number density of lepton asymmetry and is defined by $Y_{\Delta L}=Y_{L}-Y_{\bar{L}}$. Here $r_{\eta}=Y_{\eta}^{\rm eq}/Y_{l}^{\rm eq}$ and ${\bf H}(z)=\sqrt{8\pi^{3}g_{*}/90}m_{\eta}^{2}/(z^{2}M_{\rm Pl})$ is the Hubble expansion rate for the standard radiation dominated universe with $M_{\rm Pl} \simeq 1.22 \times 10^{19}$ GeV being the Planck mass. Here the $\langle \sigma v\rangle $ represents the thermally averaged cross sections for the mentioned processes. On the other hand, $\langle \sigma v\rangle^{\delta}_{\eta \eta \longrightarrow \ell \ell}$ on the right hand side of Eq. \eqref{BE1} includes the difference between $\eta \eta \longrightarrow \ell \ell$ and $\eta \eta \longrightarrow \bar{\ell} \bar{\ell}$ processes responsible for generating a net lepton asymmetry, the details of which are shown in Appendix \ref{sec:appen2}. The lepton asymmetry is converted into baryon asymmetry via sphaleron factor $Y_{\Delta B} = c_{\rm sph} Y_{\Delta L}$ where $c_{\rm sph}=-\frac{16}{39}$ for our model, as shown in Appendix \ref{sec:appen4}.


\begin{figure}[h]
\includegraphics[scale=.45]{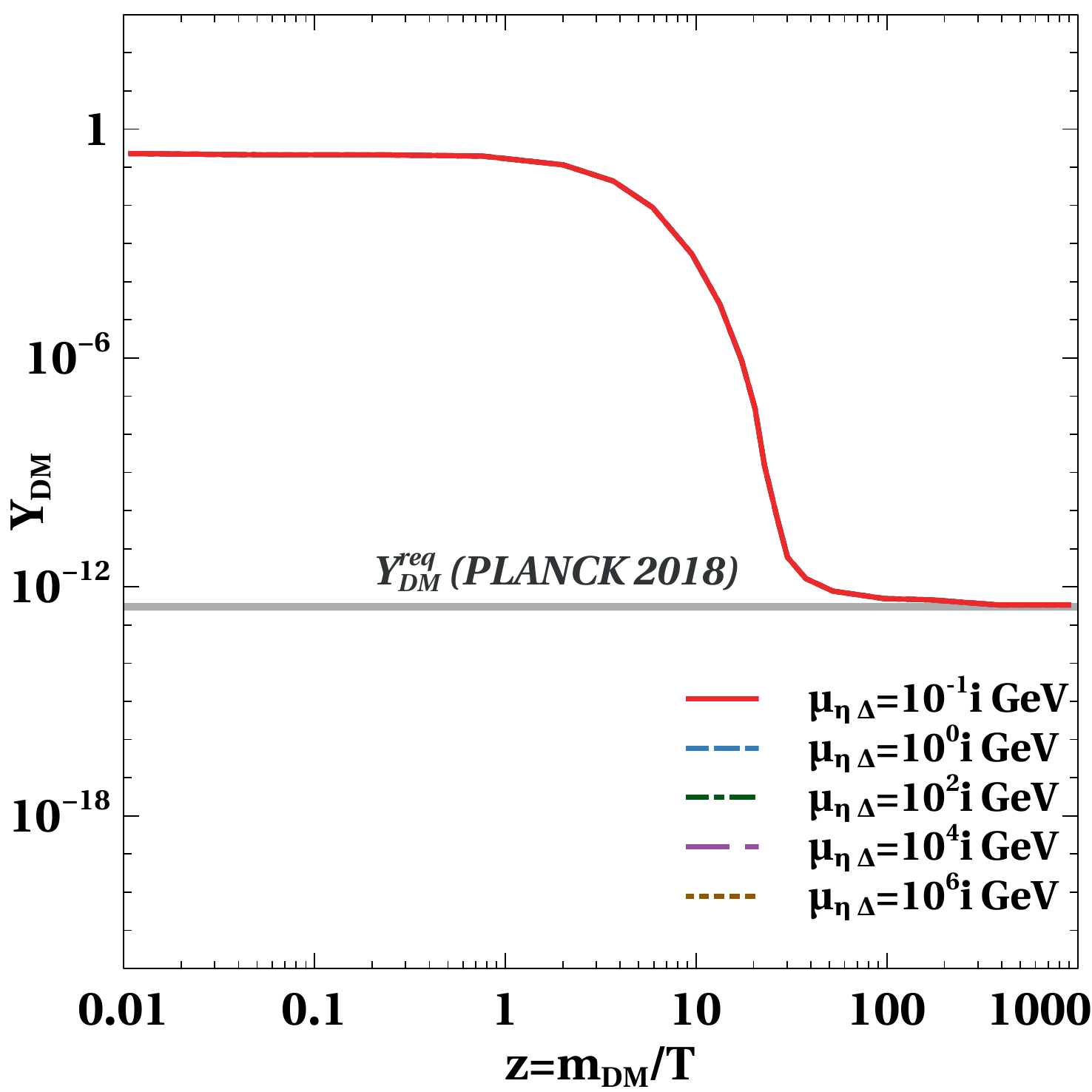}
\includegraphics[scale=.45]{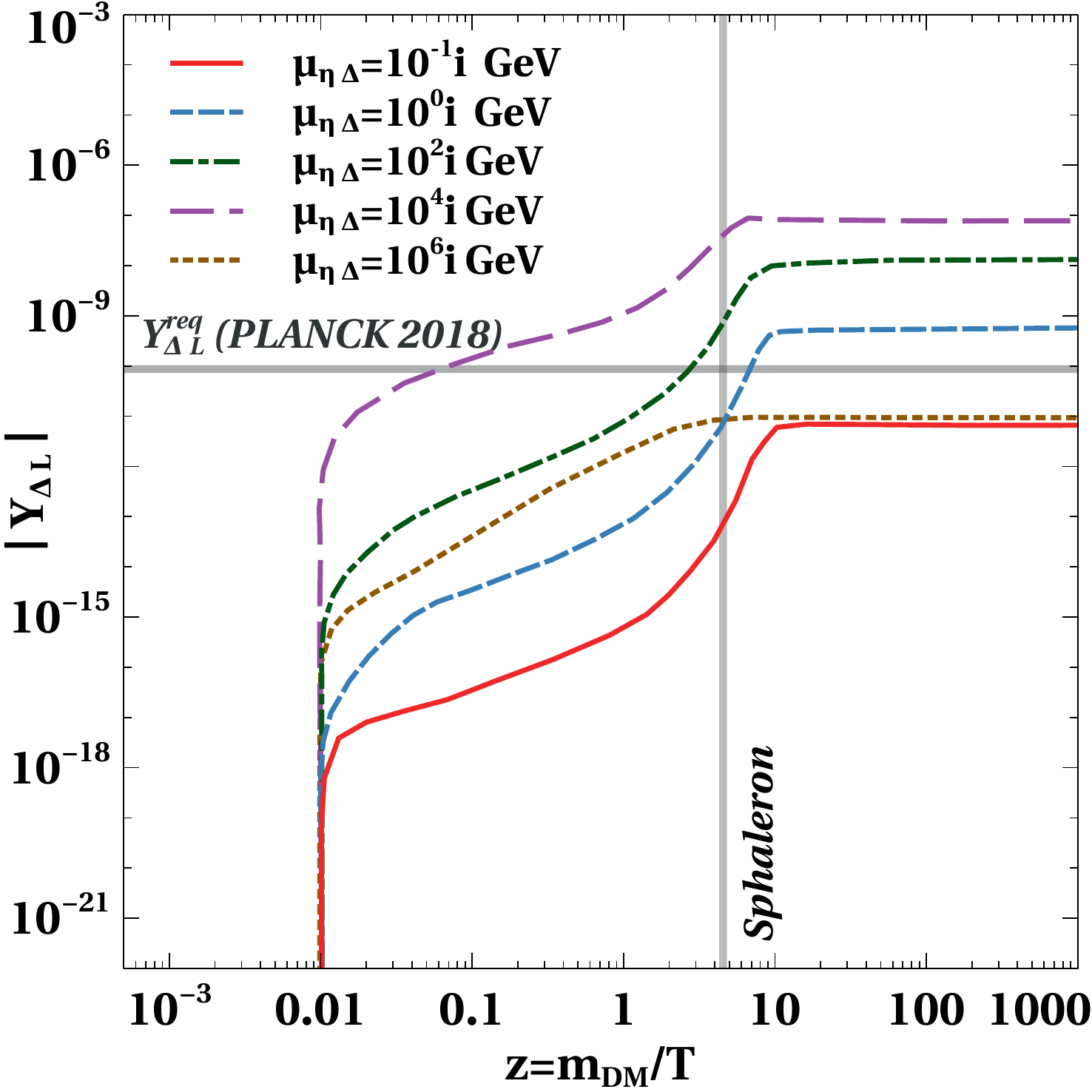}
\includegraphics[scale=.45]{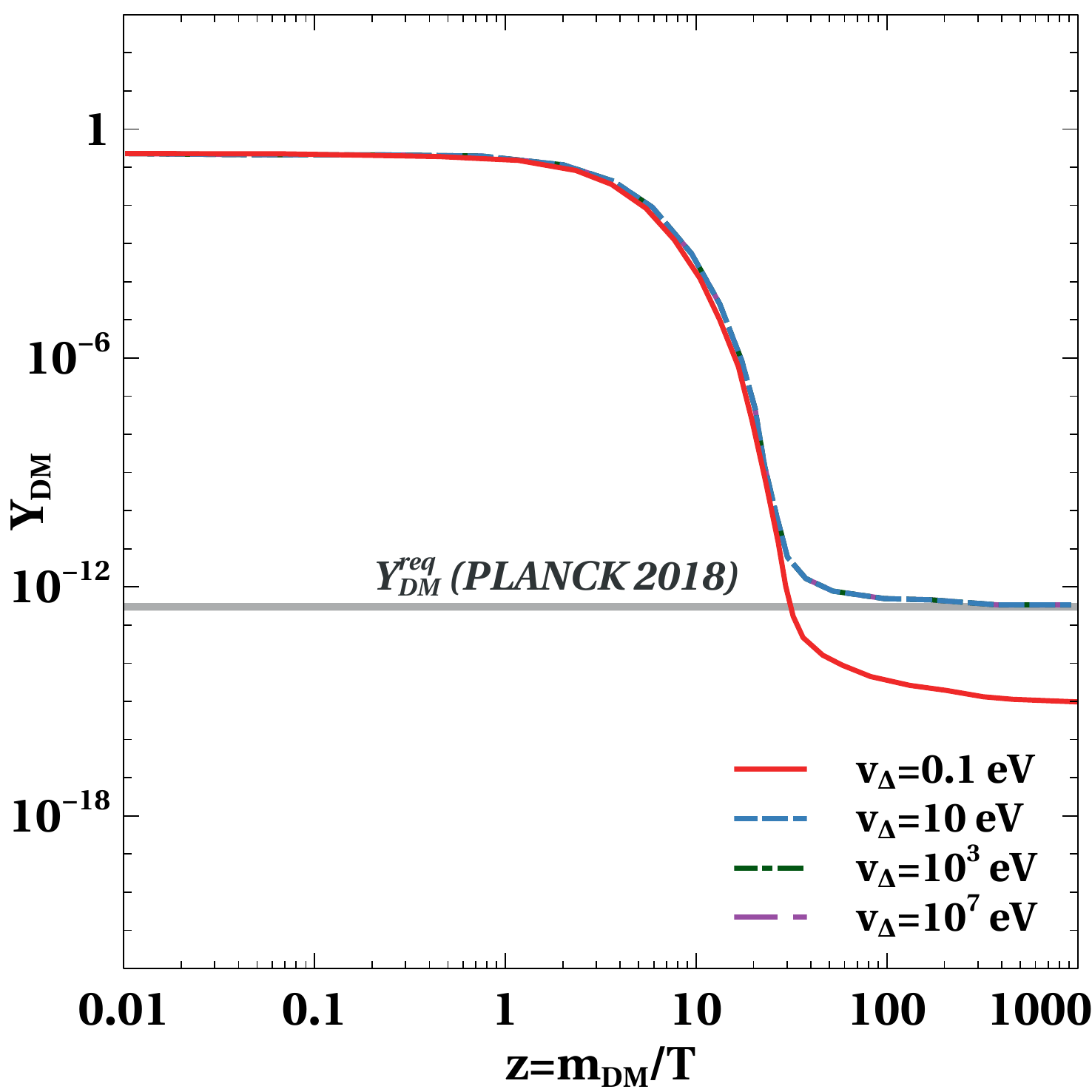}
\includegraphics[scale=.45]{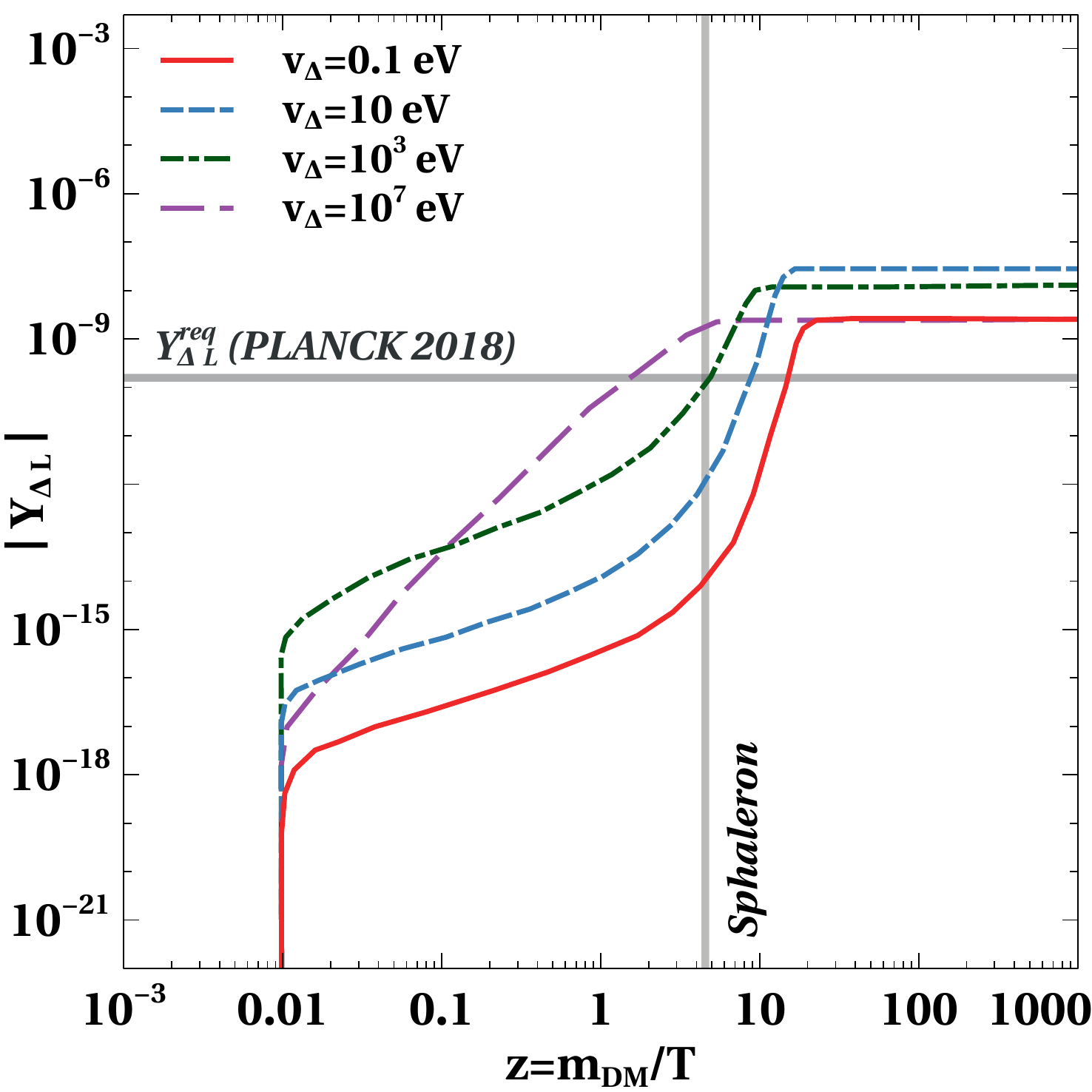}
\caption{Evolution of the comoving number density of dark matter (left panel) and $L$ asymmetry (right panel) with $z=m_{\rm DM}/T$ for different values of $\mu_{\eta \Delta}$ (top panel) and $v_{\Delta}$ (bottom panel) respectively. The other relevant parameters are set at $m_{\eta_{R} }=m_{\rm DM}=600$ GeV, $m_{\Delta^{\pm}}=m_{\Delta^{\pm \pm}}=m_{\Delta^{0}}=1.2$ TeV, $M_{1}=6$ TeV, $M_{j+1}/M_{j}=1.1$, $\lambda_{H\eta}^{''}=1\times 10^{-5}$, $v_\Delta=1$ keV (for the top panel) and $\mu_{\eta \Delta}=10i$ GeV (for the bottom panel).}
\label{standard_cosmo}
\end{figure}

In Fig. \ref{standard_cosmo} we show the evolution of the comoving number densities of DM and $L$ asymmetry with $z=m_{\rm DM}/T$ for different benchmark values of $\mu_{\eta \Delta}$ (upper panel) and $v_{\Delta}$ (lower panel). The grey vertical lines labelled as "Sphaleron" in the right panel plots of Fig. \ref{standard_cosmo} (and the subsequent evolution plots for lepton asymmetry) correspond to the sphaleron freeze-out temperature $T_{\rm Sph}=(131.7 \pm 2.3)$ GeV \cite{DOnofrio:2014rug}.  In the upper left panel plot of Fig. \ref{standard_cosmo}, it can be seen that with change in $\mu_{\eta \Delta}$ there is no change in the evolution of the comoving number density of DM, which is expected as it is governed by total annihilation rates dominated by electroweak gauge portal interactions. However, in the upper right panel of Fig. \ref{standard_cosmo}, sharp variation in comoving lepton number density is seen for different values of $\mu_{\eta \Delta}$. This is expected as this trilinear term is assumed to be the only term violating CP and hence the lepton asymmetry initially increases with increase in $\mu_{\eta \Delta}$. However, it is found that beyond a certain value of $\mu_{\eta \Delta}$ the asymmetry starts decreasing with further increase in $\mu_{\eta \Delta}$. This is because beyond a certain value of $\mu_{\eta \Delta}$, the mass difference between $\eta_{R}$ and $\eta_{I}$ starts increasing, which in turn decreases the neutrino Dirac Yukawa couplings $Y^{N}$ to maintain the radiative seesaw contribution to neutrino mass. Since the CP asymmetry depends upon both $\mu_{\eta \Delta}$ and $Y^N$, their relative increase and decrease lead to an overall decrease in lepton asymmetry at some point. On the other hand, in the lower left panel plot of Fig. \ref{standard_cosmo} we can see that the DM relic decreases beyond a certain small value of $v_{\Delta}$. The is because, beyond a certain small value of $v_{\Delta}$, the Yukawa coupling of leptons with triplet scalar namely, $Y^{\Delta} \propto v_{\Delta}^{-1} $ become large enough (for a fixed contribution of type II seesaw to neutrino mass) such that the annihilation of $\eta$ through the scalar triplet becomes much more dominant compared to the annihilations involving the the electroweak gauge bosons. Therefore, the DM relic is primarily determined by the strong annihilations involving the Yukawa $Y^{\Delta}$. From the lower right plot of Fig. \ref{standard_cosmo} it can be seen that with the increase in $v_{\Delta}$, the asymmetry first increases, but beyond a certain value of $v_{\Delta}$ the asymmetry decreases with increasing $v_{\Delta}$. As $v_{\Delta}$ increases, the Yukawa $Y^{\Delta}$ decreases (for a fixed contribution of type II seesaw to neutrino mass) and therefore the washout due to $\ell \ell \longrightarrow \eta \eta$ decreases which leads to an increase in the asymmetry. However, beyond a certain large value of $v_{\Delta}$ the washout effects become very small. At the same time due to the decrease in $Y^{\Delta}$ the generation of asymmetry itself becomes small leading to a decrease in asymmetry with further increase in $v_{\Delta}$.

\begin{figure}[h]
\includegraphics[scale=.5]{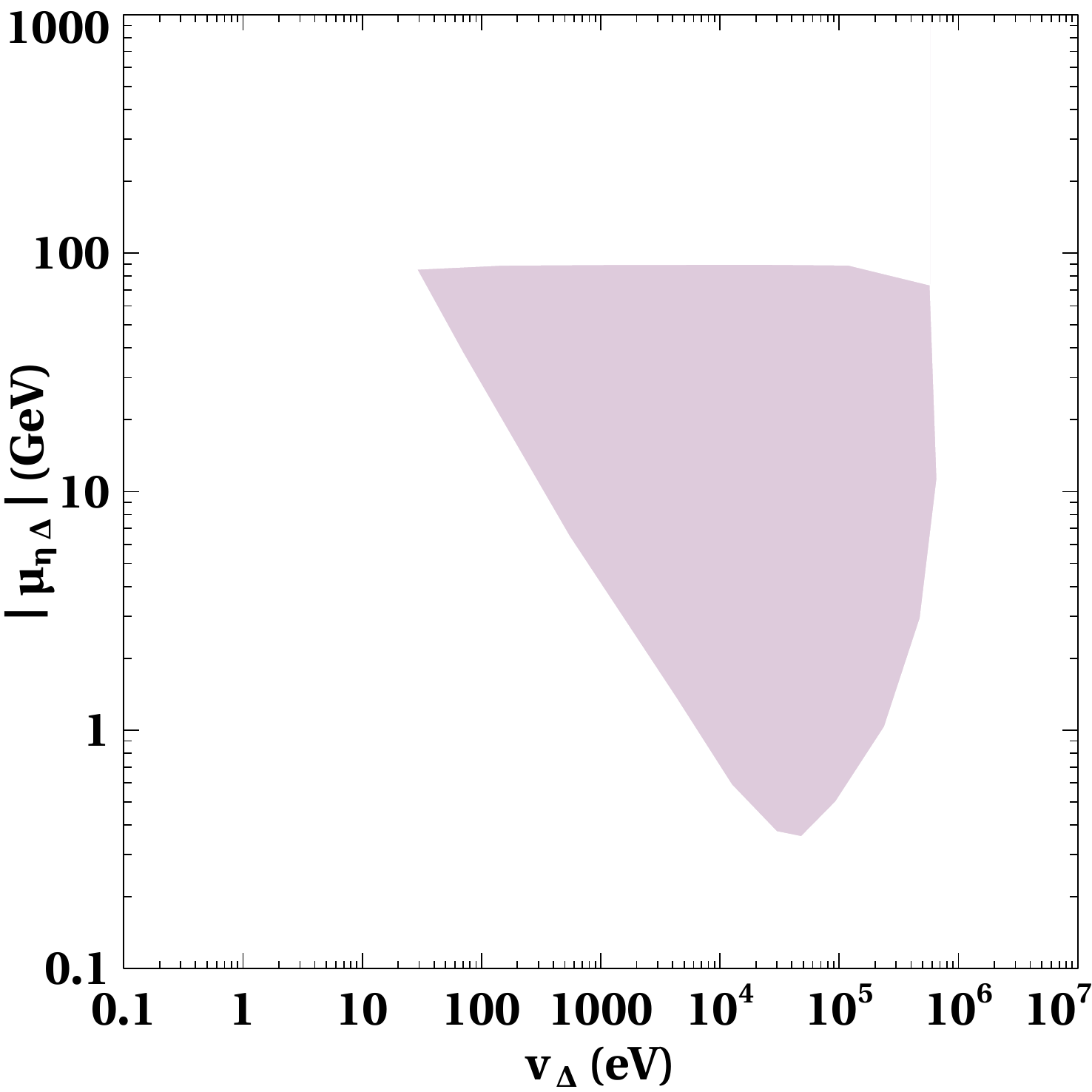}
\caption{Viable parameter space in $\mu_{\eta\Delta}$ versus $v_{\Delta}$ plane which can
generate the observed baryon asymmetry as well as correct DM relic. The other important parameters are fixed at $m_{\eta_{R}}=600$ GeV, $m_{\Delta^{\pm}}=m_{\Delta^{\pm \pm}}=m_{\Delta^{0}}=1.2$ TeV, $\lambda^{''}_{H\eta}=1 \times 10^{-5}$, $M_{1}=6$ TeV, and $M_{j+1}/M_{j}=1.1$.}
\label{fig:standard_scan}
\end{figure}

\begin{figure}[h]
 \includegraphics[scale=0.5]{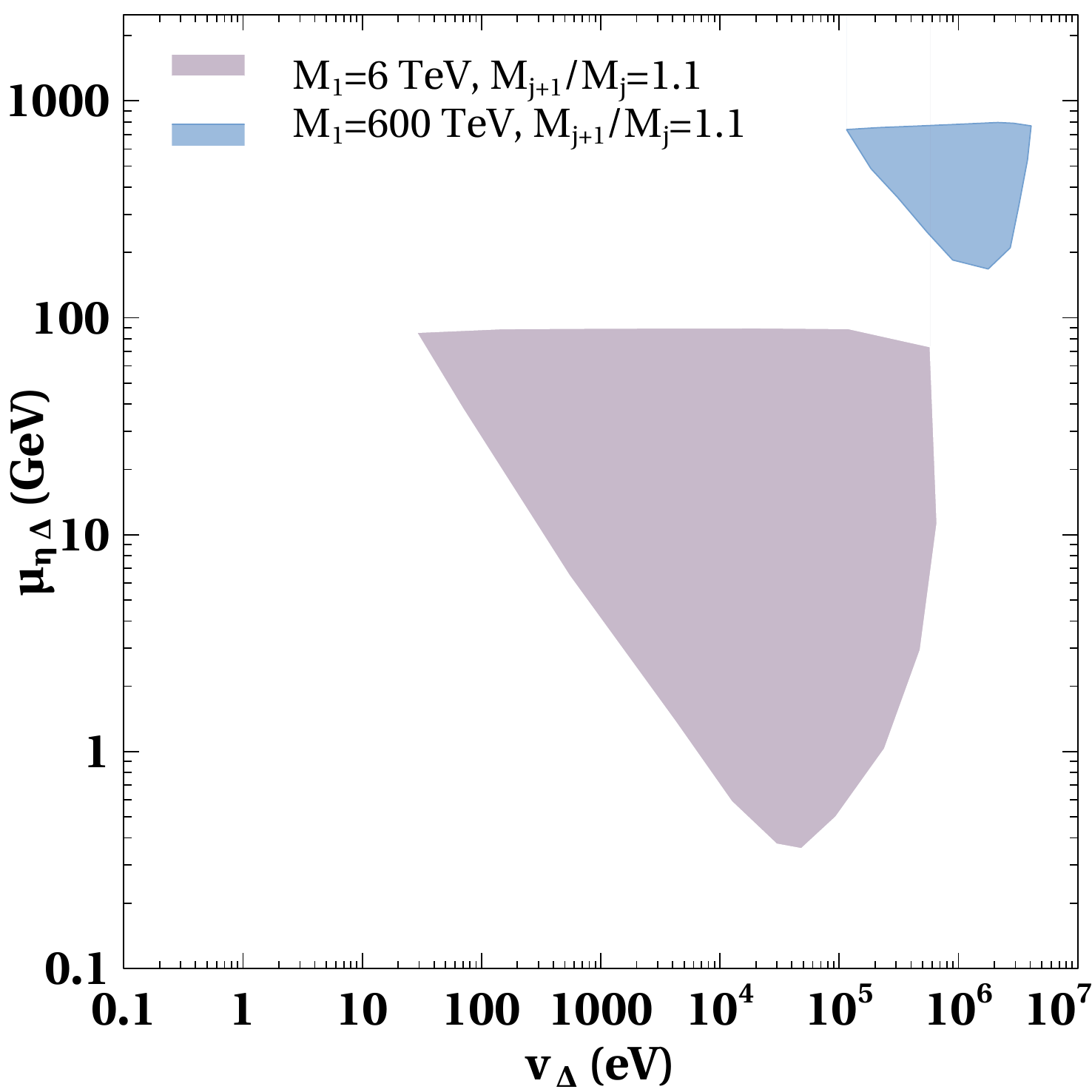}
 \includegraphics[scale=0.5]{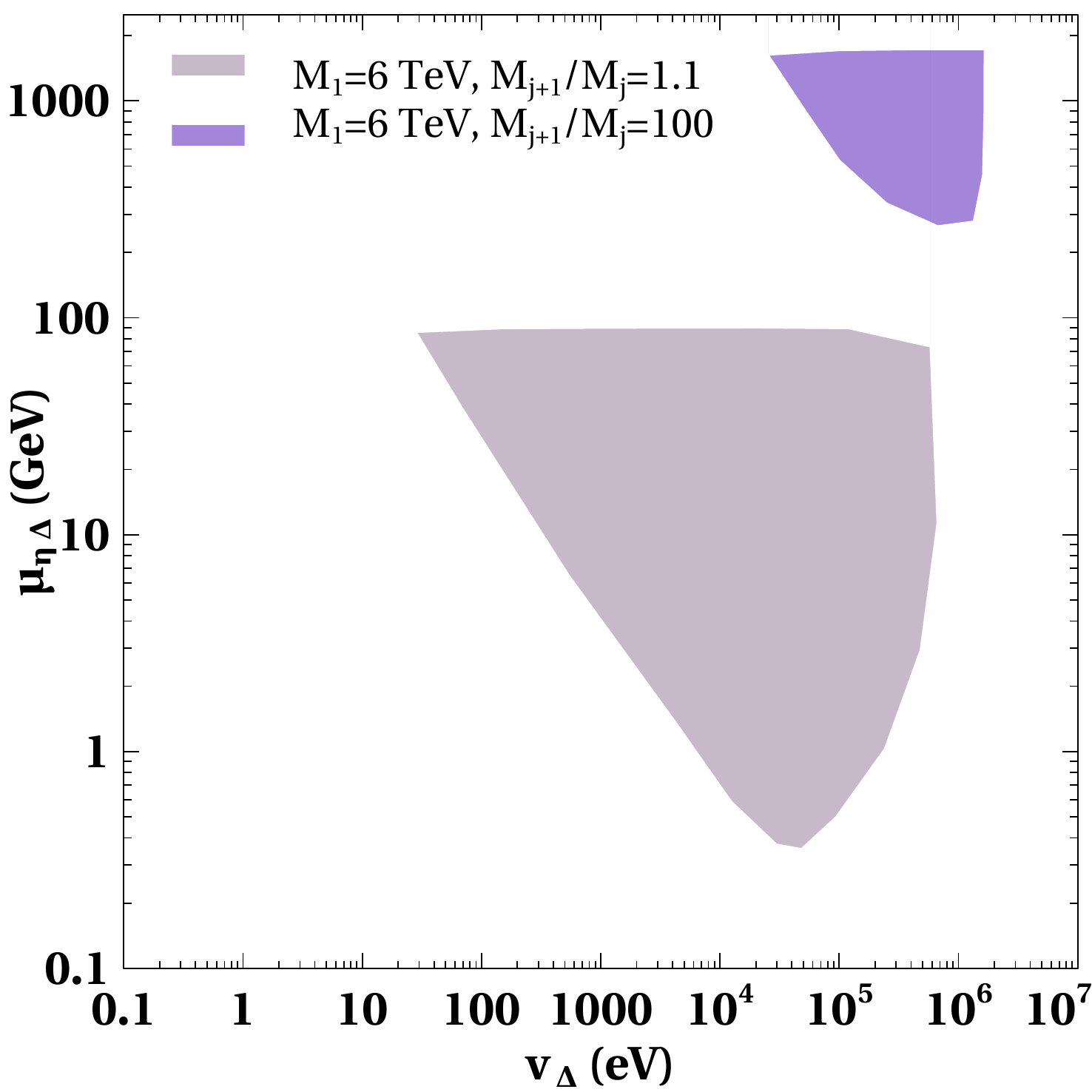}
 \caption{Viable parameter space in $\mu_{\eta \Delta}$ versus $v_{\Delta}$ plane which can generate the observed baryon asymmetry as well as correct DM relic. The other important parameters are fixed at $m_{\eta_{R}}=600$ GeV, $m_{\Delta}^{\pm}=m_{\Delta^{\pm \pm}}=m_{\Delta^{0}}=1.2$ TeV, $\lambda_{H\eta}^{''}=1 \times 10^{-5}$.}
 \label{fig:scan_comparison}
\end{figure}

In Fig. \ref{fig:standard_scan} we show the viable parameter space from successful WIMPy leptogenesis and correct DM relic in $\mu_{\eta \Delta}$ versus $v_{\Delta}$ plane while other relevant parameters are fixed at benchmark values and assuming a standard radiation dominated universe. From Fig. \ref{fig:standard_scan} one can clearly see that there is an upper limit as well as a lower limit on the allowed values of $v_{\Delta}$ and $\mu_{\eta \Delta}$. Quantitatively the bounds on $v_{\Delta}$ and $\mu_{\eta \Delta}$ are found to be $30$ eV $\lesssim v_{\Delta} \lesssim 0.7$ MeV and $0.25$ GeV $\lesssim \mu_{\eta \Delta} \lesssim$ $88$ GeV respectively. When $v_{\Delta}$ is very
small, the Yukawa coupling $Y^{\Delta} \propto v_{\Delta}^{-1}$ is so large that
the washout effects coming from the processes $\ell \ell \longrightarrow \eta \eta$ are too strong to give rise to correct asymmetry. On the other hand when $v_{\Delta}$ is large the Yukawa coupling $Y^{\Delta}$ become too small to generate the sufficient asymmetry. One would expect that the decrease in $Y^{\Delta}$ due to the increase in $v_{\Delta}$ can be compensated by increasing $\mu_{\eta \Delta}$, however, that is not to be true as increasing $\mu_{\eta \Delta}$ also increases the mass splitting between $\eta_{R}$ and $\eta_{I}$ which in turn decreases the Yukawa coupling $Y^{N}$. It should be noted that, lepton asymmetry can, in principle, be generated from $\Delta$ decay as well. However, since WIMP annihilation is generating asymmetry at a lower scale, the high scale production of lepton asymmetry remains sub-dominant, as we show in Appendix \ref{sec:appen3}.

In Fig. \ref{fig:scan_comparison}, we show how the viable parameter space in $\mu_{\eta \Delta}$ versus $v_{\Delta}$ plane changes with the change in the mass of $N_{1}$ (left panel plot) and with the change in mass hierarchy among the RHNs (right panel plot). It is observed that with the increase in $M_{1}$ as well as the mass hierarchy we require larger $\mu_{\eta \Delta}$ to generate the desired asymmetry. This is due to the propagator suppression of the CP asymmetry coming from the $t$-channel diagram shown in Fig. \ref{fig:annihilation}. This can be compensated by increasing the value of $\mu_{\eta \Delta}$. Also with the increase in $\mu_{\eta \Delta}$ the washouts increase which results in a requirement of a larger $v_{\Delta}$. There exists an interplay of these two effects as a result of which the parameter space shift towards the higher values of $\mu_{\eta \Delta}$ as well as $v_{\Delta}$.

\section{WIMPy leptogenesis in non-standard cosmology}
\label{sec:wimpy2}
In this section, we study the impact of non-standard cosmology on WIMPy leptogenesis within the framework of the minimal model mentioned above. As mentioned before, we consider three different types of such non-standard histories namely, a fast expanding universe, an early matter dominated universe and scalar tensor theory of gravity which we discuss separately below.

\subsection{Fast expanding universe}

We first study the WIMPy leptogenesis in a universe which is dominated by some scalar field $\phi$, such that its energy density falls faster than radiation, known as the fast expanding universe \cite{DEramo:2017gpl}. In this scenario, the energy density of the field $\phi$, falls with the scale factor as 
\begin{equation}
\rho_{\phi} \propto a^{-(4+n)},
\end{equation}
with $n \ge 0$. Clearly, $n=0$ corresponds to the usual radiation domination. Therefore, the field $\phi$ naturally dominates the energy density of the universe at very early epochs with the usual radiation dominated era taking over $\phi$ eventually. Thus, the total energy density of the universe in FEU scenario can, in general, be written as a sum of radiation and $\phi$ contribution
\begin{equation}
 \rho(T)=\rho_{\rm rad}(T)+\rho_{\phi}(T),
\end{equation} 
where the usual radiation density can be written as 
\begin{equation}
\rho_{\rm rad}=\dfrac{\pi^{2}}{30}g_{*}(T)T^{4}.
\end{equation}
Considering the equation of state for $\phi$ field to be $p_{\phi}=\omega_{\phi} \rho_{\phi}$, one can find $n=3\omega_{\phi}-1$. It should be noted that $\phi$ does not have any interactions with the SM bath and it only plays the role of a spectator by contributing to the energy density (and not to the entropy density) of the universe. In order to reproduce a radiation dominated universe during BBN era, the equality between the energy density of $\phi$ and radiation must happen at a temperature $T_{\rm r}\gtrsim T_{\rm BBN}$. The energy density of $\phi$ can be written as 
\begin{equation}
\rho_{\phi}(T)=\rho_{\phi}(T_{\rm r})\left( \dfrac{g_{*s}(T)}{g_{*s}(T_{\rm r})} \right)^{(4+n)/3} \left( \dfrac{T}{T_{\rm r}} \right)^{4+n},
\end{equation}
which also leads to the total energy density as 
\begin{equation} \label{eq:54}
\rho(T)=\rho_{\rm rad}(T)+\rho_{\phi}(T)=\rho_{\rm rad}(T) \left[ 1+\dfrac{g_{*}(T_{r})}{g_{*}(T)} \left( \dfrac{g_{*s}(T)}{g_{*s}(T_{\rm r})} \right)^{(4+n)/3}  \left( \dfrac{T}{T_{\rm r}}  \right) ^{n} \right].
\end{equation}
Considering $g_{*s}(T)=g_{*}(T)$ for most of the history of the universe the Hubble parameter can be calculated to be 
\begin{equation} \label{eq:55}
{\bf H}(T)\simeq \dfrac{\pi g_{*}^{1/2}(T)T^{2}}{3 \sqrt{10} M_{\rm Pl}} \left[ 1+ \left( \dfrac{g_{*}(T)}{g_{*}(T_{\rm r})} \right)^{(1+n)/3} \left( \dfrac{T}{T_{\rm r}} \right) ^{n} \right]^{1/2}.
\end{equation}

In FEU scenario, the Boltzmann equation for WIMP type DM, in terms of its comoving number density, reads as\cite{DEramo:2017gpl}

\begin{equation} \label{eq:56}
\dfrac{dY}{dz}=-A \dfrac{ \langle \sigma v_{\rm rel} \rangle}{z^{3}L \left[n,z,z_{r} \right]}\left[ Y^{2}-Y_{\rm eq}^{2} \right] ,
\end{equation} 
where, $A=\dfrac{s(z=1)}{{\bf H}_{\rm rad}(z=1)}=\dfrac{2\sqrt{2}\pi}{3\sqrt{5}}g_{*}^{1/2}m_{\rm DM} M_{\rm Pl}$ and the function $L \left[n,z,z_{r} \right]$ has the form 
\begin{equation}
L \left[n,z,z_{r} \right]=\left({n+4}\right)\left[\dfrac{1}{z^4}+\left(\dfrac{g_{*}(z)}{g_{*}(z_{r})} \right)^{(1+n)/3}\dfrac{z_{r}^n}{z^{n+4}} \right] ^{3/2}\left[ \dfrac{4}{z^5}+(4+n)\left( \dfrac{g_{*}(z)}{g_{*}(z_{r})}\right)^{(1+n)/3}\dfrac{ z_{r}^n}{z^{n+5}} \right]^{-1}.
\end{equation}
In WIMPy leptogenesis scenario, we need to write the Boltzmann equation for $L$ asymmetry too along with the one for DM. Assuming the universe to be dominated by the $\phi$ field only till the WIMP freeze-out that is, $T\gg T_{\rm r}$, we can simply write the relevant Boltzmann equations as

\begin{eqnarray}
\dfrac{dY_{\eta}}{dz} & = & -\dfrac{s(z=1)}{z^{2-n/2}z_{r}^{n/2}{\bf H}_{\rm rad}(z=1)}\langle \sigma v\rangle_{\eta \eta \longrightarrow {\rm SM \; SM}} \left[  Y_{\eta}^{2}-(Y_{\eta}^{\rm eq})^{2} \right] \\
\dfrac{dY_{\Delta L}}{dz} & = & \dfrac{s(z=1)}{z^{2-n/2}z_{r}^{n/2}{\bf H}_{\rm rad}(z=1)} \langle \sigma v \rangle^{\delta}_{\eta \eta \longrightarrow \ell \ell} \left[  Y_{\eta}^{2}-(Y_{\eta}^{\rm eq})^{2} \right] \nonumber \\ & & -\dfrac{s(z=1)}{z^{2-n/2}z_{r}^{n/2}{\bf H}_{\rm rad}(z=1)} Y_{\Delta L} Y_{l}^{\rm eq} r_{\eta}^{2} \langle \sigma v\rangle_{\eta \eta \longrightarrow \ell \ell}   \nonumber \\ & & -\dfrac{s(z=1)}{z^{2-n/2}z_{r}^{n/2}{\bf H}_{\rm rad}(z=1)} Y_{\Delta L}Y_{\eta}^{\rm eq} \langle \sigma v \rangle_{\eta \bar{\ell} \longrightarrow \eta \ell}.
\end{eqnarray}
While we write the simplified form of equations here, we consider the complete function $L(n, z, z_r)$ in the numerical calculations.

\begin{figure}[h]
\begin{center}
\includegraphics[scale=.45]{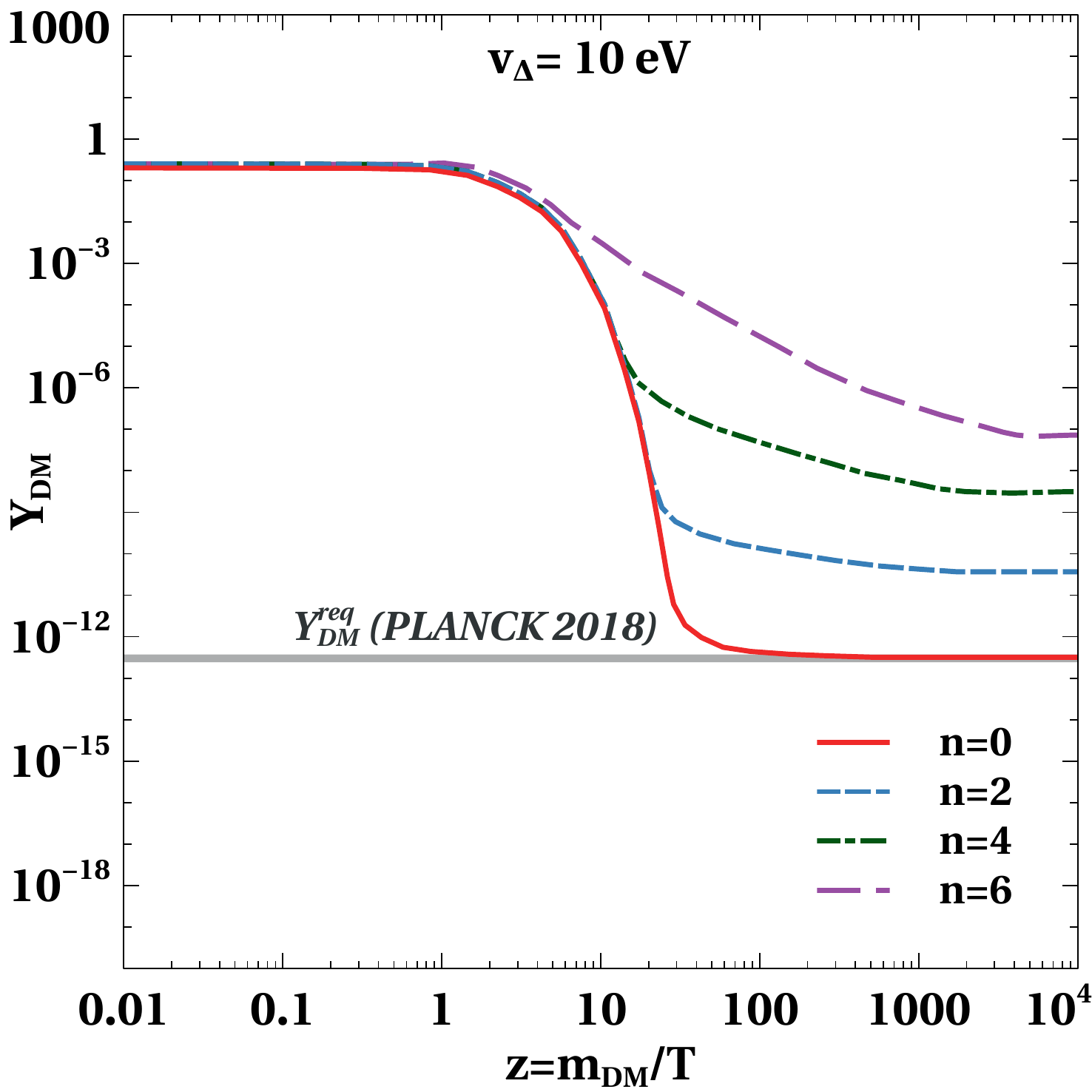}
\includegraphics[scale=.45]{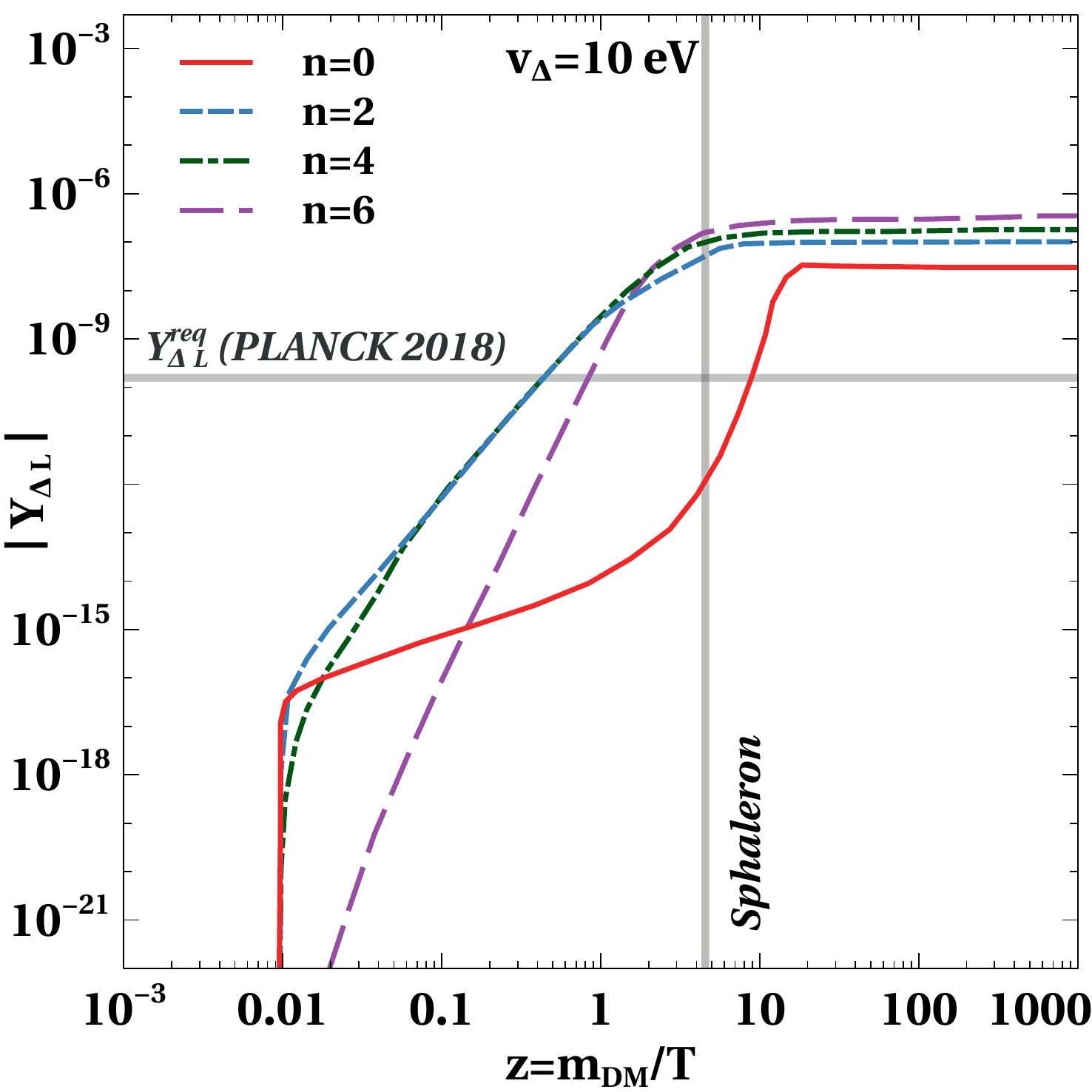}
\includegraphics[scale=.45]{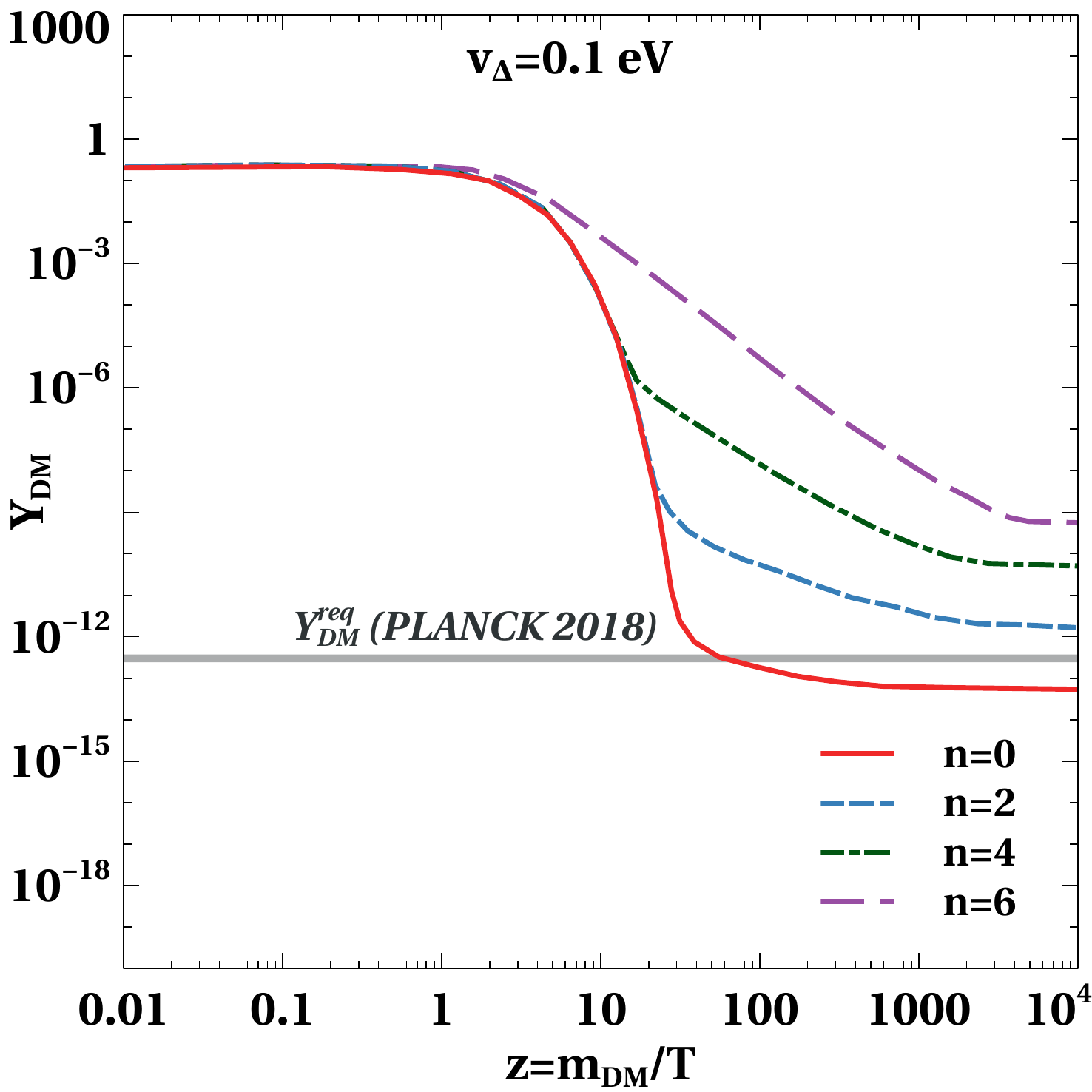}
\includegraphics[scale=.45]{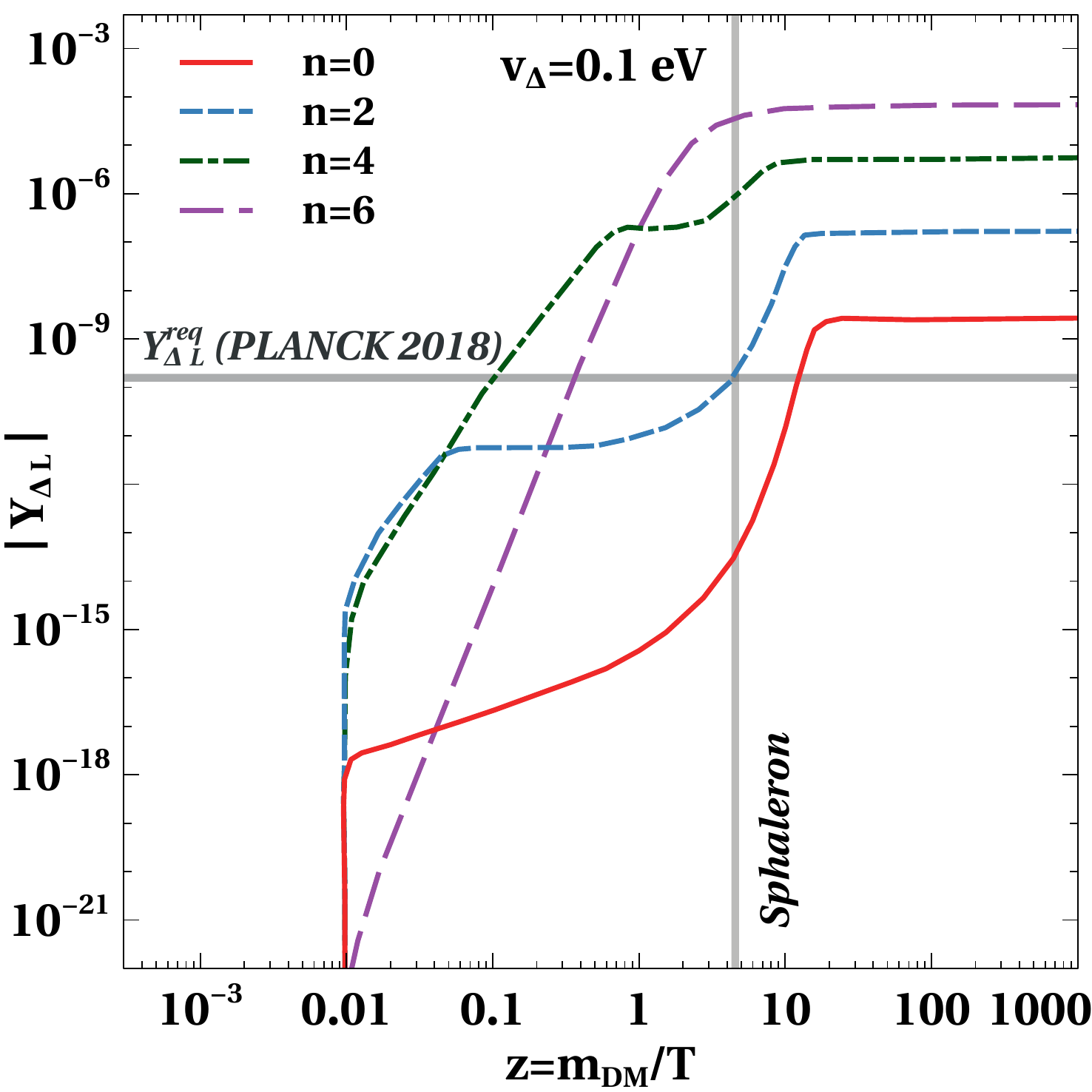}
\caption{Evolution of the comoving number densities of dark matter (left panel) and $L$ asymmetry (right panel) with $z=m_{\rm DM}/T$ for different values of FEU parameter n. The relevant parameters are set at $m_{\eta R}=600$ GeV,  $m_{\Delta^{\pm}}=m_{\Delta^{\pm \pm}}=m_{\Delta^{0}}=1.2$ TeV,$M_{1}=6$ TeV, $M_{j+1}/M_{j}=1.1$, $\mu_{\eta \Delta}=10i$ GeV, $\lambda_{H\eta}^{''}=1\times 10^{-5}$ and $v_{\Delta}=10$ eV (upper panel) and $v_{\Delta}=0.1$ eV (lower panel). Here we have taken $T_{r}=30$ MeV ($z_{r} \simeq 2\times 10^{4}$).   }
\label{fig:asym_DM_NSC}
\end{center}
\end{figure}


\begin{figure}
\includegraphics[scale=.45]{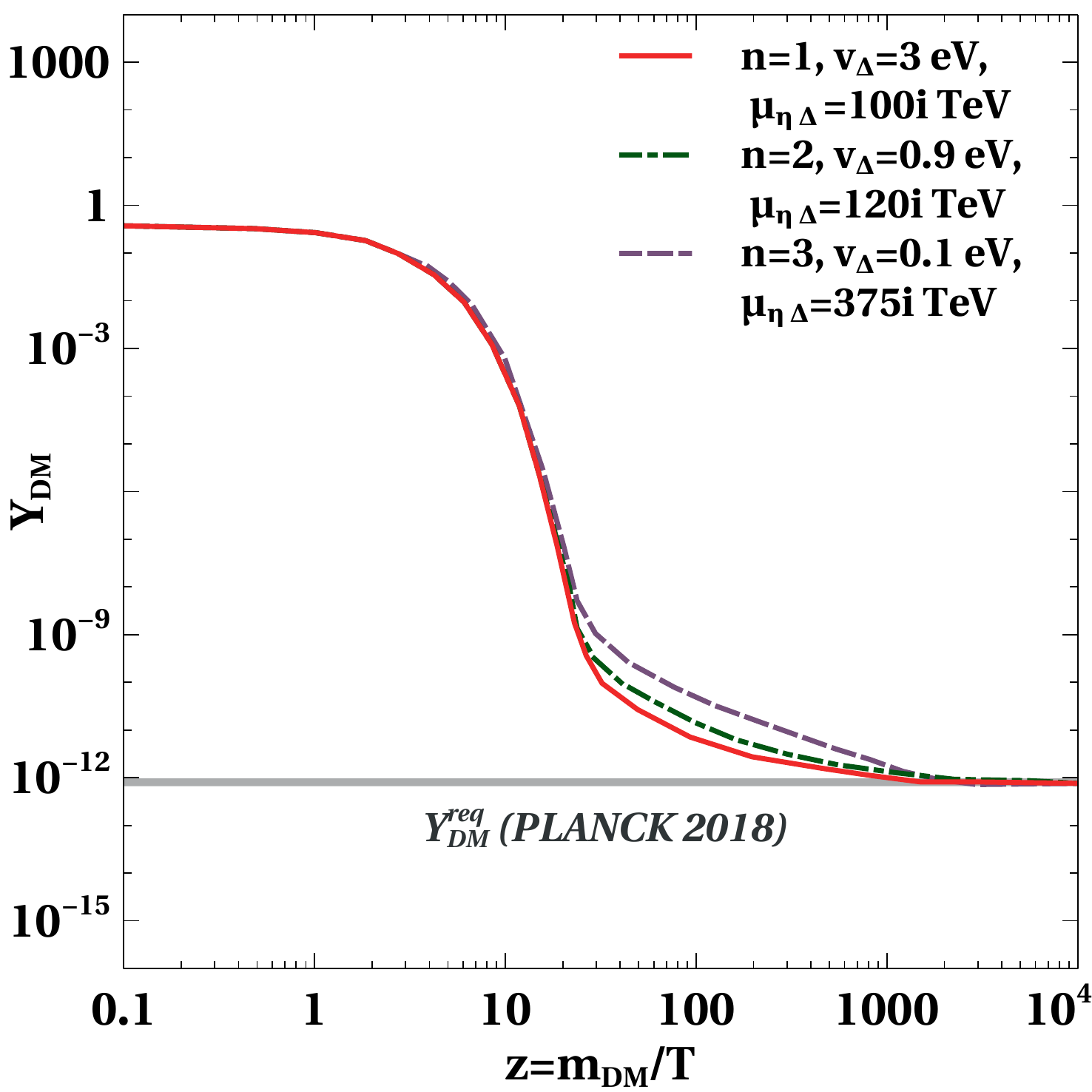}
\includegraphics[scale=.45]{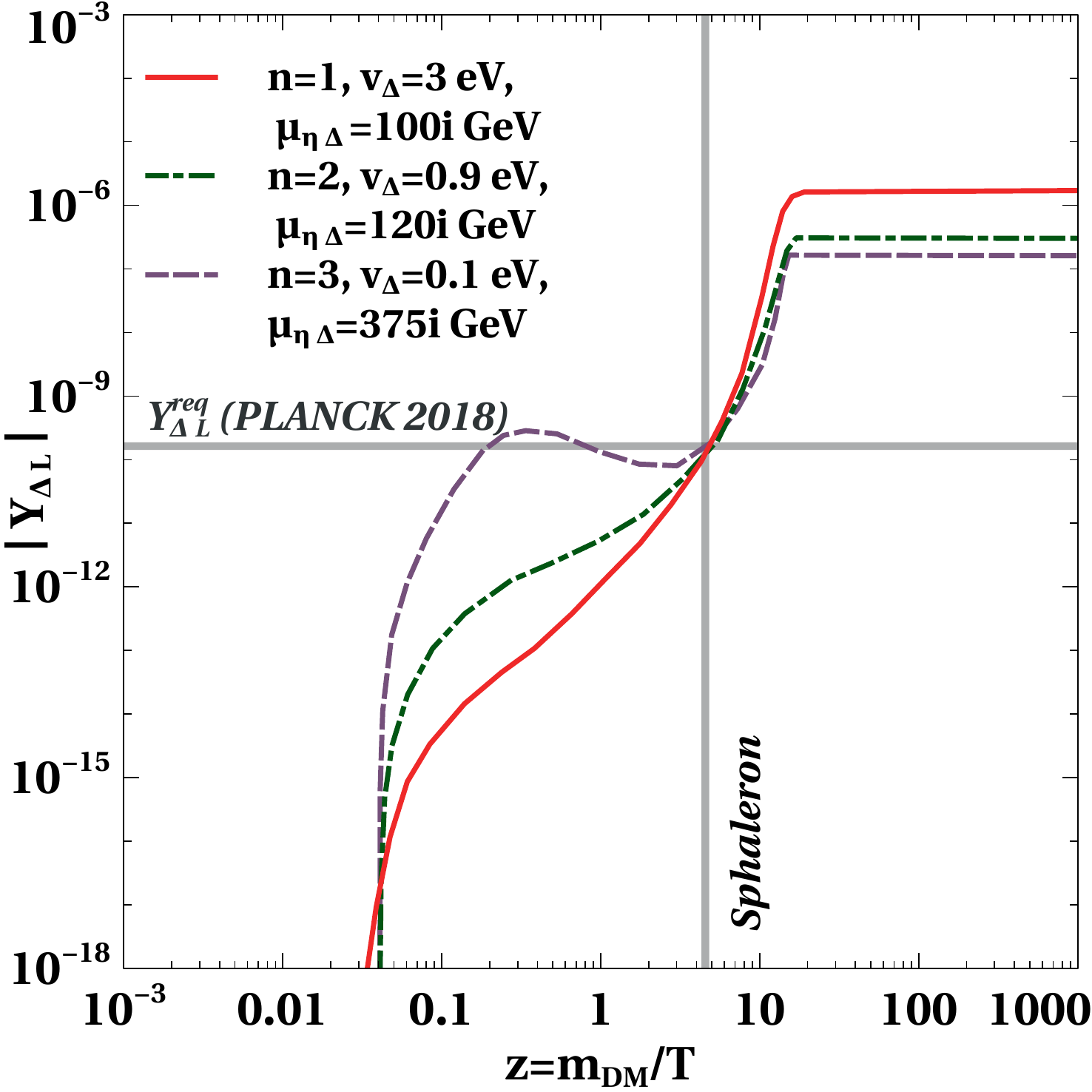}
\caption{Evolution of the comoving number densities of DM and $L$ asymmetry for few benchmark points which can generate the correct DM relic and the observed asymmetry. The other relevant parameters are fixed at $m_{\eta_{R}}=600$ GeV, $\lambda_{H\eta}^{''}=10^{-5}$ ,$m_{\Delta^{0}}=m_{\Delta^{\pm}}=m_{\Delta^{\pm \pm}}=1.2$ TeV,$M_{1}=6$ TeV, $M_{j+1}/M_{j}=1.1$ and $z_{r}=2\times 10^{4}$.}
\label{fig:benchmark_correct}
\end{figure}

In Fig. \ref{fig:asym_DM_NSC} we show the evolution of the comoving number density of dark matter and $L$ asymmetry for different values of the FEU parameter $n$ keeping the other important parameters fixed. From the DM relic plots one can see that with increasing values of $n$, the DM abundance also increases. It is because, larger the value of $n$, larger is the expansion rate and therefore the decoupling of the DM particles happen much earlier. However, since the decoupling occurs at a time when the rate of annihilations are sufficiently large, a few DM particles keep on annihilating upto a much later time giving rise to a relentless nature of the DM abundance, as pointed out in \cite{DEramo:2017gpl}. Due to the early deviation of $\eta$ from its equilibrium abundance the asymmetry also increases with the increase in $n$. Also the rates of washout processes become relatively suppressed with the increase in $n$ because of the faster expansion. This effect is clearly visible in the lower right panel plot of Fig. \ref{fig:asym_DM_NSC} where the washouts are relatively strong compared to the upper right panel plot. In the lower panel plots of Fig. \ref{fig:asym_DM_NSC} we choose relatively smaller value of $v_{\Delta}=0.1$ eV which leads to very strong annihilations of $\eta$ by the process $\eta \eta \longrightarrow ll$. This leads to two distinct effects on DM abundance as well as in the asymmetry. For such value of $v_{\Delta}$ the DM relic is primarily determined by the process $\eta \eta \longrightarrow ll$ involving $Y^{\Delta}$. For small $v_{\Delta}$, the DM relic is less than the observed value in standard cosmolgy as can be seen in the lower left plot of Fig. \ref{fig:asym_DM_NSC} (keeping in mind that $n=0$ leads to standard cosmological history). Since faster expansion leads to an increase in DM relic we expect to achieve the correct relic by increasing the value of $n$. Similarly smaller value of $v_{\Delta}$ increases the washout coming from the process $ll \longrightarrow \eta \eta$. This results in a decrease of the asymmetry for standard history ($n=0$) as can be seen in the lower right panel plot of Fig. \ref{fig:asym_DM_NSC}. Therefore, increasing the value of $n$ can open up new regions of parameter space consistent with the correct DM relic and the observed baryon asymmetry.  In Fig. \ref{fig:benchmark_correct} we show evolution plot of comoving number densities of DM and $L$ for three benchmark points which can satisfy the observed relic density of DM and also the correct $L$ asymmetry in a FEU.

\begin{figure}
\includegraphics[scale=.45]{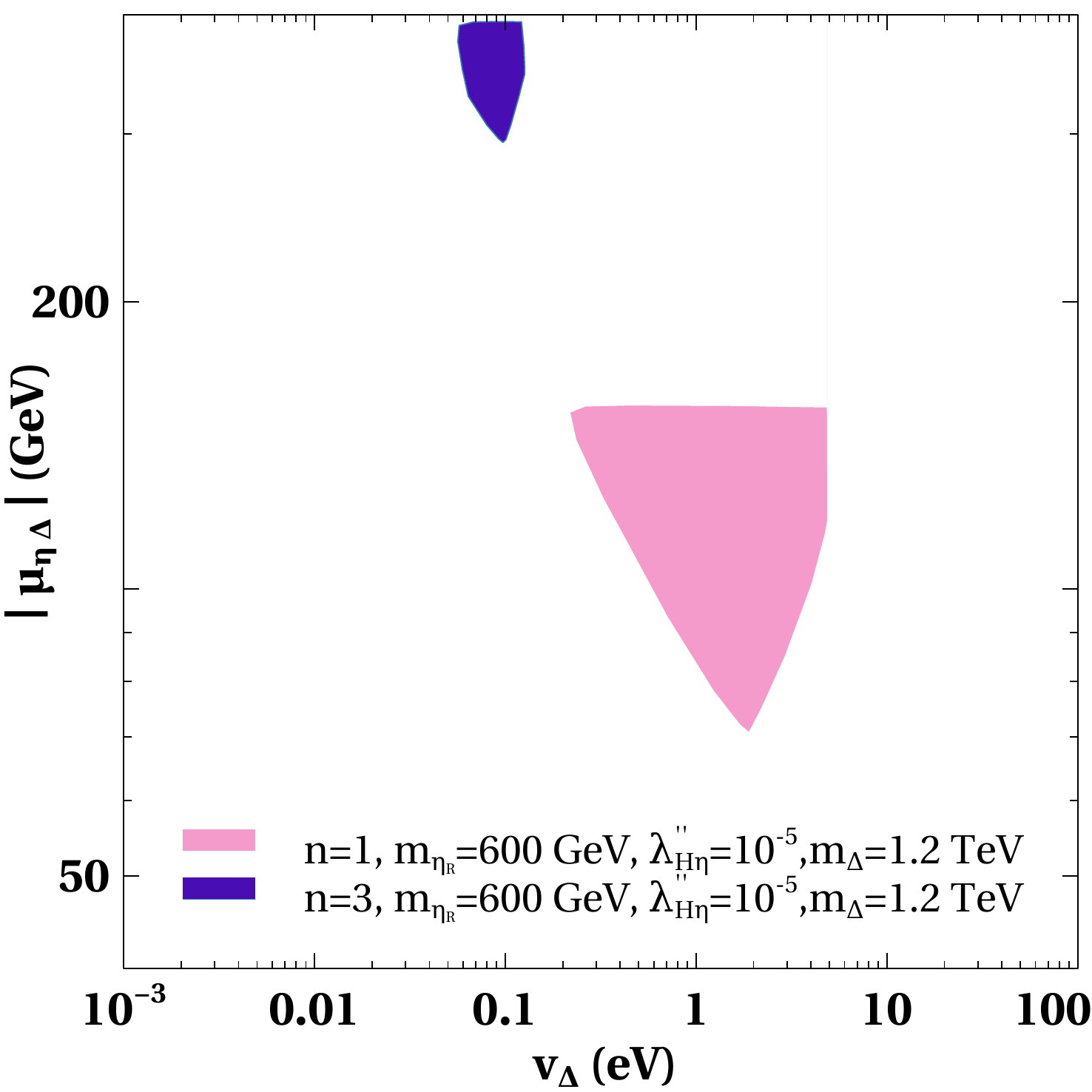}
\caption{Scan plot showing the available parameter space in $\mu_{\eta \Delta}$ versus $v_{\Delta}$ plane from the requirement of observed baryon asymmetry and the correct DM relic. $z_{r}$ is fixed at $2\times 10^{4}$. The other important parameters are fixed at $m_{\eta_{R}}=600$ GeV, $m_{\Delta^{\pm}}=m_{\Delta^{\pm \pm}}=m_{\Delta^{0}}=1.2$ TeV, $\lambda^{''}_{H\eta}=1 \times 10^{-5}$, $M_{1}=6$ TeV, and $M_{j+1}/M_{j}=1.1$. }
\label{fig:NSC_scan}
\end{figure}

In Fig. \ref{fig:NSC_scan} we show the viable parameter space in $\mu_{\eta \Delta}$ versus $v_{\Delta}$ plane in a FEU scenario. We consider two different values of the FEU parameter namely, $n=1$ and $n=3$. One can see that the viable parameter space changes in FEU from the standard radiation case. More specifically, the parameter space gets squeezed in this plane compared to the same in standard cosmology shown in Fig. \ref{fig:standard_scan}. This is due to the different interplay of these two parameters on DM relic as well as washout processes discussed earlier. While we choose $T_r=30$ MeV, choosing larger values will reduce the effect of non-standard cosmology and bring the results closer to the ones in standard radiation dominated scenario discussed earlier.
Finally we found the limits on $v_{\Delta}$ and $\mu_{\eta \Delta}$ to be $0.2$ eV $\lesssim v_{\Delta}\lesssim$ $4$ eV and $67$ GeV $\lesssim \mu_{\eta \Delta} \lesssim $ 143 GeV for $m_{\rm DM}=600$ GeV in a FEU with $n=1$. Similarly the limits are found to be $0.068$ eV $\lesssim v_{\Delta} \lesssim$ $0.13$ eV, $285$ GeV $\lesssim \mu_{\eta \Delta} \lesssim$ $390$ GeV respectively for $m_{\rm DM}=600$ GeV in a FEU with $n=3$.

\begin{figure}
\includegraphics[scale=.5]{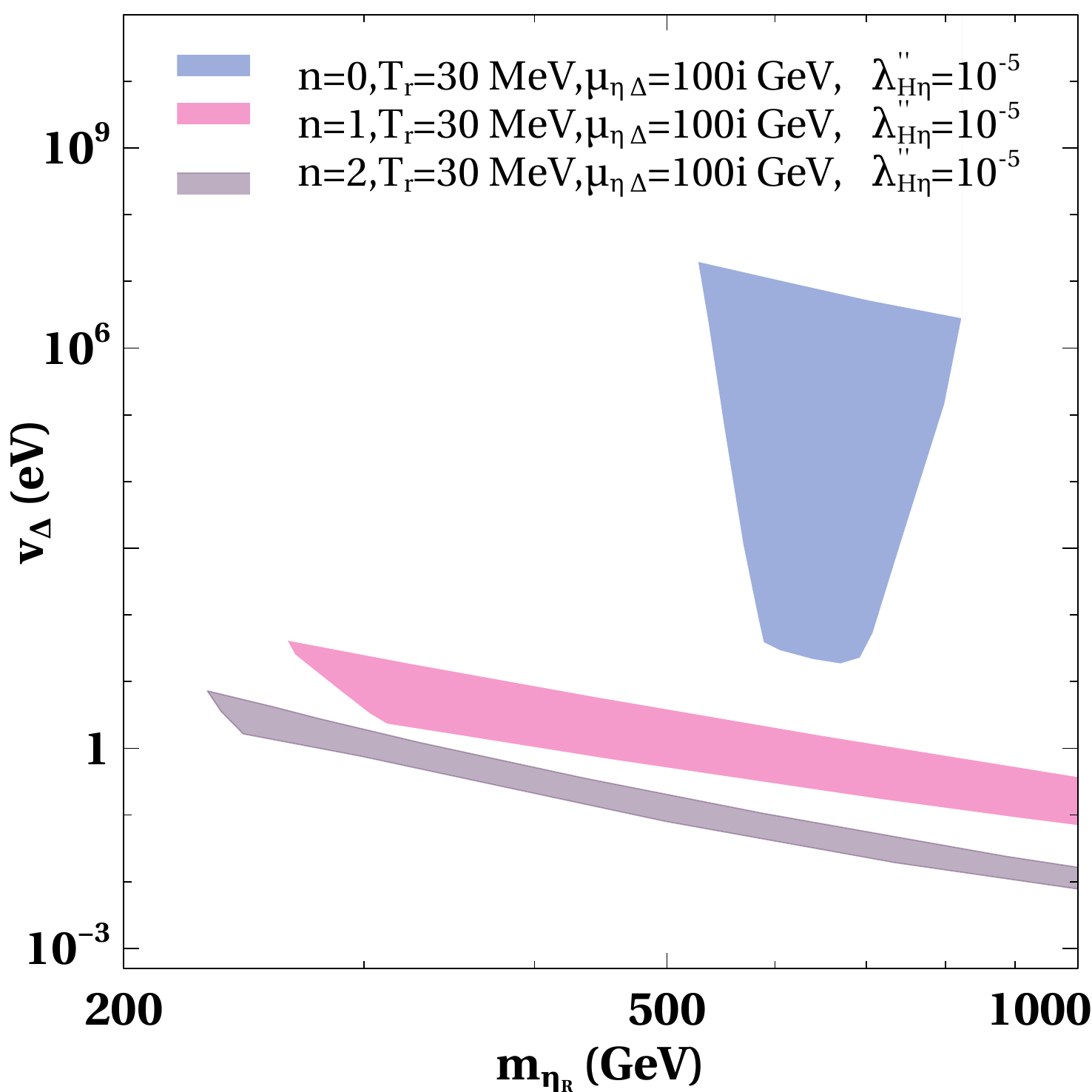}
\caption{Scan plot showing the viable parameter space in $m_{\eta_{R}}$ versus $v_{\Delta}$ plane for different possible FEU. The other important parameters are fixed at $m_{\Delta^{\pm}}=m_{\Delta^{\pm \pm}}=m_{\Delta^{0}}=1.2$ TeV, $M_{1}=6$ TeV, and $M_{j+1}/M_{j}=1.1$.}
\label{fig:mdm_NSC}
\end{figure}

In Fig. \ref{fig:mdm_NSC} we show the the viable parameter space in $m_{\eta_{R}}$ versus $v_{\Delta}$ plane from the requirement of correct DM relic and the observed baryon asymmetry for different FEU parameter $n$. In this figure the region shown by the blue colour corresponds to the $n=0$ case or equivalently to the standard radiation dominated universe. For $n=1,2$ we can see that the scale of leptogenesis can be lower than the standard radiation case. For the standard radiation dominated universe the DM mass required to satisfy the correct relic is very specific due to strong gauge portal annihilations and therefore the viable parameter space is appearing around a specific value of $m_{\eta_{R}}$. However, as seen from Fig. \ref{fig:asym_DM_NSC} for FEU, the DM relic and asymmetry both increases with increase in $n$ and therefore with increase in $n$ we need stronger DM annihilations to satisfy DM relic and baryon asymmetry. Hence for FEU the viable parameter space is appearing with smaller values of $v_{\Delta}$, which make the Yukawa mediated annihilations stronger. This also opens up parameter space with smaller $m_{\eta_{R}}$ compared to the standard case as smaller $m_{\eta_{R}}$ leads to stronger annihilations. Also we observed that with increase in $v_{\Delta}$ the required DM mass decreases, which is expected. However, we can not increase the value of $v_{\Delta}$ arbitrarily to lower the DM mass, because, beyond a certain large value of $v_{\Delta}$ the Yukawa mediated annihilations of DM become subdominant and we have to rely of the standard gauge sector to achieve the correct relic. We found that the lowest possible DM mass is $m_{\eta_{R}}=230$ GeV for $n=1$ and is $m_{\eta_{R}}=275$ GeV for $n=2$ case keeping the other particle physics parameters fixed as shown in Fig. \ref{fig:mdm_NSC}. This is significantly lower than the standard radiation case with the same benchmark parameters. Such low DM mass as well as low scale of leptogenesis can have promising detection prospects from colliders to direct and indirect DM detection experiments.

\subsection{Early matter domination}
In EMD scenario, a matter field is assumed to dominate the energy density of the pre-BBN universe for a certain duration. It can be in the form of a scalar field $\phi$ behaving like ordinary pressure-less matter. The presence of this field in addition to the standard radiation bath changes the expansion rate of the EMD universe compared to the radiation dominated universe of standard cosmology. This is equivalent to the fact that the energy density of the matter field $\rho_{\phi}$ falls with the expansion of the universe at a slower rate compared to the radiation energy density $\rho_{\rm rad}$ as long as $\phi$ does not decay. In principle $\phi$ can decay to both SM radiation and dark sector particles like DM. Here we assume $\phi$ to decay into radiation reproducing the standard cosmological phase while releasing entropy. Since entropy is not conserved, we do not consider ratio of number density to entropy density while writing the relevant Boltzmann equations. Instead, we write in terms of $N = n a^3$ where $n, a$ denote the number density and scale factor respectively. Accordingly, we track the variation in terms of scale factor only. The coupled Boltzmann equations for WIMPy leptogenesis in an EMD universe can be written as follows.

\begin{align}
\dfrac{d N_{\eta}}{da} & = -\dfrac{\langle \sigma v\rangle_{\eta \eta \longrightarrow {\rm SM \, SM}}}{{\bf H} a^{4}} \left[  N_{\eta}^{2}-(N_{\eta}^{\rm eq})^{2} \right], \\ 
\dfrac{d N_{\Delta L}}{da} & = \dfrac{\langle \sigma v\rangle^{\delta}_{\eta \eta \longrightarrow ll} }{{\bf H} a^{4}} \left[ N_{\eta}^{2}-(N_{\eta}^{\rm eq})^{2} \right] - \dfrac{N_{\Delta L}}{{\bf H} a^{4}}N_{l}^{\rm eq}r_{\eta}^{2} \langle \sigma v\rangle_{\eta \eta \longrightarrow ll} -\dfrac{N_{\Delta L}}{{\bf H} a^{4}} N_{\eta}^{\rm eq}\langle \sigma v\rangle_{\eta \bar{l} \longrightarrow \eta l}, \\
\dfrac{d \rho_{\phi}}{da} & = -\dfrac{\Gamma_{\phi}\rho_{\phi}}{{\bf H} a}-\dfrac{3(1+\omega)\rho_{\phi}}{a},  \\ 
\dfrac{dT}{da} & = \left( 1+\dfrac{T}{g_{*s}} \dfrac{dg_{*s}}{dT}  \right)^{-1} \left[-\dfrac{T}{a}+ \dfrac{\Gamma_{\phi}\rho_{\phi}}{3{\bf H}sa} \left ( 1 - b\frac{E}{m_\phi} \right) + \dfrac{2}{3} \dfrac{E \langle \sigma v \rangle_{\eta \eta \longrightarrow {\rm SM \, SM}}}{{\bf H}s a^{7}} \left[  N_{\eta}^{2}-(N_{\eta}^{\rm eq})^{2} \right] \right], \\  
\dfrac{ds}{da}+\dfrac{3s}{a} & = \dfrac{\Gamma_{\phi}\rho_{\phi}}{T {\bf H}a}+\dfrac{2E}{T {\bf H} a} \dfrac{\langle \sigma v\rangle_{\eta \eta \longrightarrow {\rm SM \, SM}}}{a^{6}} \left[ N_{\eta}^{2}-(N_{\eta}^{\rm eq})^{2}  \right]. 
\end{align}

Here, $r_{\eta}$ is defined by $r_{\eta}=N_{\eta}^{\rm eq}/N_{l}^{\rm eq}$. Here, $\omega$ is the equation of state parameter for the new species $\phi$ which under the assumption of being a pressure-less matter field yields $\omega=0$. The Hubble parameter and the relativistic degrees of freedom contributing to the entropy density of the universe are represented by the usual symbols ${\bf H}$ and $g_{*s}$ respectively. The Hubble parameter, in general, is given by 
\begin{equation}
{\bf H}(a)=\sqrt{\dfrac{\rho_{\phi}(a)+\rho_{\rm rad}(a)}{3M_{\rm Pl}^{2}}}
\end{equation} 
E represents the average thermal energies of the DM particles and is given by $E=\sqrt{m_{\rm DM}^{2}+3T^{2}}$.  Here, $b$ is twice the branching ratio of $\phi$ decaying into a couple of DM particles, which is assumed to be zero in our setup. The decay width of the matter field is parametrised as \cite{Arias:2019uol}
\begin{equation}
\Gamma_{\phi}=\sqrt{\dfrac{\pi^{2}g_{*}(T_{\rm end})}{90M_{\rm Pl}^{2}}}T_{\rm end}^{2},
\end{equation} 
where $T_{\rm end}$ is the temperature at which the matter field decays into the radiation. There are two important cosmological parameters in this scenario which are the ratio of $\phi$ energy density to that of the radiation at the initial temperature $k=\rho_{\phi}^{\rm in}/\rho_{\rm rad}^{\rm in}$ and the  $T_{\rm end}$. Therefore we study the impact of these parameters on the asymmetry and DM relic. 

\begin{figure}[h]
\includegraphics[scale=.45]{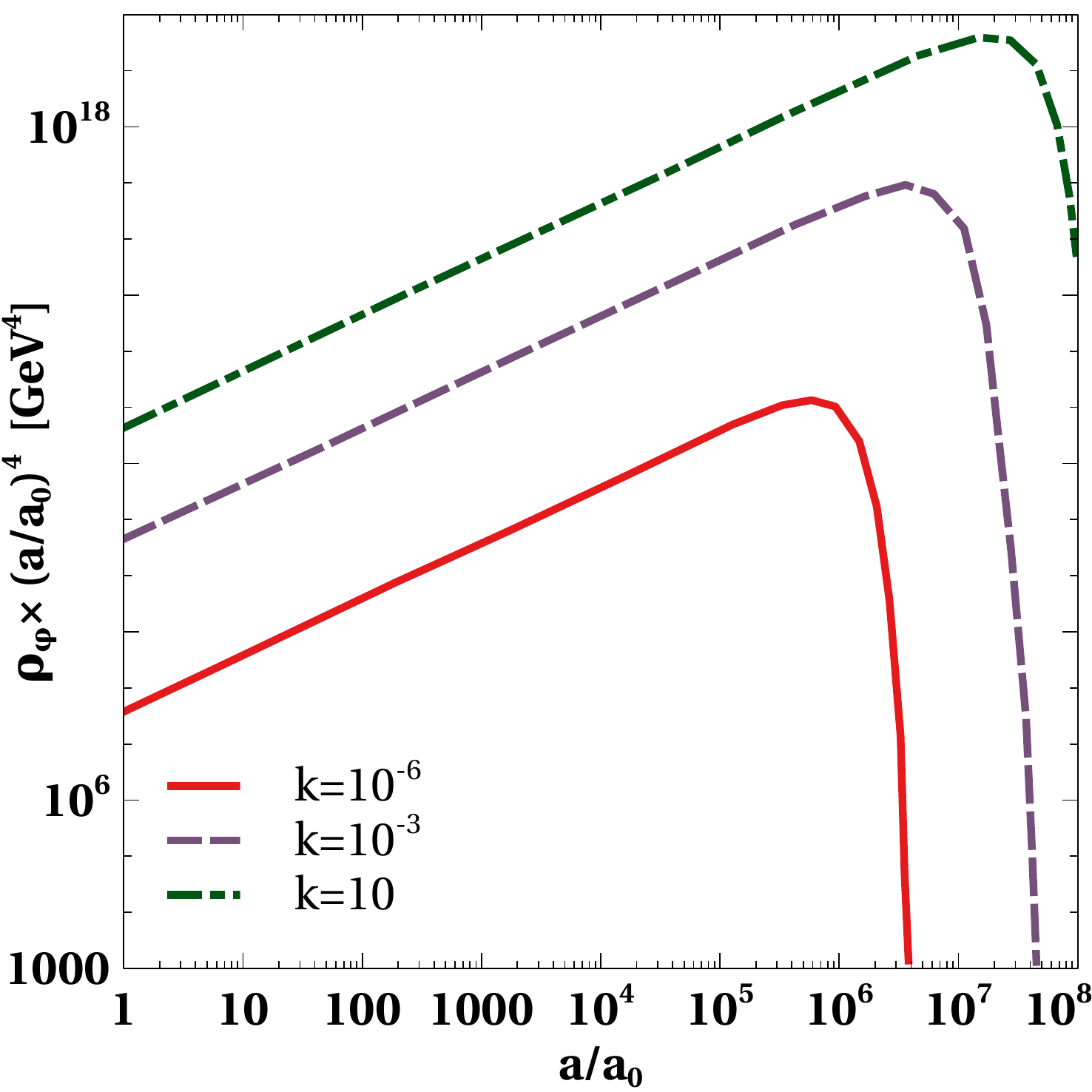}
\includegraphics[scale=.45]{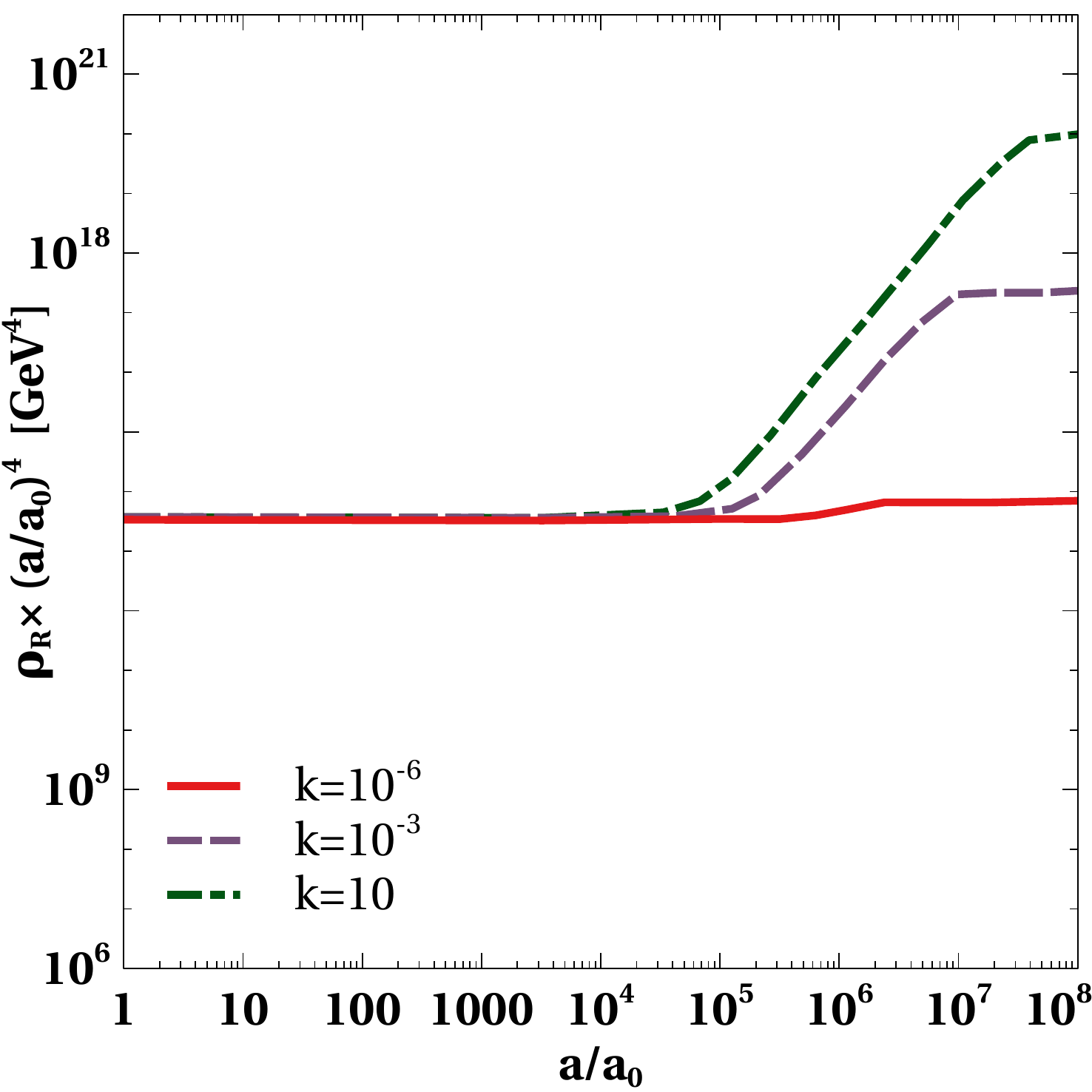}
\includegraphics[scale=.45]{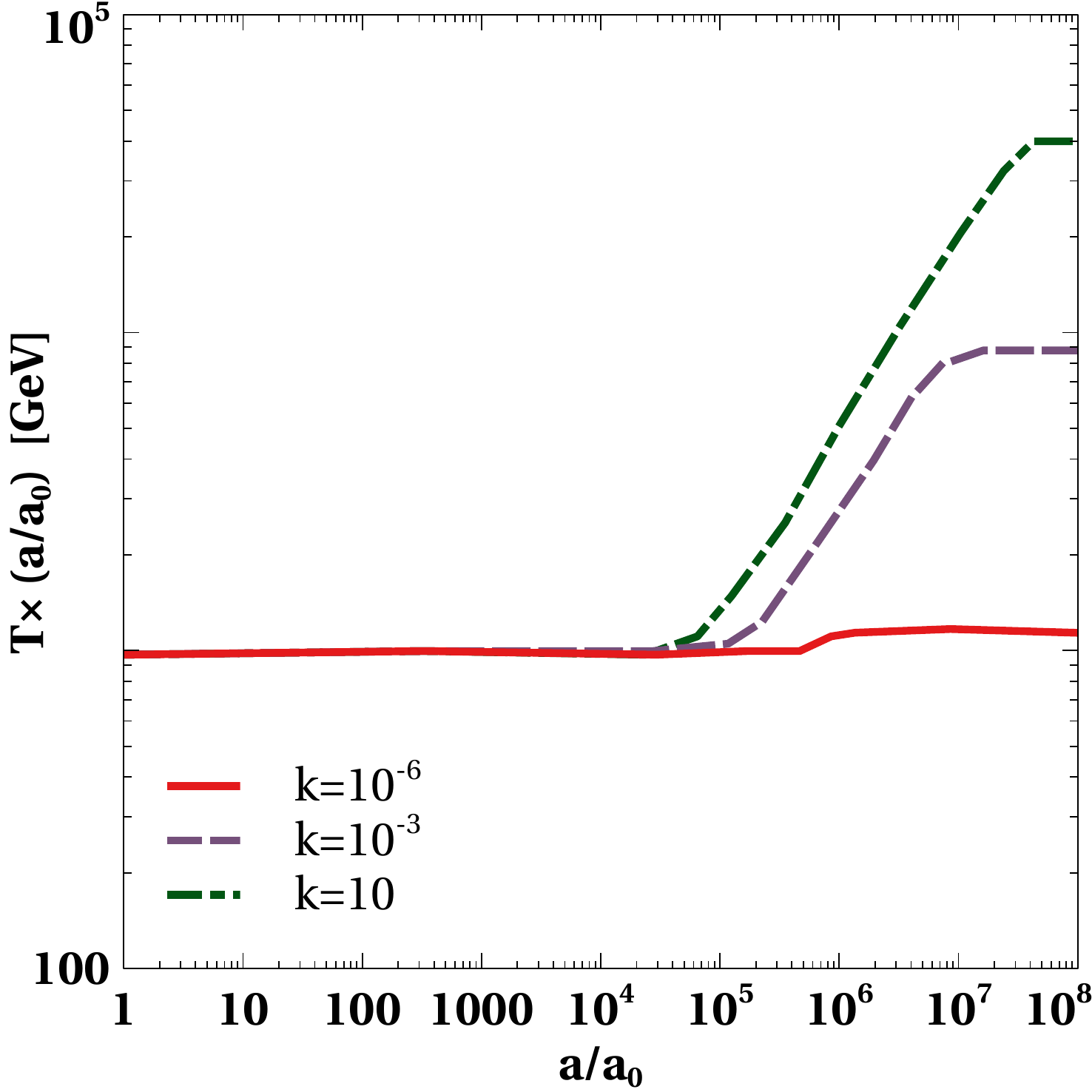}
\caption{The evolution of the energy densities of $\phi$, radiation and temperature with scale factor for different values of $k=\rho_{\phi}^{\rm in}/\rho_{R}^{\rm in}$. $T_{\rm end}$ is fixed at $2$ MeV for all the plots. The scale factor at initial temperature is denoted by $a_0$.}
\label{fig:energy_temp_EMD}
\end{figure}

In Fig. \ref{fig:energy_temp_EMD} we show the evolution of $\rho_{\phi}$, $\rho_{\rm rad}$ and $T$ with the scale factor. From the upper left panel of Fig. \ref{fig:energy_temp_EMD} it can be seen that the energy density of the $\phi$ field falls like $\rho_{\phi} \propto a^{-3}$ before it decays near $T=T_{\rm end}$. Similarly from the upper right panel plot of Fig. \ref{fig:energy_temp_EMD} it is observed that the radiation energy density falls as $\rho_{\rm rad} \propto a^{-4}$ and finally it gets a push when the $\phi$ field decays into radiation. From the lower panel plot of Fig. \ref{fig:energy_temp_EMD} we can see that the temperature falls as $T \propto a^{-1}$ initially and finally it also gets a push when the $\phi$ field decays into radiation. Also, it can be seen that larger the value of $k=\rho_{\phi}^{\rm in}/\rho_{\rm rad}^{\rm in}$, larger is the push appearing in $\rho_{\phi}$ and $T$. This is expected as large quantity of $\phi$ field energy density will inject greater entropy into the plasma when it decays. In addition to $T_{\rm end}$ another temperature relevant for our analysis is the sphaleron freeze-out temperature $T_{\rm Sph} \sim 130$ GeV. Depending upon $T_{\rm end}$, we study three different cases below. Note that, we have not discussed the case where $ T_{\rm end} \ll T_{\rm }$ as this is very similar to leptogenesis in a radiation dominated universe (upto a subsequent entropy dilution). Alternately, if $T_{\rm end}$ are much larger than the scale of leptogenesis $T=M_1$, then also it resembles the usual scenario as $\phi$ decays way before the scale of leptogenesis.

\subsubsection{Case 1: $T_{\rm end}< T_{\rm Sph}$}
\begin{figure}[h]
\includegraphics[scale=.44]{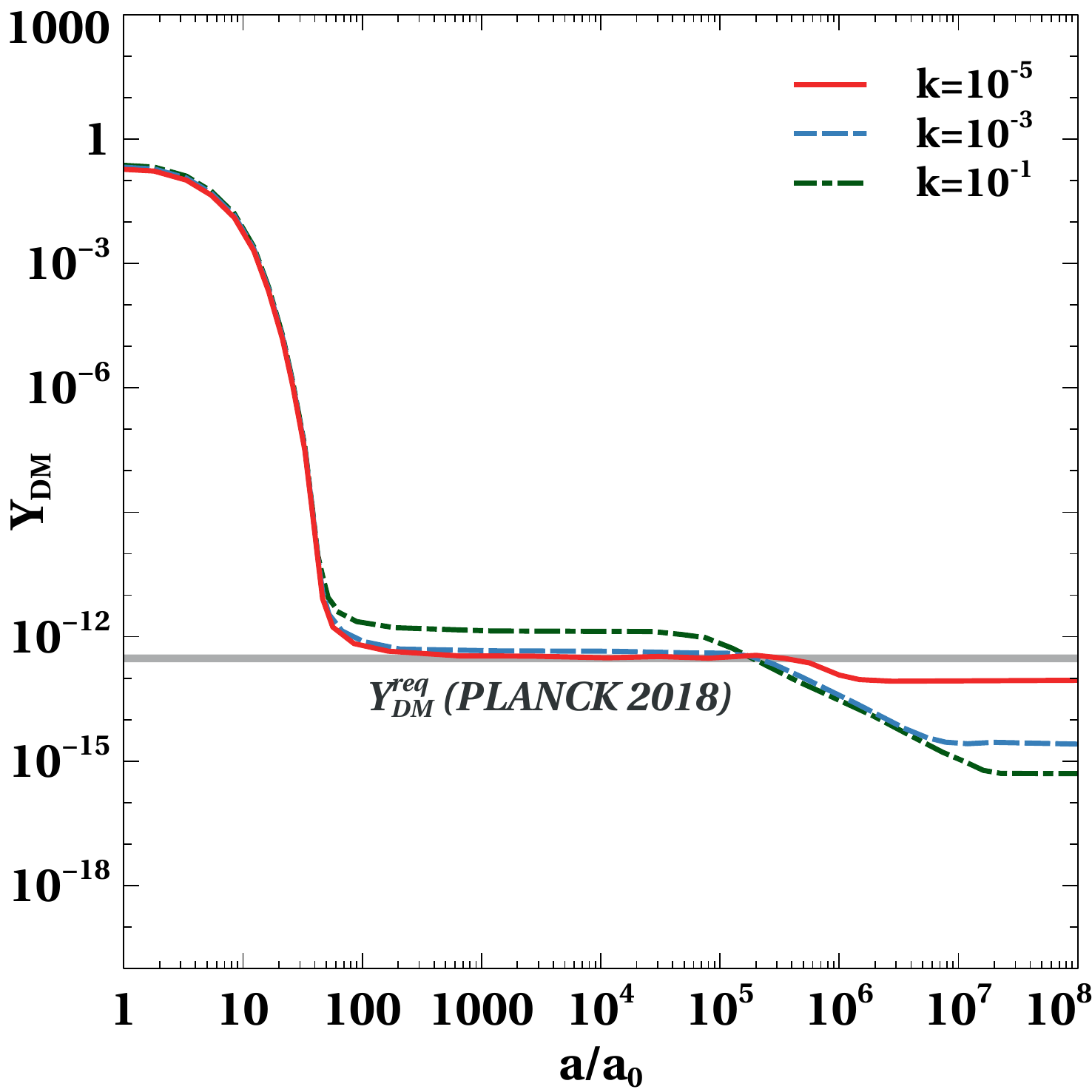}
\includegraphics[scale=.44]{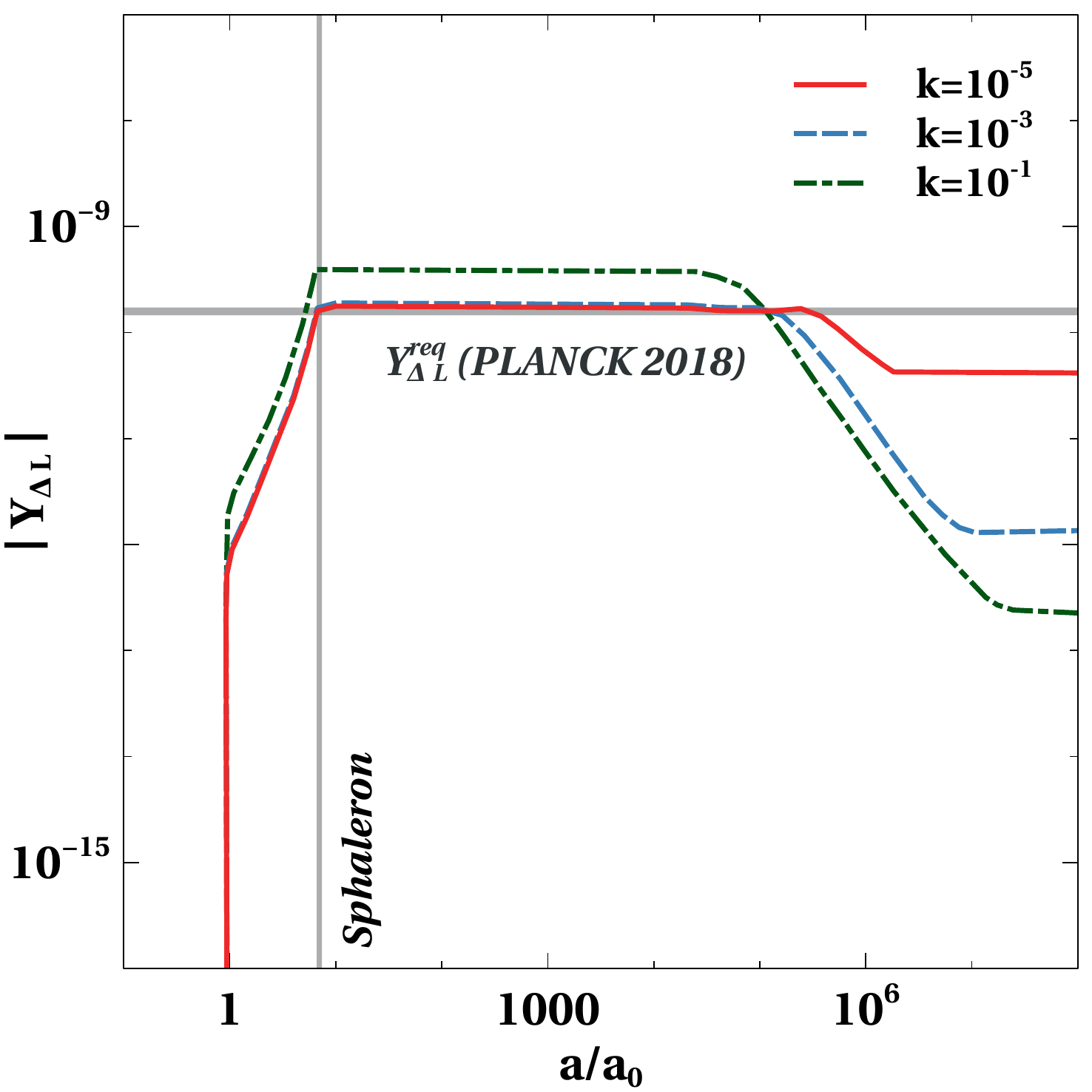}
\includegraphics[scale=.44]{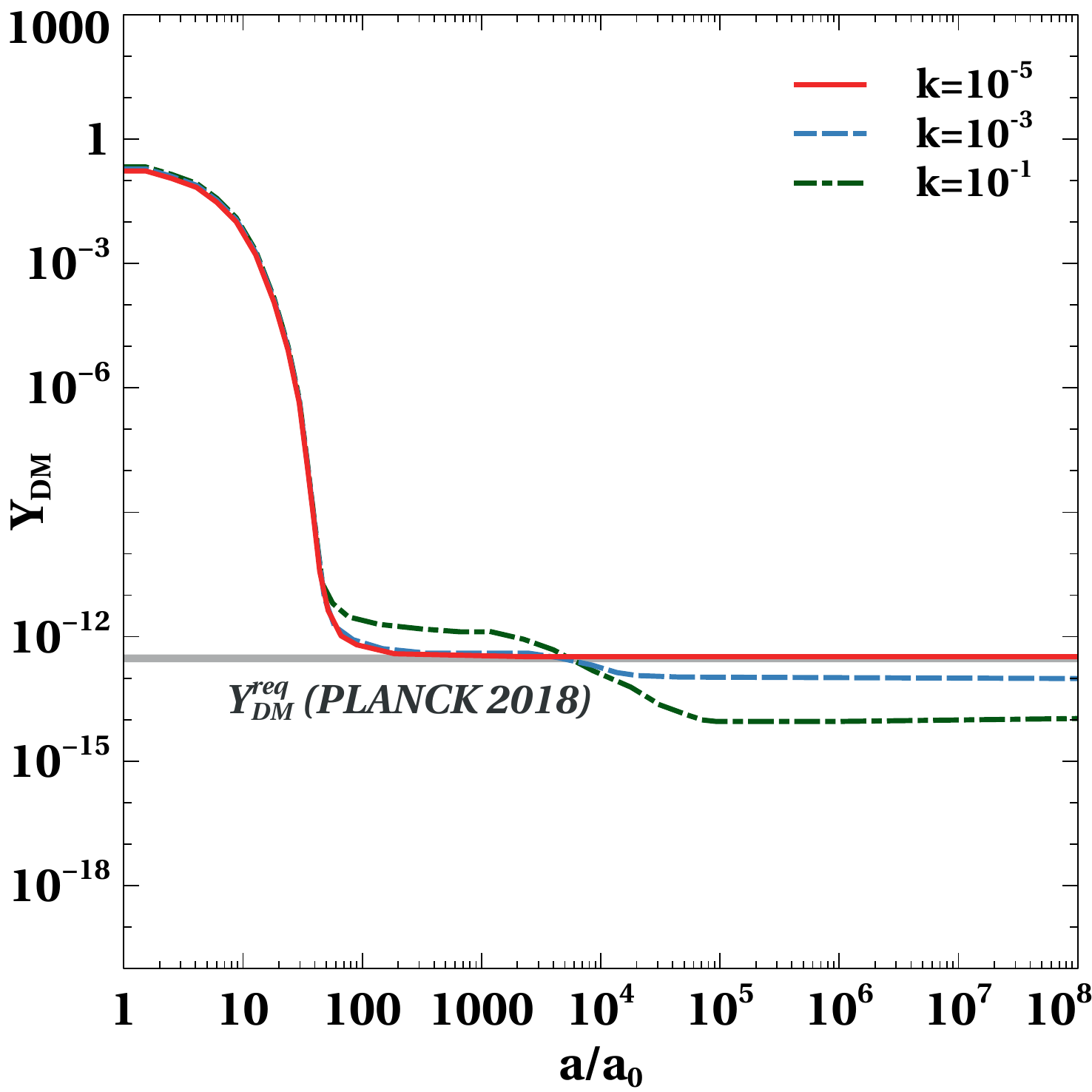}
\includegraphics[scale=.44]{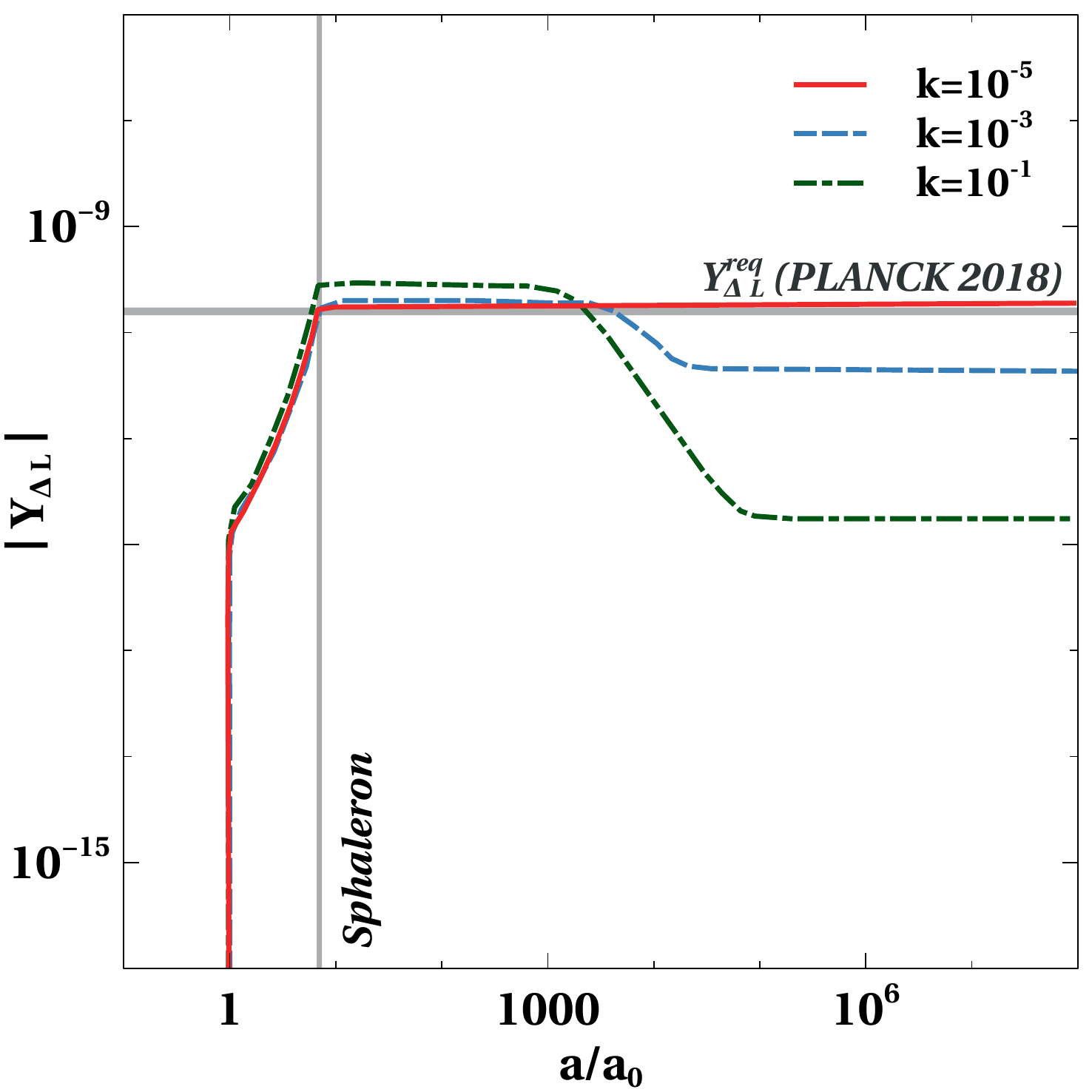}
\caption{The comoving number densities of DM and $L$ asymmetry with the relative scale factor $a/a_{0}$ for different values of $k$. The model parameters are fixed at $m_{\eta_{R}}=600$ GeV, $v_{\Delta}=1$ keV, $\lambda_{H\eta}^{''}=1\times 10^{-5}$, $\mu_{\eta \Delta}=10i$, $M_{1}=6$ TeV, $M_{j+1}/M_{j}=1.1$, and $m_{\Delta^{0}}=m_{\Delta^{\pm}}=m_{\Delta^{\pm \pm}}=1.2$ TeV. Here $T_{\rm end}$ is fixed at $2$ MeV for the upper panel plots, and at $200$ MeV for the lower panel plots.}
\label{fig:Asym_DM_EMD1}
\end{figure}

Here we consider the evolution of the DM relic and the $L$ asymmetry by taking two different values of $T_{\rm end}$ such that $T_{\rm end}<T_{\rm Sph}$. In Fig. \ref{fig:Asym_DM_EMD1} we show the evolution of the comoving number densities of DM and the $L$ asymmetry with the scale factor for the case when $T_{\rm end}=2$ MeV and $T_{\rm end}=200$ MeV respectively. While we solve the equations in terms of $N$ as written above, we convert it to $Y=n/s = N/(a^3 s(a))$ for both DM and $L$ asymmetry, shown in these plots. The entropy dilution effect on DM abundance and baryon asymmetry is clearly visible in Fig. \ref{fig:Asym_DM_EMD1}. In an EMD universe, larger value of $k$ makes the expansion rate larger since the Hubble expansion rate is determined by both $\rho_{\phi}$ and $\rho_{\rm rad}$. Therefore, the DM abundance starts deviating from the equilibrium abundance at earlier epochs for larger $k$, as can be seen in the left panel plots of Fig. \ref{fig:Asym_DM_EMD1}. As a consequence, the generated asymmetry also increases slightly. However, for larger $k$ the late entropy dilution effect on DM abundance and lepton asymmetry is much more dominant compared to the enhancement coming due to the change in the expansion rate of the universe and therefore with the increase in $k$ both final DM abundance as well as the  asymmetry decreases. It is important to note that the entropy dilution effect in the lower panel plots of Fig. \ref{fig:Asym_DM_EMD1} is less compared to that in the upper panel plots. The is because, we have chosen $T_{\rm end}=200$ MeV for the lower panel plot whereas $T_{\rm end}=2$ MeV for the upper panel plots, implying that the $\phi$ field decays when the universe is much hotter for the lower panel plots compared to the upper panel plots. Therefore, upon decay of the scalar field the entropy injection into the plasma is small compared to the entropy the plasma has. Hence it is expected that as we increase $T_{\rm end}$ the entropy dilution effect will become weaker. 

We run a parameter scan over the parameter $v_{\Delta}$ and $\mid \mu_{\eta \Delta} \mid$,  by taking $T_{\rm end}=2$ MeV and $T_{\rm end}=200$ MeV respectively while keeping the other particle physics parameters fixed. Due to the strong entropy dilution effect, we found no available parameter space that can satisfy the correct DM relic and the observed baryon asymmetry irrespective of the value of $k$ (for $k\geq 10^{-5}$) with $T_{\rm end}=2$ MeV and $T_{\rm end}=200$ MeV. Increasing $T_{\rm end}$ to 2 GeV allows some parameter space in $v_{\Delta}-\mu_{\eta\Delta}$ plane, to be discussed below.

\subsubsection{Case 2: $T_{\rm end} \simeq T_{\rm Sph}$}

\begin{figure}[h]
\includegraphics[scale=.44]{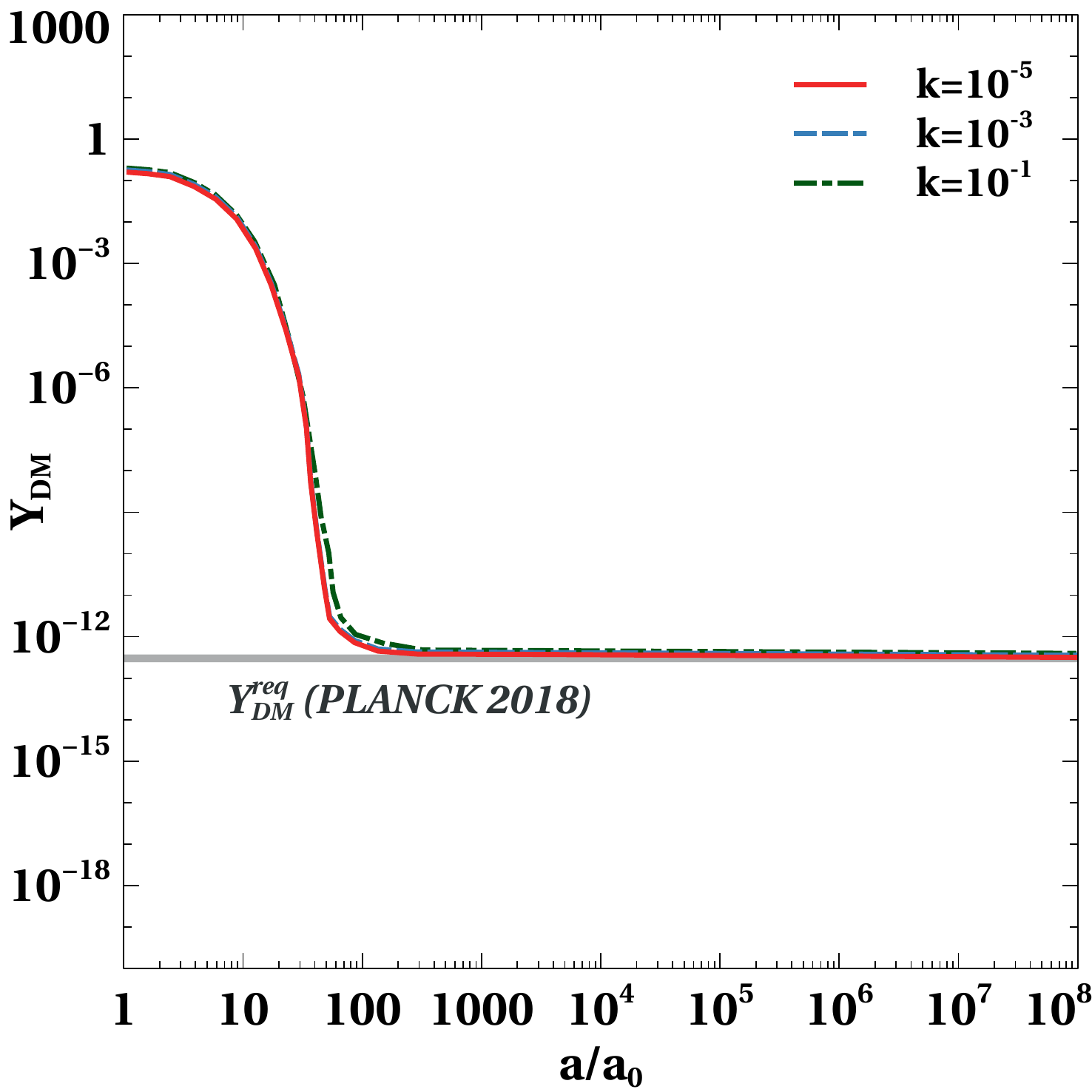}
\includegraphics[scale=.44]{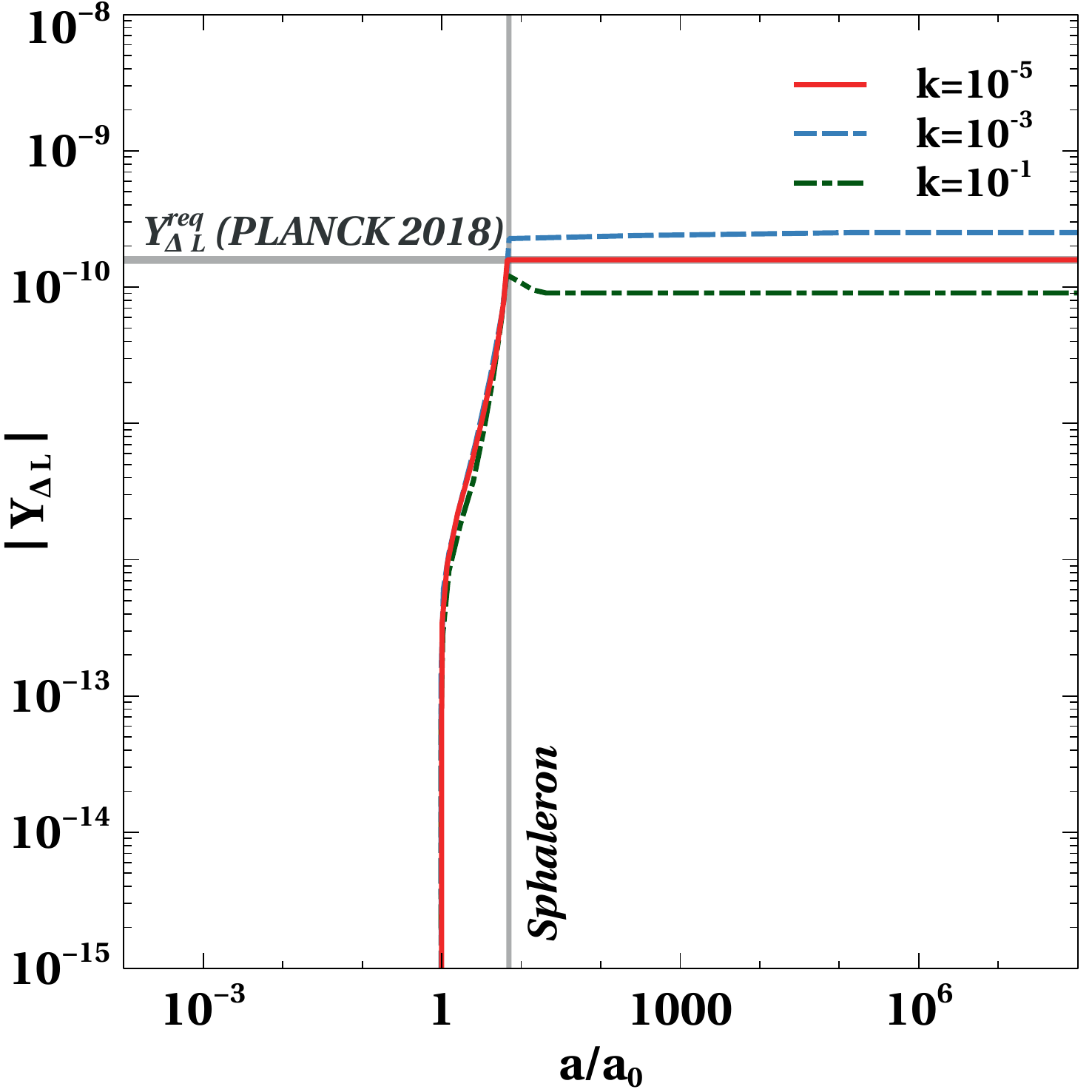}
\caption{The comoving number densities of DM and $L$ asymmetry with the relative scale factor $a/a_{0}$ for different values of $k$. The model parameters are fixed at $m_{\eta_{R}}=600$ GeV, $v_{\Delta}=1$ keV, $\lambda_{H\eta}^{''}=1\times 10^{-5}$, $\mu_{\eta \Delta}=10i$, $M_{1}=6$ TeV, $M_{j+1}/M_{j}=1.1$, and $m_{\Delta^{0}}=m_{\Delta^{\pm}}=m_{\Delta^{\pm \pm}}=1.2$ TeV. Here $T_{\rm end}$ is fixed at $150$ GeV.}
\label{fig:EMD3}
\end{figure}

Now we redo the analysis for DM abundance and $L$ asymmetry by considering $T_{\rm end} \simeq T_{\rm Sph}$. 
Here we have chosen $T_{\rm end}=150$ GeV to illustrate
the phenomenology near $T_{\rm end} \simeq T_{\rm }$. The results for the evolution of DM abundance and the $L$ asymmetry are shown in Fig. \ref{fig:EMD3}. In this case we can see that the entropy dilution effect is very feeble. It is observed that near a particular value of $k \simeq10^{-3}$ the generated asymmetry can be even more than the required asymmetry.

\subsubsection{$T_{\rm end}>T_{\rm Sph}$}

\begin{figure}[h]
\includegraphics[scale=.44]{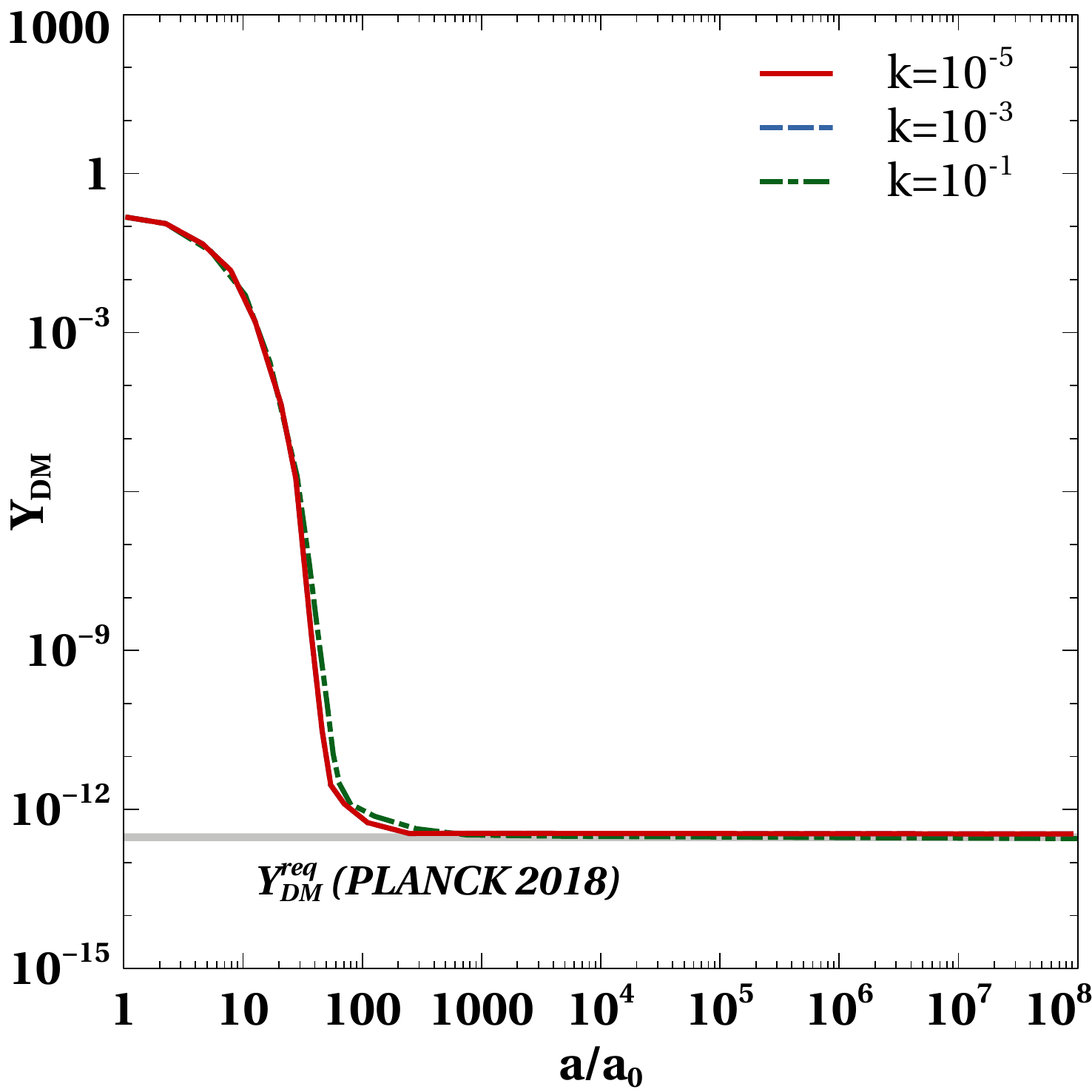}
\includegraphics[scale=.44]{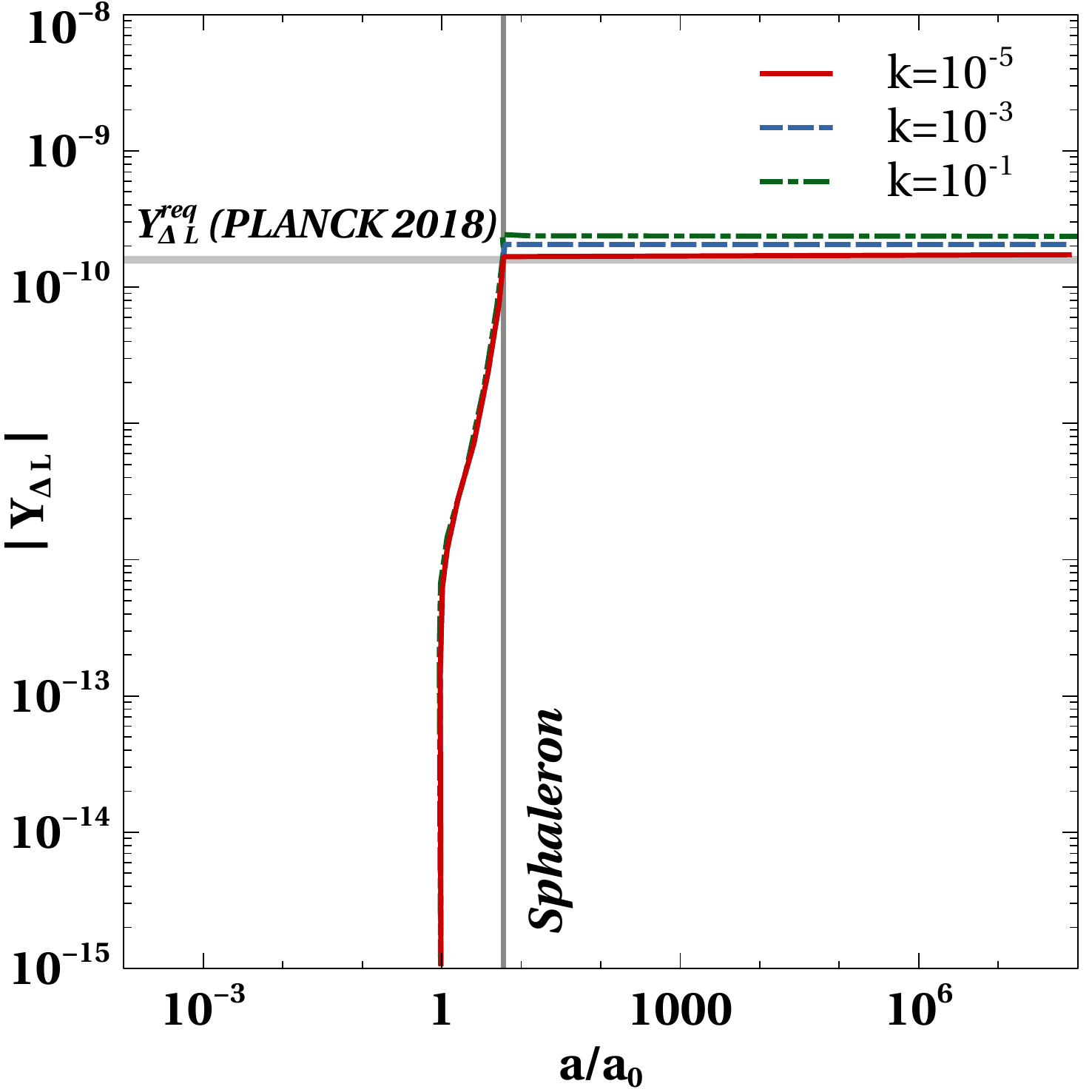}
\caption{The comoving number density of DM and $L$ asymmetry with the relative scale factor $a/a_{0}$ for different values of $k$. The model parameters are fixed at $m_{\eta_{R}}=600$ GeV, $v_{\Delta}=1$ keV, $\lambda_{H\eta}^{''}=1\times 10^{-5}$, $\mu_{\eta \Delta}=10i$, $M_{1}=6$ TeV, $M_{j+1}/M_{j}=1.1$, and $m_{\Delta^{0}}=m_{\Delta^{\pm}}=m_{\Delta^{\pm \pm}}=1.2$ TeV. Here $T_{\rm end}$ is fixed at $250$ GeV.}
\label{fig:EMD4}
\end{figure}

Finally, we consider $T_{\rm end}=250$ GeV such that $T_{\rm end}>T_{\rm Sph}$. The evolution of the comoving number densities of DM and the $L$ asymmetry are shown in Fig. \ref{fig:EMD4}. It is clearly seen from the figure that the entropy dilution effect is not evident in this case even for large $k$ namely, $k=10^{-1}$. On the other hand, it is observed that at least upto $k \simeq 10^{-1}$, the $L$ asymmetry increases with the increase in $k$.

\begin{figure}
\includegraphics[scale=.5]{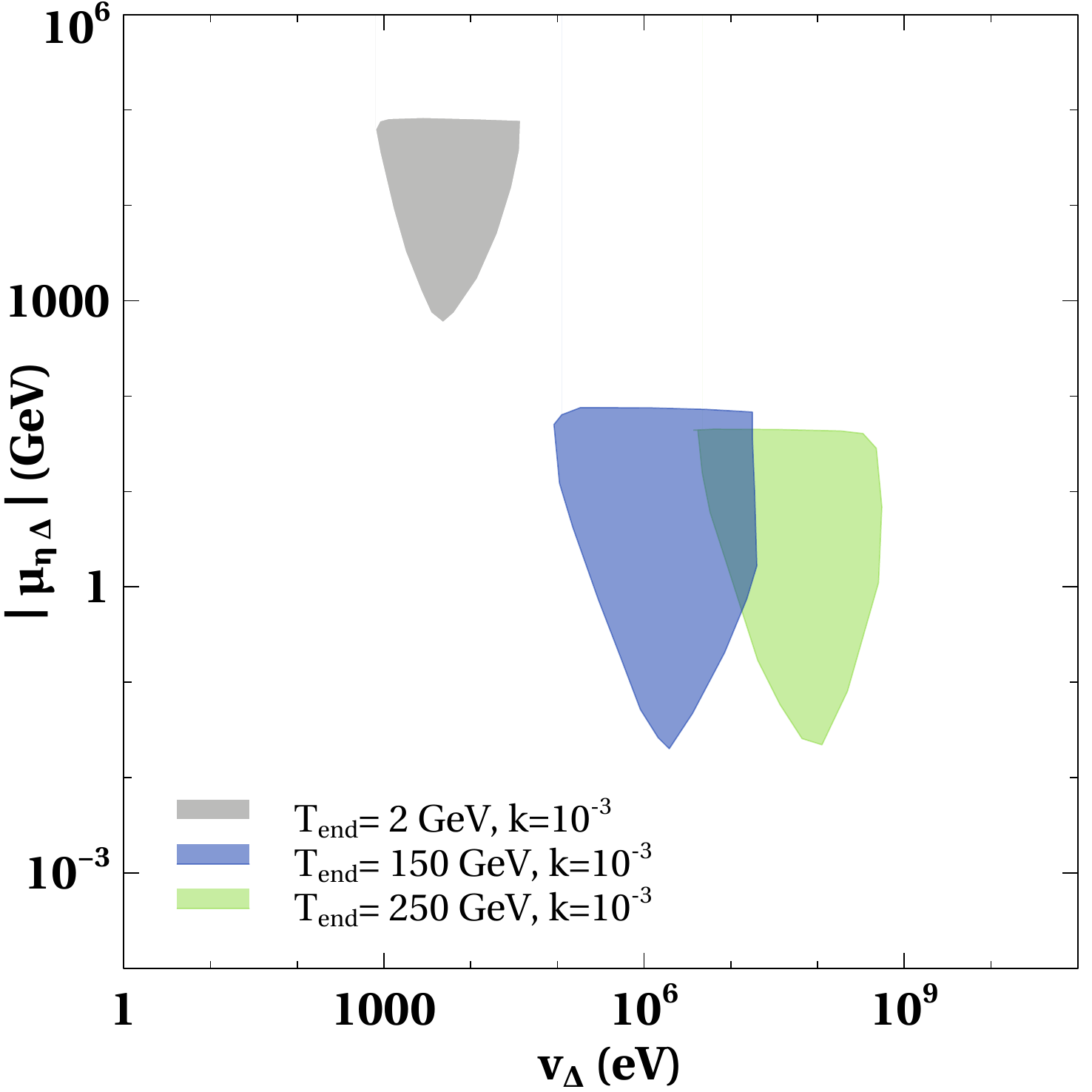}
\caption{Scan plot showing the available parameter space in $\mu_{\eta \Delta}$-$v_{\Delta}$ plane from the requiremnet of correct DM relic and the observed baryon asymmetry for two different possible scenario of early matter dominated universe. The other relevant parameters are set at $m_{\eta_{R}}=600$ GeV, $\lambda_{H\eta}^{''}=10^{-5}$, $m_{\Delta^{0}}=m_{\Delta^{\pm}}=m_{\Delta^{\pm \pm}}=1.2$ TeV, $M_{1}=6 $ TeV, $M_{j+1}/M_{j}=1.1$.}
\label{fig:EMD_scan}
\end{figure}

After studying the evolution of number densities for three different cases, we do a parameter scan over the parameters $v_{\Delta}$ and $\mid \mu_{\eta \Delta}\mid$ by keeping $T_{\rm end}$ fixed for each of these cases. The resulting parameter space in Case 1, for $T_{\rm end}=2$ GeV is shown by the grey coloured patch in Fig. \ref{fig:EMD_scan}. The limits on the parameter are found to be 0.008 MeV $ \lesssim v_{\Delta} \lesssim$  0.05 MeV, 700 GeV $\lesssim \lvert \mu_{\eta \Delta} \rvert \lesssim$ $9\times 10^{4}$ GeV in an EMD universe with $k=10^{-3}$ and $T_{\rm end}=2$ GeV for $m_{\eta_R}=600$ GeV.

The same in Case 2, for $T_{\rm end}=150$ GeV is shown by the blue coloured patch in Fig. \ref{fig:EMD_scan}. For Case 2, the limits on the $v_{\Delta}$ and $\mu_{\eta \Delta}$ are found to be 0.07 MeV $ \lesssim v_{\Delta} \lesssim $ 26 MeV, 0.02 GeV $\lesssim \lvert \mu_{\eta \Delta} \rvert \lesssim$ 85 GeV respectively while keeping DM mass fixed at $m_{\eta_{R}}=600$ GeV. Repeating the same for Case 3 while keeping $T_{\rm end}=250$ GeV results in the light green coloured patch in Fig. \ref{fig:EMD_scan}. In this case the bounds on $v_{\Delta}$ and $\mu_{\eta\Delta}$ are found to be $2$ MeV $\lesssim v_{\Delta} \lesssim 260$ MeV and $0.35$ GeV $\lesssim \lvert \mu_{\eta \Delta} \rvert \lesssim$ 80 GeV. One can clearly see that the viable parameter space changes for the EMD universe compared to the standard radiation dominated universe. We also found that for larger value of $k$ we need larger $T_{\rm end}$ to satisfy the correct asymmetry. In particular, we found that the correct baryon asymmetry and DM relic can not be achieved for $k \gtrsim 10^{-2}$ due to the strong entropy dilution coming from the matter field $\phi$, irrespective of $T_{\rm end}$. As discussed above, there is a range of $k$ near $10^{-3}$ where the asymmetry can be even more than the standard radiation case for sufficiently large $T_{\rm end} \thicksim 100$ GeV. Therefore, we have shown the viable parameter space in $\mu_{\eta \Delta}$ versus $v_{\Delta}$ plane keeping $k=10^{-3}$ for all the three cases. It is clearly visible that the allowed parameter space reduces for $T_{\rm end}=2$ GeV compared to $T_{\rm end}=150, 250$ GeV, because for higher $T_{\rm end}$, the entropy dilution effect is very minimal on both comoving number densities of DM and $L$ asymmetry. 

\begin{figure}
\includegraphics[scale=.5]{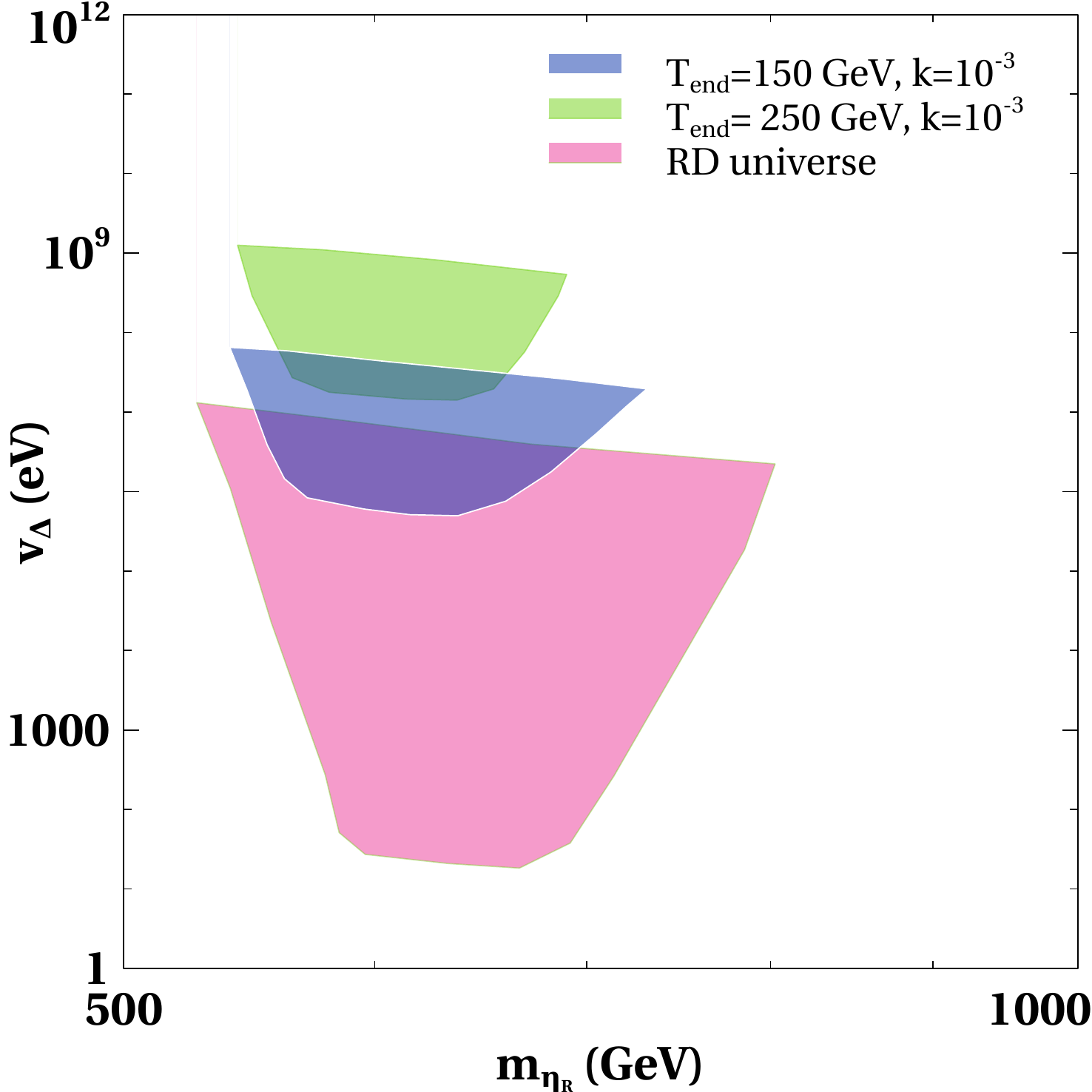}
\caption{Scan plot showing the allowed parameter space in $m_{\eta_{R}}$ versus $v_{\Delta}$ plane. The relevant parameters are fixed at $\mu_{\eta \Delta}=80i$ GeV, $\lambda^{''}_{H\eta}=1\times 10^{-5}$, $M_{1}=6$ TeV, $M_{j+1}/M_{j}=1.1$.}
\label{fig:mdm_EMD}
\end{figure}

We then do another scan over the parameters $v_{\Delta}$ and $m_{\eta_{R}}$ by keeping the other relevant parameters fixed. The result is shown in Fig. \ref{fig:mdm_EMD} for case 2, 3 along with the result for standard cosmology. We found that no parameter space is allowed for smaller values of $T_{\rm end}\leq \mathcal{O}(100 \, {\rm MeV})$ where the entropy dilution effect is very dominant. However, near $T_{\rm end} \sim T_{\rm Sph}$, when the entropy dilution effect is very minimal we found available parameter space as shown in Fig. \ref{fig:mdm_EMD}. The region shown by pink colour in Fig. \ref{fig:mdm_EMD} is for $k=0$ which represents the standard radiation dominated universe. We can clearly see an upward shift of the parameter space along $v_{\Delta}$ axis in case of EMD universe from the standard
radiation dominated scenario. This is expected because for case 2, 3 with less effects of entropy dilution, we get more asymmetry than the standard radiation case. Therefore a larger $v_{\Delta}$ is required to satisfy the correct asymmetry. However, the mass of DM needed to
satisfy the correct relic does not change much from that in standard radiation domination.
This is because, for such large values of $v_{\Delta}$ , required to satisfy the correct asymmetry the
annihilations of DM is mainly through the gauge interactions. Therefore the DM relic is
determined by the usual gauge interactions and hence the lower limit on DM mass remains same. The upper bound on DM mass however, gets reduced visibly in the EMD scenario compared to the standard case. From these results we can conclude that an early matter domination can not lower the scale of WIMPy leptogenesis unlike in the FEU scenario discussed before.

\subsection{Scalar-Tensor theory of gravity}

In this section, we study the impact of modified expansion rate of the universe on WIMPy leptogenesis within the framework of a class of Scalar-Tensor theories of gravity (STG) \cite{Fierz:1956zz, Brans:1961sx}. In STG, gravity is described by both metric and a scalar field. In such theories the cosmological expansion rate deviate from the standard general relativity (GR) and an attractor mechanism relaxes it to the standard expansion era prior to the onset of the BBN. STG are often formulated in either Einstein frame or in Jordan frame with the most general transformation between the two frames is given by 
\begin{eqnarray}
\tilde{g}_{\mu \nu} & =& C(\phi)g_{\mu \nu}+ D(\phi) \partial_{\mu}\phi \partial_{\nu} \phi,
\end{eqnarray}
where $\tilde{g}_{\mu \nu}$ and $g_{\mu \nu}$ represent the metrics in Jordan frame and in Einstein frame respectively. Here $C(\phi)$ and $D(\phi)$ are so called the conformal and disformal couplings respectively. In Jordan frame, the matter fields $\Psi$ are directly coupled to the metric, $\tilde{g_{\mu \nu}}$ and therefore the action of the matter sector can be written as $S_{\rm Matter}=S_{\rm Matter} (\tilde{g}_{\mu \nu},\Psi)$. The effect of modified gravity enters only through the expansion rate of the universe and usual particle physics observables remain unchanged. However, in Einstein frame the scenario becomes completely opposite. Therefore the Jordan frame is considered throughout this work. Leptogenesis in scalar-tensor theories of gravity was studied in \cite{Dutta:2018zkg} while the implications of such non-standard cosmology on DM relic were studied in several earlier works including \cite{Catena:2004ba, Catena:2009tm, Meehan:2015cna, Dutta:2016htz, Dutta:2017fcn}

Following the procedure of \cite{Dutta:2016htz,Dutta:2018zkg} the master equation for the evolution of the scalar field in the conformal limit ($D(\phi)=0$) while ignoring its potential energy is found to be 
\begin{eqnarray} \label{eq:master}
\dfrac{2}{3B \left[1-\alpha(\varphi)\varphi^{'}  \right]^{3}}\left( \varphi^{''}+\dfrac{d\alpha}{d\varphi} (\varphi^{'})^{3}\right)+\dfrac{1-\tilde{\omega}}{\left[ 1-\alpha(\varphi)\varphi^{'} \right]}\varphi^{'}+2(1-3\tilde{\omega})\alpha(\varphi)=0. 
\end{eqnarray}
Here $\varphi=\kappa \phi$ is a dimensionless scalar introduced for convenience and $B=1-\dfrac{1}{6}\dfrac{\varphi^{'}}{\left(1-\alpha(\varphi)\varphi^{'} \right)^{2}}$. The derivatives are taken with respect to the number of e-folds ($d\tilde{N}=\tilde{\bf H}dt$) in Jordan frame, which is defined with a prime $f^{'}=df/d\tilde{N}$. The $\tilde{\bf H}$ is the modified Hubble expansion rate in the Jordan frame. Here the function $\alpha(\varphi)=\dfrac{d \ln C^{1/2}}{d\varphi}$, where $C(\varphi)$ is the conformal coupling. We have considered the conformal coupling to be $C(\varphi)=(1+0.1 {\rm exp}(-8\varphi))^{2}$. The choice of such a conformal function is motivated from earlier works \cite{Catena:2004ba,Dutta:2016htz,Dutta:2018zkg}. The number of e-folds can be expressed as a function of the Jordan frame temperature as follows,

\begin{eqnarray}
\tilde{N} & = & \ln \left[ \dfrac{\tilde{T_{0}}}{\tilde{T}} \left(  \dfrac{g_{*s}(\tilde{T_{0}})}{g_{*s}(\tilde{T})} \right)^{1/3} \right].
\end{eqnarray}

The modified expansion rate $\tilde{\bf H}$ in the Jordan frame can be written as \cite{Dutta:2016htz,Dutta:2018zkg}

\begin{eqnarray} \label{eq:Hubble}
\tilde{\bf H}^{2} & = & \dfrac{k^{2}}{k_{\rm GR}^{2}}\dfrac{C(1+\alpha(\phi)\phi^{'})^{2}}{1-\varphi^{'2}/6} H^{2},
\end{eqnarray}

where ${\bf H}^{2}=\dfrac{k_{\rm GR}^{2}}{3}\tilde{\rho}$, $k^{2} \simeq k_{\rm GR}^{2}=8 \pi G$ and $\tilde{\rho} \sim g(\tilde{T})\tilde{T}^{4}$ for the radiation dominated era. From Eq. \eqref{eq:Hubble} the ratio of the new expansion rate to the standard GR expansion rate is defined as the speed-up parameter, 

\begin{eqnarray}
\xi & = & \dfrac{\tilde{\bf H}}{\bf H} =\left[ \dfrac{k^{2}}{k_{\rm GR}^{2}}\dfrac{C(1+\alpha(\phi)\phi^{'})^{2}}{1-\phi^{'2}/6} \right]^{1/2}. 
\end{eqnarray}

The third term of the master equation shown in Eq. \eqref{eq:master} behaves like an effective potential term which is given by $V_{\rm eff}=\ln C^{1/2}$. During the radiation dominated era, $\tilde{\omega}=1/3$ due to which the effective potential term disappears. Later, when the particles start becoming non-relativistic, $\tilde{\omega}$ slightly deviates from $1/3$ and the effective potential kicks in. In absence of the effective potential, the master equation predicts a solution of the type $\varphi^{'}\propto e^{-\tilde{N}}$. This means that any initial value of velocity will instantly become zero. As discussed in \cite{Dutta:2016htz,Dutta:2018zkg}, positive initial value of $\varphi$ and negative initial value of $\varphi^{'}$ lead to a very interesting scenario where the field moves to negative values until its velocity becomes zero and then becomes positive again as the field rolls back down the effective potential. This change in pattern in the evolution of the scalar field leads to a peak in the conformal function C which gives rise to a non-trivial change in the Jordan's frame Hubble expansion rate. Later the conformal factor becomes one and the standard GR expansion rate is recovered. 

\begin{figure}[h] 
\includegraphics[scale=0.55]{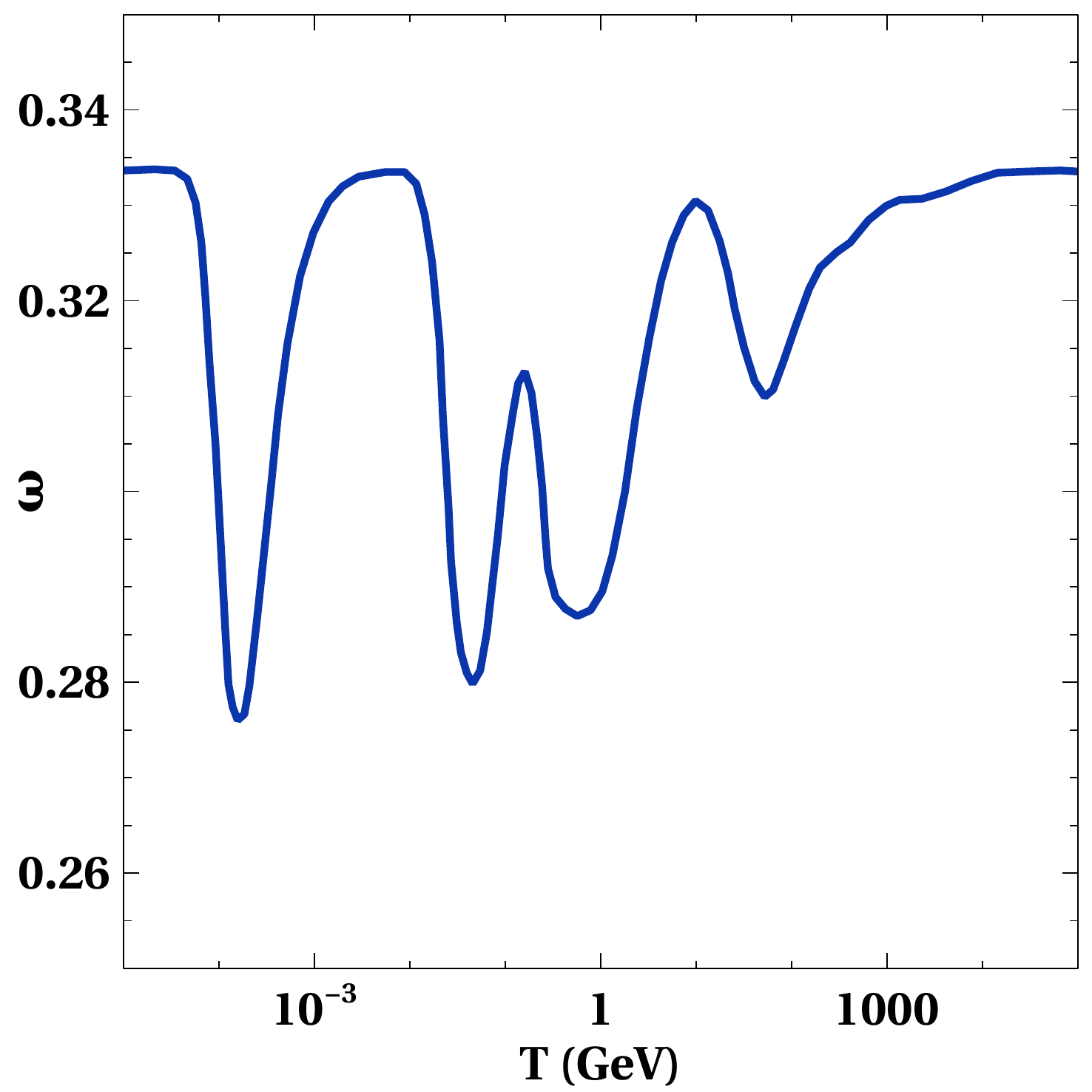}
\caption{Equation of state parameter as a function of  temperature. Here $T$ in x-axis is equivalent to $\tilde{T}$, the temperature in the Jordan frame.}
\label{fig:omega}
\end{figure}

To calculate the equation of state parameter $\tilde{\omega}$ throughout the early stages of the universe we start writing  \cite{Catena:2004ba,Meehan:2015cna}
\begin{eqnarray}
1-3\tilde{\omega} = \dfrac{\tilde{\rho}-3\tilde{p}}{\tilde{\rho}} = \sum_{A} \dfrac{\tilde{\rho_{A}}-3\tilde{p_{A}}}{\tilde{\rho}}+ \dfrac{\tilde{\rho_{m}}}{\tilde{\rho}},
\end{eqnarray}
where the sum runs over all the particles in thermal equilibrium during the early stages of the universe. The $\tilde{\rho_{m}}$ is the contribution from the non-relativistic pressureless matter. During the radiation dominated era the $\tilde{\rho_{m}}$ is negligible. Therefore the equation of state parameter becomes 
\begin{eqnarray}
\tilde{\omega} & = & \dfrac{1}{3} \left( 1- \sum_{A} \dfrac{\tilde{\rho_{A}}-3\tilde{p_{A}}}{\tilde{\rho}} \right). 
\end{eqnarray}
The energy density and pressure of any particle A are given by 
\begin{eqnarray}
\tilde{\rho_{A}}(\tilde{T}) & = & \dfrac{g_{A}}{2\pi^{2}}\int_{m _{A}}^{\infty} \dfrac{(E^{2}-m_{A}^{2})^{1/2}}{{\rm exp}(E/\tilde{T})\pm 1}E^{2}dE \\
\tilde{p}_{A}(\tilde{T}) & = & \dfrac{g_{A}}{6\pi^{2}} \int_{m_{A}}^{\infty} \dfrac{(E^{2}-m_{A}^{2})^{3/2}}{{\rm exp}(E/\tilde{T}\pm 1)}dE,
\end{eqnarray}
where $g_{A}$ is the number of internal degrees of freedom of species A and the plus (minus) sign inside the integrals correspond to fermions (bosons). In the  calculation of $\tilde{\omega}$ we have taken all the relevant particles in our model into consideration. Apart from the particles considered in \cite{Dutta:2016htz}, we have three additional Majorana neutrinos, one inert doublet scalar and one triplet scalar. We show the result for the evolution of $\tilde{\omega}$ in Fig. \ref{fig:omega}, between temperature $1 \, {\rm MeV}\leq \tilde{T} \leq 100$ TeV. As the individual particles are becoming non-relativistic at different temperatures, the kicks are appearing in $\tilde{\omega}$. We then solve the Boltzmann equations for Wimpy leptogenesis. The relevant BEs are given below.

\begin{eqnarray}\label{eq:BE_ST}
  \dfrac{dY_{\eta}}{dz} & = &-\dfrac{s(z)}{z\xi(z) {\bf H}(z)}\langle \sigma v\rangle_{\eta \eta \longrightarrow {\rm SM\, SM}}\left[ Y_{\eta}^{2}-(Y_{\eta}^{\rm eq})^{2} \right] \\
  \dfrac{dY_{\Delta L}}{dz} & = & \dfrac{s(z)}{z\xi(z) {\bf H}(z)}\langle \sigma v\rangle_{\eta \eta \longrightarrow ll}^{\delta}  \left[ Y_{\eta}^{2}-(Y_{\eta}^{\rm eq})^{2} \right]\nonumber 
 \\ &&  -\dfrac{s(z)}{z\xi(z){\bf H}(z)}Y_{\Delta L}Y_{l}^{\rm eq}r_{\eta}^{2} \langle \sigma v\rangle_{\eta \eta \longrightarrow ll} \nonumber \\ && -\dfrac{s(z)}{z\xi(z){\bf H}(z)} Y_{\Delta L} Y_{\eta}^{\rm eq} \langle \sigma v\rangle_{\eta \bar{l} \longrightarrow \eta l}.
\end{eqnarray}
Here the speed-up parameter can be written in terms of the parameter $z$ as follows
\begin{eqnarray} \label{eq:speedup}
\xi(z) & = &  \dfrac{C^{1/2}(\varphi)}{C^{1/2}(\varphi_{0})}\dfrac{1}{\left( 1-\alpha(\varphi)\,z \, d\varphi / dz \right)\sqrt{B} \sqrt{1+\alpha^{2}(\varphi_{0})}}. 
\end{eqnarray}
where $B=1-\dfrac{1}{6}\dfrac{1}{(1-\alpha(\varphi) \, z\, d\varphi /dz )^{2} }\left( z\dfrac{d\varphi}{dz}\right)^{2}$. Here $s(z)$ is the entropy density defined in the Jordan frame. In terms of the Jordan frame temperature it is given by $s(\tilde{T})=\dfrac{2\pi}{45}g_{*s}(\tilde{T})\tilde{T^{3}}$.

\begin{figure}[h] 
\includegraphics[scale=.45]{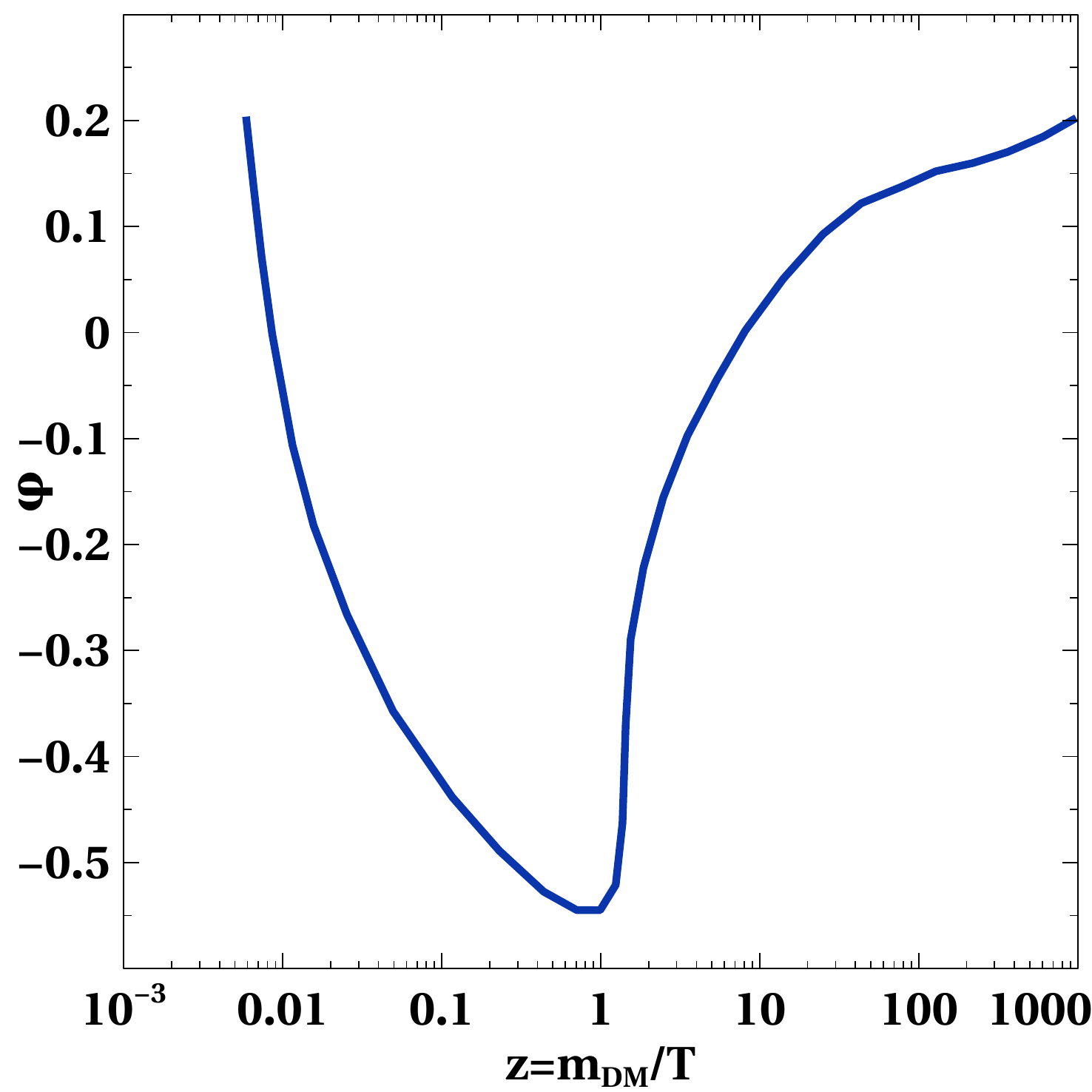}
\includegraphics[scale=.45]{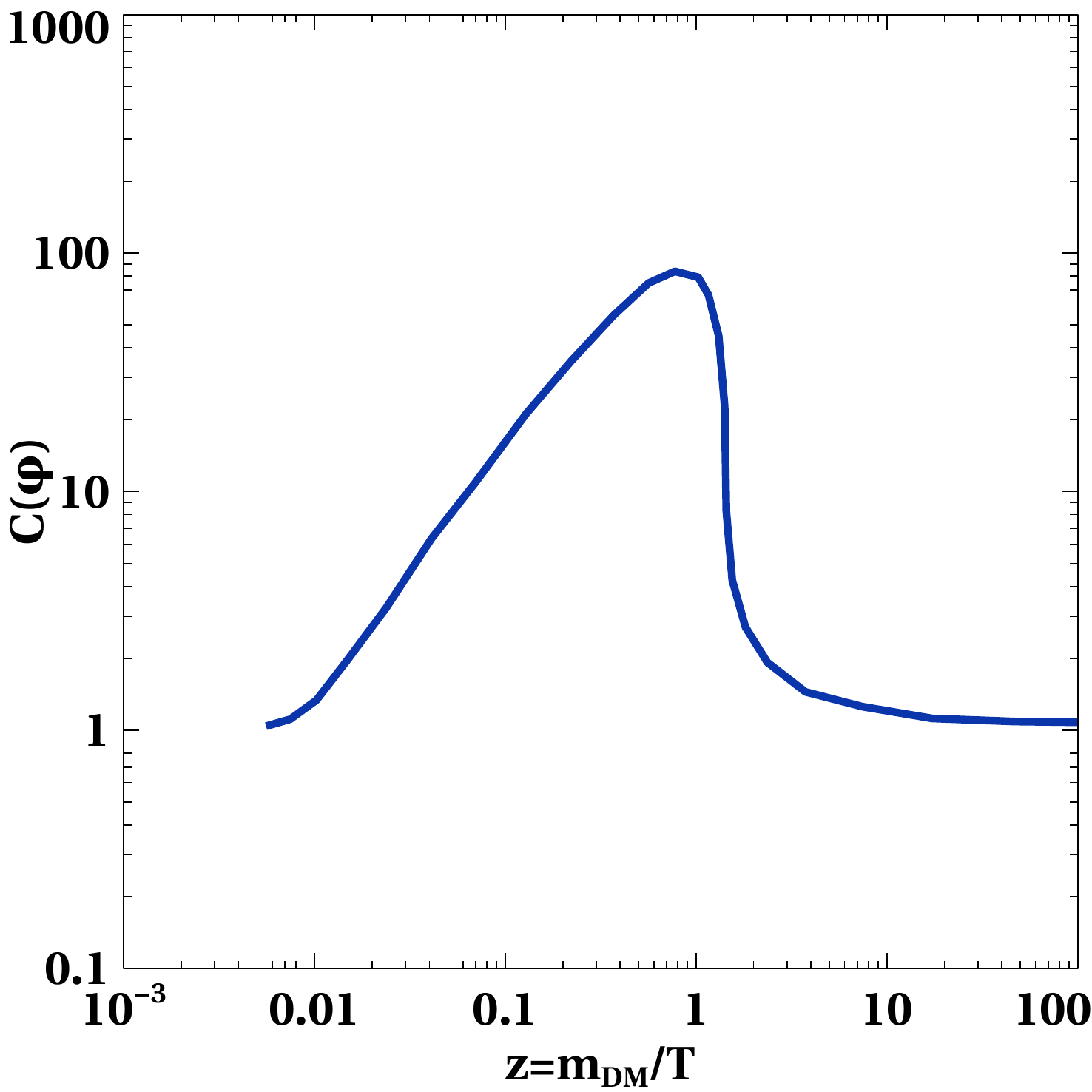}
\includegraphics[scale=.45]{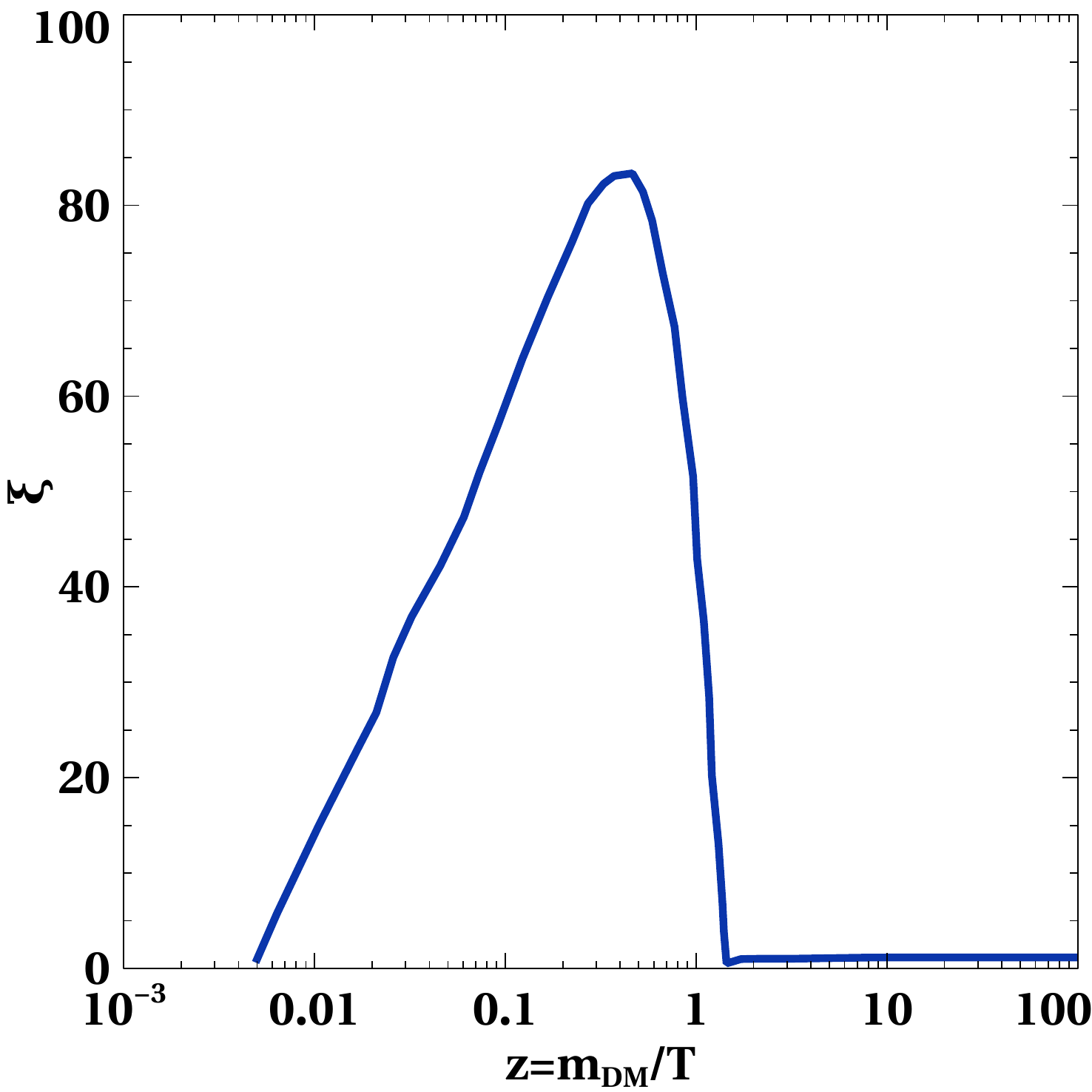}
\includegraphics[scale=.45]{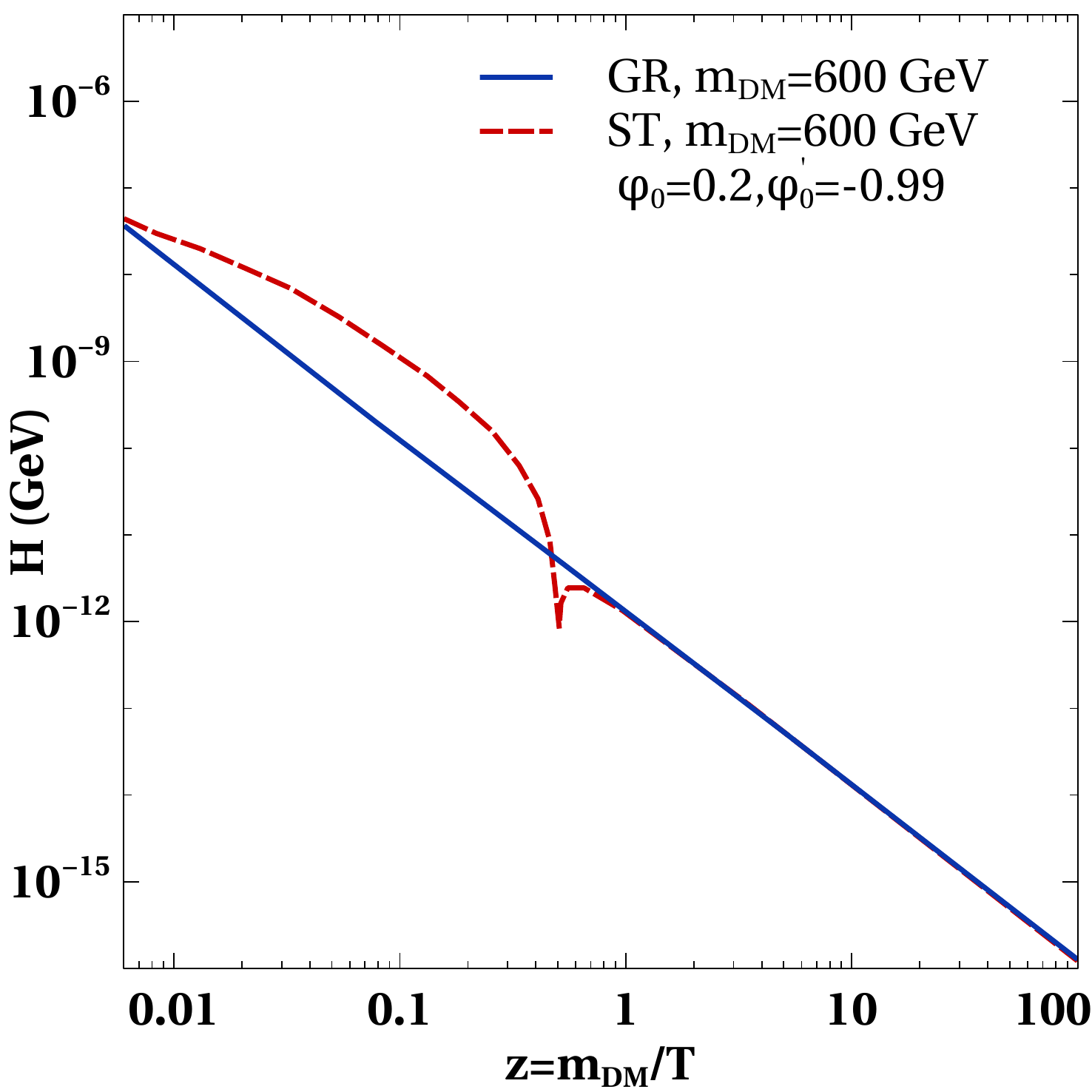}
\caption{Plots showing the evolution of the field (upper left plot), the conformal factor (upper right plot), speed-up parameter $\xi$ (lower left plot) and Hubble expansion rate (lower right plot) with $z=m_{\rm DM}/\tilde{T}$. Here $m_{\rm DM}=600$ GeV and the initial conditions are  chosen to be $(\varphi_{0},\varphi_{0}^{'})=(0.2,-0.72)$ and $\tilde{T}_{0}=100$ TeV. Here $T$ in x-axis is equivalent to $\tilde{T}$, the temperature in the Jordan frame.}
\label{fig:field_C}
\end{figure}

\begin{figure}[h] 
\includegraphics[scale=.45]{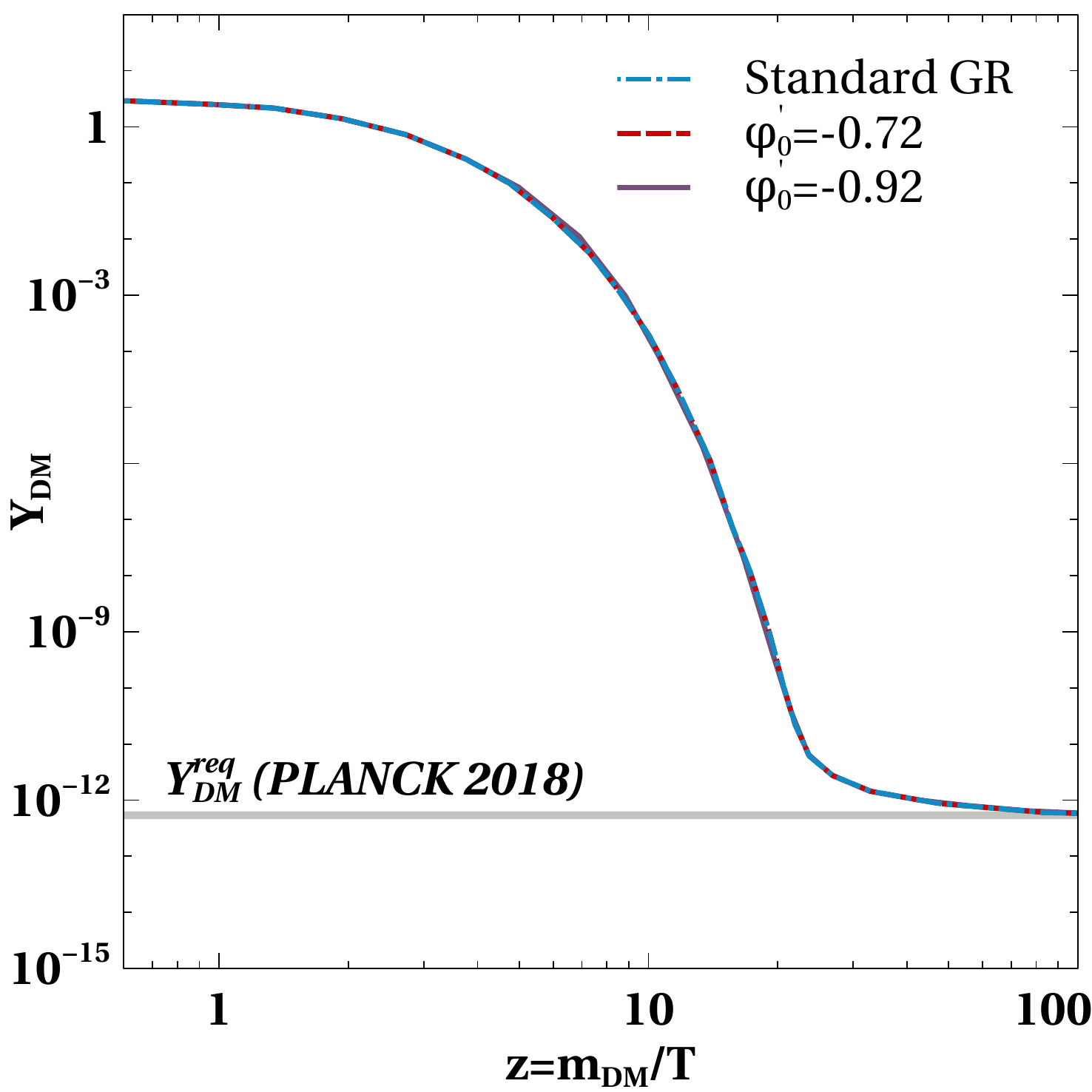}
\includegraphics[scale=.45]{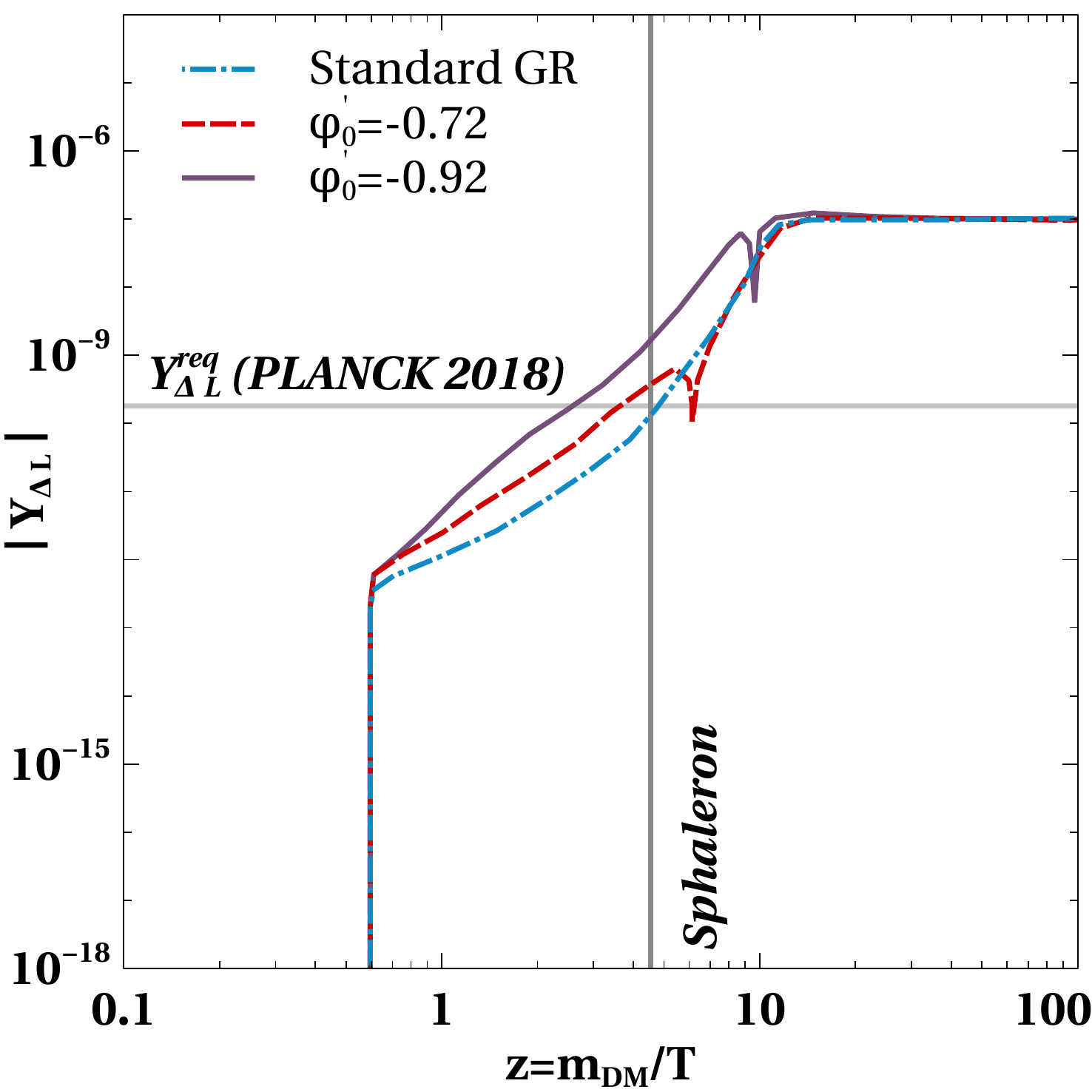}
\includegraphics[scale=.45]{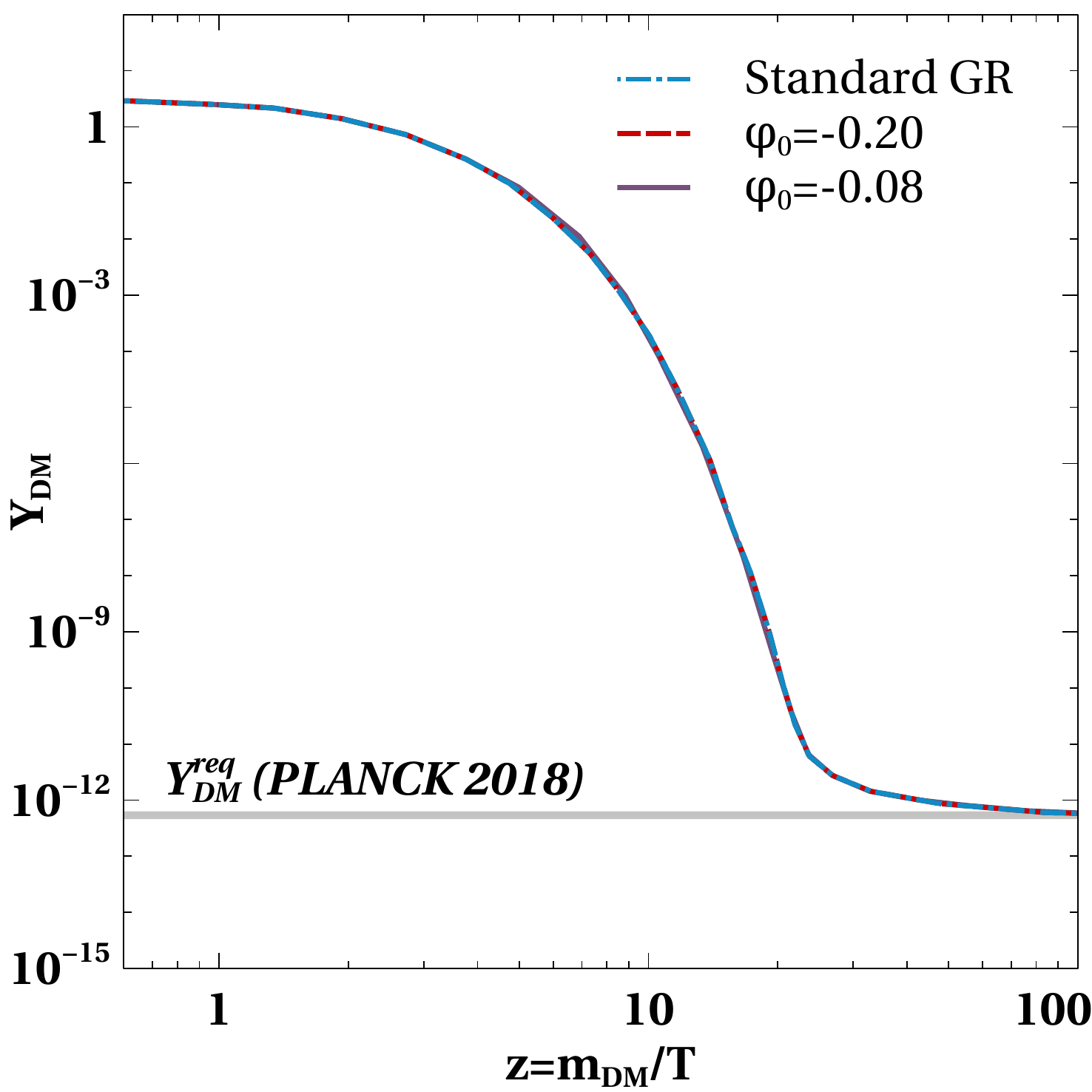}
\includegraphics[scale=.45]{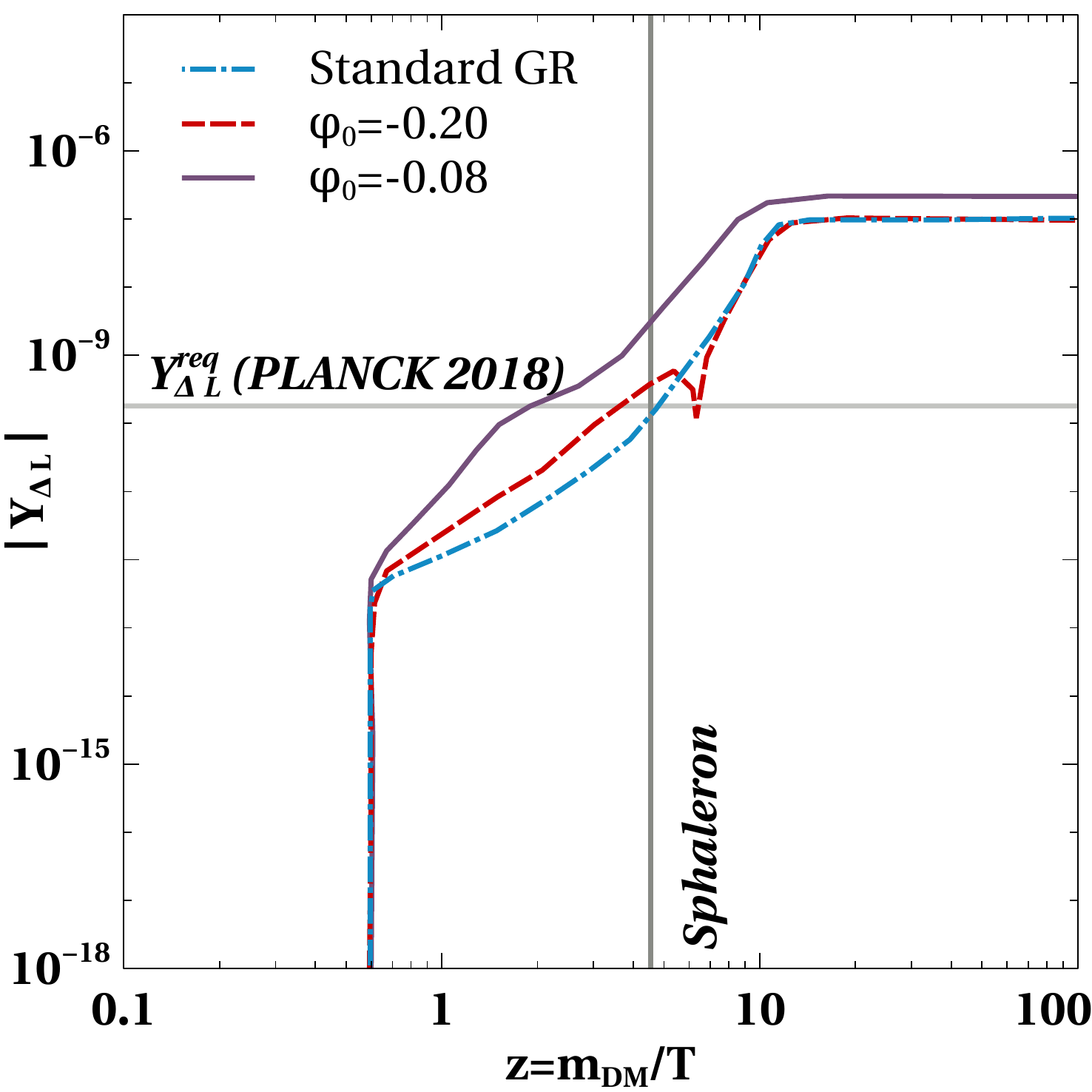}
\caption{Evolution plot for the comoving number density of DM and $L$ asymmetry with different value of for $\varphi^{'}_{0}$ with $\varphi_{0}=0.2$ (left panel plots) and for different $\varphi_{0}$ with $\varphi_{0}^{'}=-0.72$ (right panel plot) with $\tilde{T}_{0}=1$ TeV. The particle physics parameter are fixed at $m_{\eta_{R}}=m_{\rm DM}=600$ GeV, $\lambda_{H\eta}^{''}=1\times 10^{-5}$, $\mu_{\eta \Delta}=10i$ GeV, $v_{\Delta}=1$ keV, $m_{\Delta^{0}}=m_{\Delta^{\pm}}=m_{\Delta^{\pm \pm}}=1.2$ TeV, $M_{1}=6$ TeV, and $M_{j+1}/M_{j}=1.1$. Here $T$ in x-axis is equivalent to $\tilde{T}$, the temperature in the Jordan frame. }
\label{fig:DM_B-L_ST}
\end{figure}

Solving the $\varphi$ equation we show the evolution of the field $\varphi$, the conformal factor $C(\varphi)$, the speed-up parameter and the Hubble $\tilde{H}$ with $z=m_{\rm DM}/\tilde{T}$ in Fig. \ref{fig:field_C}. It can be seen that the field $\varphi$ changes the nature of its evolution at a particular value of $\tilde{T}$ depending on the initial conditions chosen. This change in $\varphi$ brings a non-trivial change in the conformal factor as well as in $\tilde{H}$ and hence the speed-up parameter. From the evolution plot it can be seen that as the field moves down the potential there is an increase in the Hubble $\tilde{H}$ compared to the standard GR expansion rate, as shown in the lower right panel of Fig. \ref{fig:field_C}. A sharp enhancement in the the speed-up parameter can also be seen. But as the field rolls back and settles at a positive value, both the conformal factor and the speed-up parameter become one relaxing back to the standard GR case. Thereafter the Hubble $\tilde{H}$ settles with standard GR expansion rate. We then solve the BEs together with the $\varphi$ equation to look for the possible changes in DM relic and $L$ asymmetry due to the enhancement of the expansion rate. From the $L$ asymmetry evolution plots in Fig. \ref{fig:DM_B-L_ST}, it can be seen that the asymmetry increases in such a scenario compared to the standard case. However, it is observed that the DM relic is unaffected for $m_{\rm DM}=600$ GeV. The reason for this is the following. For $\tilde{T}_{0}=1$ TeV, the enhancement in $\xi$ is occurring near $z\simeq 1$ and during that period the DM particles are in equilibrium and the freeze-out of DM occurs near $20 \lesssim z \lesssim 30$. Hence their abundance does not change much. But, with the increase in $\xi$ the washout processes of the asymmetry get relatively weaker and a rise in the asymmetry results. Also it can be seen that with the increase in $\varphi^{'}_{0}$ and with the decrease in $\varphi_{0}$ the asymmetry increases. It is because, with the increase in $\varphi^{'}_{0}$ and with the decrease in $\varphi_{0}$, the conformal factor and the $\xi$ increases. However, one can not increase the $\varphi^{'}_{0}$ arbitrarily. There is an upper limit on $\varphi^{'}_{0}$ ( $\varphi^{'}_{0} \lesssim \pm \sqrt{6}$) coming from the Friedman equation \cite{Dutta:2016htz}. 

\begin{figure}
\includegraphics[scale=.45]{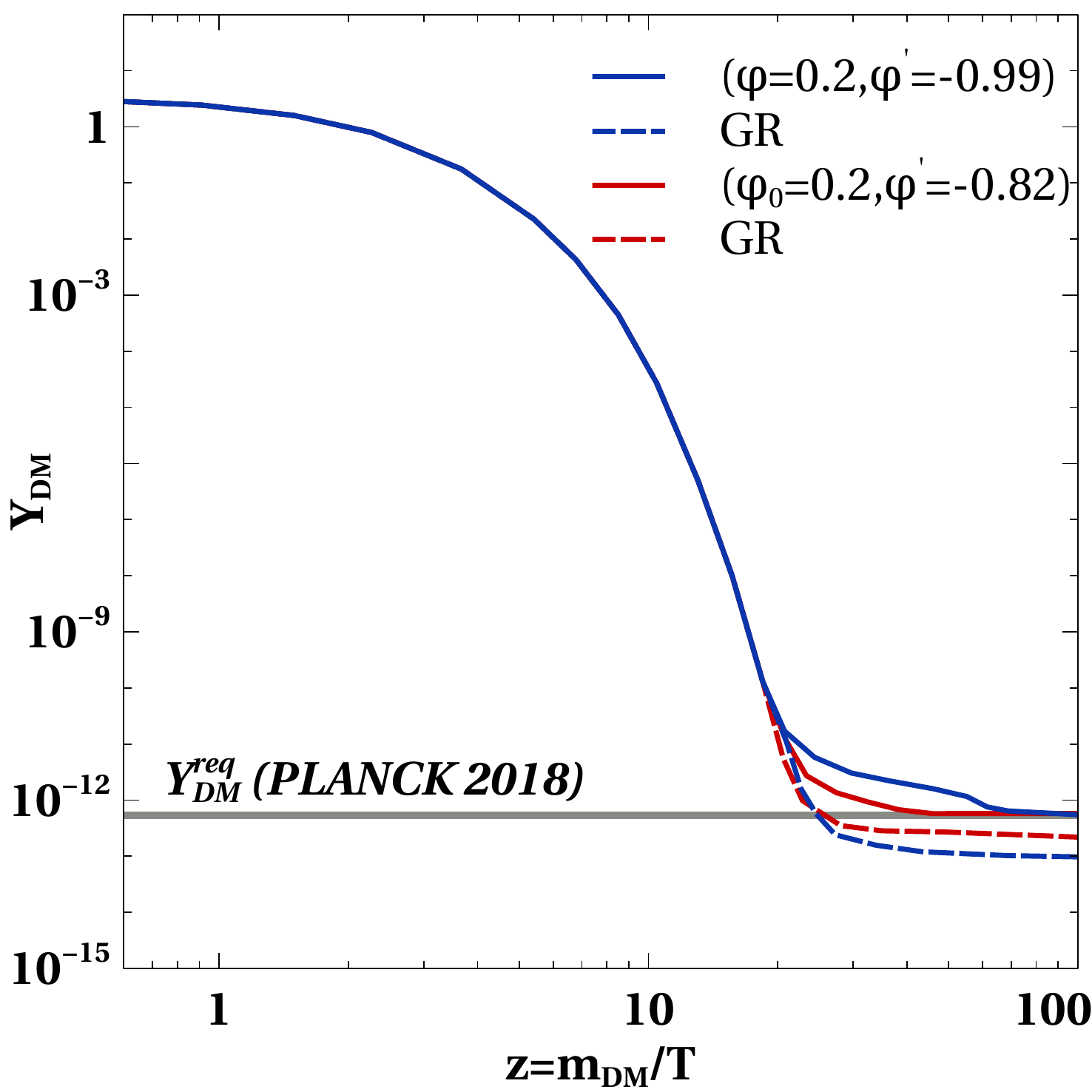}
\includegraphics[scale=0.45]{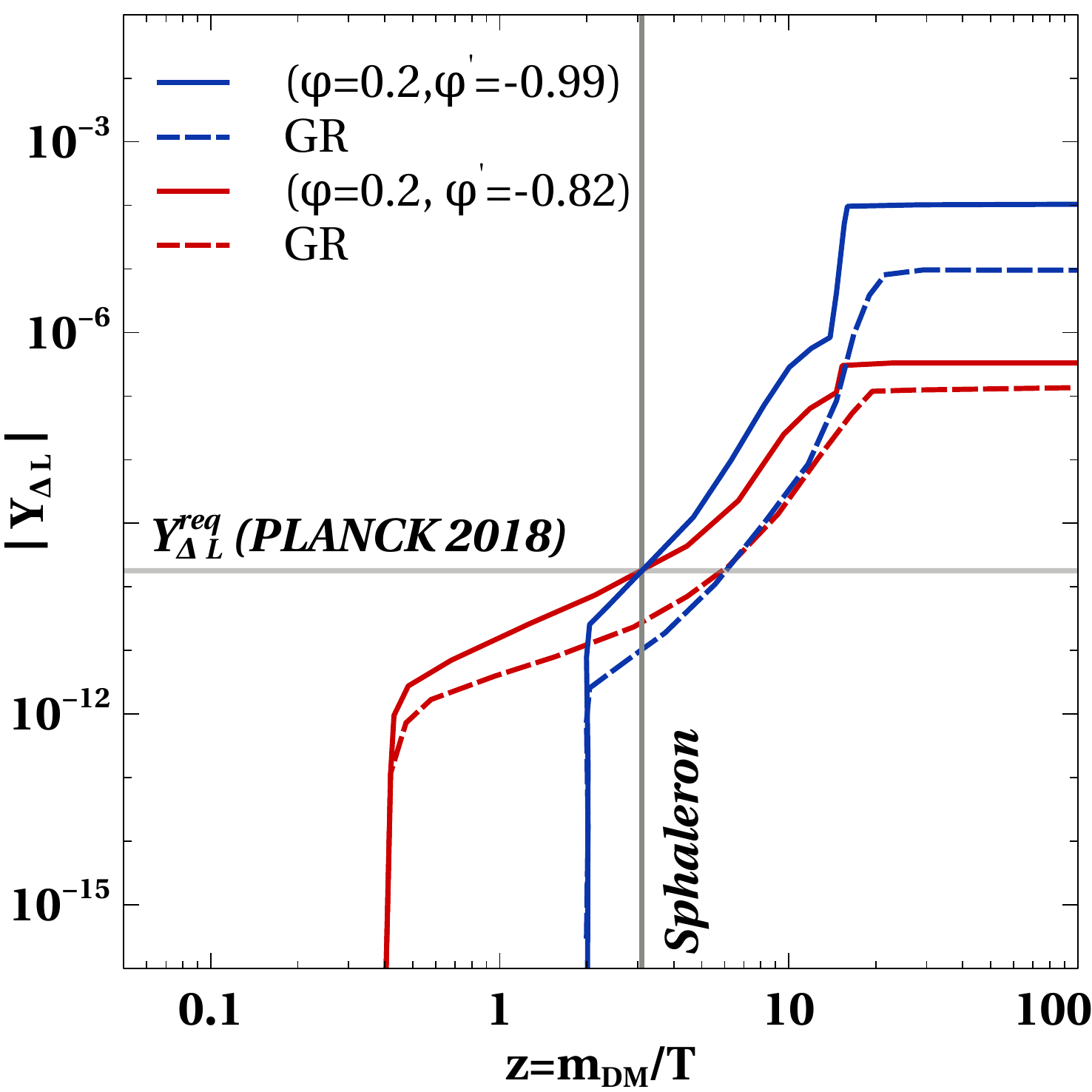}
\caption{Evolution plot of the comoving number density of $DM$ and $L$ asymmetry with $z=m_{\rm DM}/\tilde{T}$ for different benchmark values of the cosmological parameters. Here the particle Physics parameters are fixed at $m_{\rm DM}=400$ GeV, $\lambda_{H\eta}^{''}=1\times 10^{-5}$, $m_{\Delta^{0}}=m_{\Delta^{\pm}}=m_{\Delta^{\pm \pm}}=800$ GeV, $M_{1}=6$ TeV, $M_{j+1}/M_{j}=1.1$, $\mu_{\eta \Delta}=100i$ GeV and $v_{\Delta}=1$ eV. The initial temperature ($\tilde{T_{0}}$) are taken to be $1$ TeV (for the red lines) and $200$ GeV (for the blue lines). Here $T$ in x-axis is equivalent to $\tilde{T}$, the temperature in the Jordan frame.}
 \label{fig:bp_ST}
\end{figure}

We then look for the possibility of lowering the scale of leptogenesis by choosing the appropriate cosmological parameters. We fix the mass of the DM to be $m_{\rm DM}=400$ GeV (keeping in mind that it is not allowed with standard GR expansion due to suppressed relic). Keeping the the other particle physics parameters at benchmark values as described in Fig. \ref{fig:bp_ST}, we change the cosmological parameters $\varphi_{0},\varphi^{'}_{0}$ and  $\tilde{T}_{0}$. From the evolution plot of DM shown on the left panel of Fig. \ref{fig:bp_ST}, it can be seen that both the benchmark values give rise to relic lower than the observed one in the case of standard GR expansion (dashed lines). Apart from the strong gauge portal annihilations, the chosen value of $v_{\Delta}=1$ eV, also leads to strong DM annihilations via the processes $\eta \eta \longrightarrow ll$ leading to further suppression in relic. However, with the enhancement of the Hubble expansion rate in STG, both these benchmark points give rise to the correct relic (solid lines). Similarly, from the evolution plot of the $L$ asymmetry on the right panel of Fig. \ref{fig:bp_ST}, we can see that both these benchmark points lead to insufficient $L$ asymmetry at the time of sphaleron decoupling. But with the modified expansion rate, they generate the correct $L$ asymmetry. Therefore, we can conclude that in scalar-tensor theory of gravity a new viable parameter space will open up with lower scales of WIMPy leptogenesis.

\begin{figure} 
\includegraphics[scale=.45]{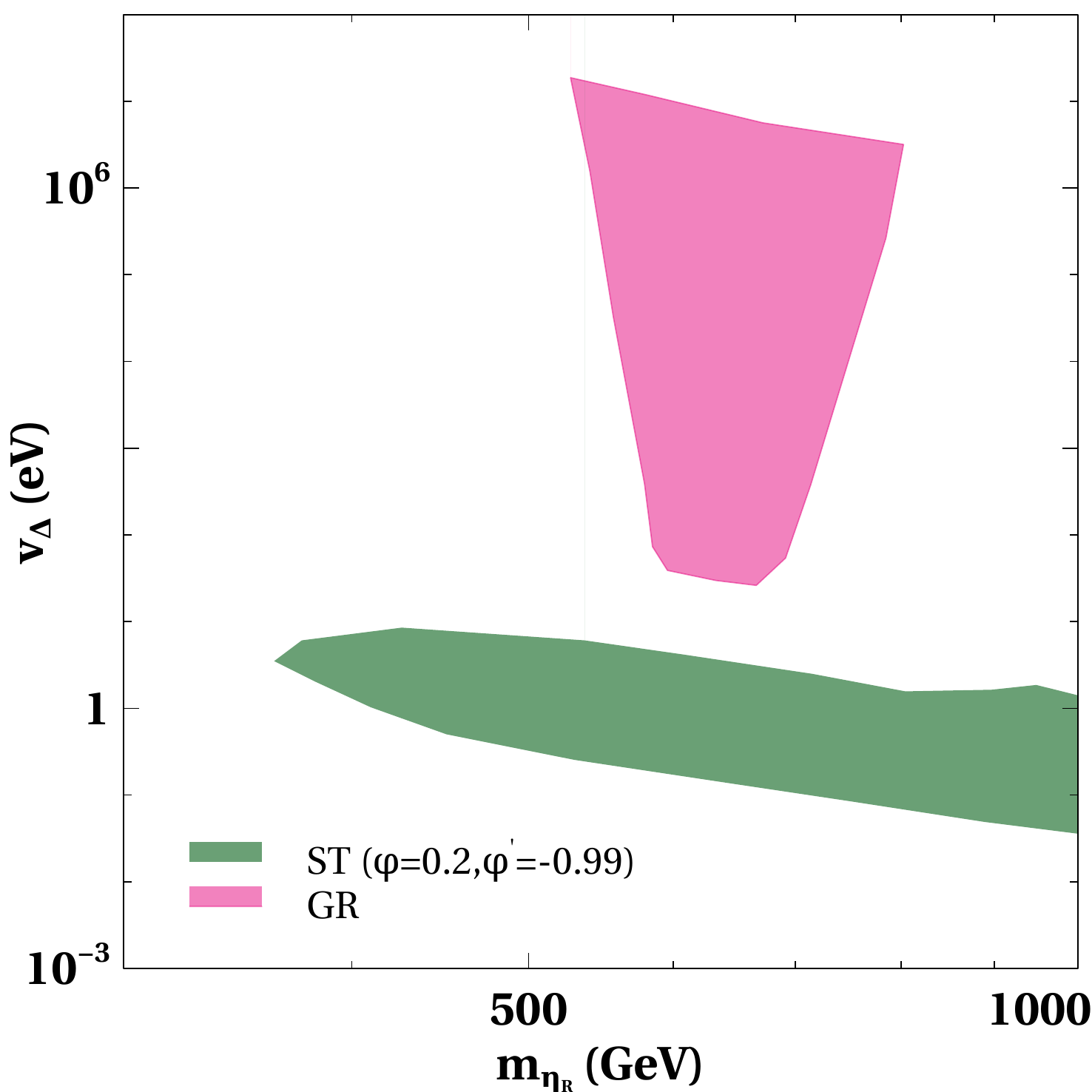}
\caption{Scan plot showing the viable parameter space in $m_{\eta_R}-v_{\Delta}$ plane by keeping the other parameters fixed. The important cosmological parameters are fixed at $\varphi_{0}=0.2$, $\varphi^{'}=-0.99$ and $T_{0}=200$ GeV. The other relevant particle physics parameters were set at $\mu_{\eta \Delta}=80i$ GeV, $\lambda_{H\eta}^{''}=1\times 10^{-5}$, $M_{1}=6$ TeV, $M_{j+1}/M_{j}=1.1$.}
\label{fig:ST_mdm}
\end{figure}

Finally we do a parameter scan over the parameters $m_{\rm DM} \equiv m_{\eta_R}$ and $v_{\Delta}$ by keeping the other parameters fixed. We show the result in Fig. \ref{fig:ST_mdm}. From Fig. \ref{fig:ST_mdm} we can see that the viable parameter space changes significantly from that in standard GR case. This can be understood from the fact that enhanced background expansion rate in STG leads to an increase in DM relic as well as the $L$ asymmetry. This leads to a requirement of large $Y^{\Delta}$ such that the annihilation of DM through the Yukawa interaction ($ \eta \eta \longrightarrow ll$) is more compared to the usual gauge portal channels. This also makes the washout processes stronger which help in reducing the asymmetry. Hence the parameter space can be seen to be shifted towards smaller values of the $v_{\Delta}$. Also since we are no longer relying on the standard gauge portal interactions to satisfy the DM relic, a broader range of $m_{\rm DM}$ is allowed compared to the standard GR case. For the specific values of other parameters mentioned in the caption of Fig. \ref{fig:ST_mdm} the limits on $m_{\eta_R}$ and $v_{\Delta}$ are found to be $320$ GeV $\lesssim m_{\eta_R} \lesssim 1.55$ TeV and $0.02$ eV $\lesssim v_{\Delta} \lesssim 9.6$ eV.

\section{Conclusion}
\label{sec:conclude}
We have studied the impact of three different non-standard cosmology scenarios on leptogenesis while considering the lepton asymmetry to be originating from WIMP type DM annihilations. We choose a minimal particle physics setup with both tree level and radiative contribution to light neutrino mass originating from type-II and scotogenic seesaw respectively. The neutral component of the $Z_2$-odd scalar doublet is considered to be the WIMP DM having both lepton number violating and lepton number conserving annihilation processes. After reproducing the known results for WIMPy leptogenesis in standard cosmology, we implement it in three different non-standard cosmology scenario namely, fast expanding universe, early matter domination and scalar-tensor theory of gravity. Apart from the shift in parameter space compared to the standard cosmology scenario, we also found the scale of leptogenesis or equivalently the WIMP DM mass to be lower in FEU and STG scenarios. This can open up interesting detection prospects of such lighter DM at both direct as well as indirect detection experiments together with colliders. This is in sharp contrast with canonical seesaw or leptogenesis scenarios where the scale remains out of reach from direct experimental probes. In addition, the non-standard cosmological scenarios can be probed due their impact on primordial gravitational wave (GW) spectrum \cite{Cui:2018rwi, Bernal:2019lpc}. Such complementary probes of our setup keeps it verifiable in near future. In fact, such non-standard cosmological era lead to tilts in GW spectrum with amplification of amplitude in high or low frequency regime depending upon the equation of state \cite{Giovannini:1998bp, Saikawa:2018rcs, Bernal:2019lpc}. Accordingly, such non-standard era can be constrained from its effect on primordial GW spectra generated by inflationary dynamics. Firstly, this can be constrained from non-observation of primordial GW at existing experiments. Secondly, if GW starts dominating the radiation energy density of the universe it can give rise to additional $\Delta N_{\rm eff}$ which is constrained from BBN and CMB measurements. Such bound on $\Delta N_{\rm eff}$ can be used to put upper bound $\Omega_{\rm GW} h^2 \lesssim 10^{-6}$. In order to keep the amplified GW amplitude in non-standard cosmology within these bounds, the inflationary scale has to be lower than the maximum allowed value in standard cosmology ${\bf H}_{\rm inf} \leq 10^{-6} M_{\rm Pl}$. These bounds can be strong in scenarios where the equation of state parameter is much stiffer than radiation, like in FEU. Since we are agnostic about inflationary scale and origin of primordial GW here, we leave such studies to future works.

\acknowledgements
DB would like to thank Arnab Dasgupta for useful discussions related to the CP asymmetry calculations. The work of DB is supported by SERB, Government of India grant MTR/2022/000575.

\appendix
\section{Particle Spectrum and Relevant Cross Section}
\label{sec:appen1}
The scalar potential for the model can be identified to be 
\begin{eqnarray}\label{Potential}
V & = & -\mu_{H}^{2}\left( H^{\dagger} H \right)+\mu_{\eta}^{2} \left( \eta^{\dagger} \eta  \right)+\mu_{\Delta}^{2} {\rm Tr}[\Delta^{\dagger} \Delta]+(\mu_{H\Delta} \tilde{H}^{\dagger}\Delta H+ \mu_{\eta \Delta} \eta^{\dagger} \Delta^{\dagger}\tilde{\eta}+{\rm h.c.}) \nonumber \\ & & \lambda_{H}(H^{\dagger} H)^{2} + \lambda_{\eta} \left( \eta^{\dagger} \eta \right)^{2}+ \lambda_{\Delta} \left( {\rm Tr}[\Delta^{\dagger}\Delta]  \right)^{2} + \lambda_{\Delta}^{'} {\rm Tr}[\Delta^{\dagger}\Delta \Delta^{\dagger} \Delta]+ \lambda_{H \eta} \mid H^{\dagger} \eta \mid^{2} \nonumber \\  & & +\lambda_{H \eta}^{'} \left( H^{\dagger} H\right) \left( \eta^{\dagger} \eta \right) + \lambda_{H \eta}^{''} \left( (H^{\dagger} \eta)^{2}+{\rm h.c.} \right) + \lambda_{H \Delta} \left( H^{\dagger}H \right) {\rm Tr}\left[ \Delta^{\dagger} \Delta \right] \nonumber \\ & & +\lambda^{'}_{H \Delta} {\rm Tr} [H^{\dagger} \Delta \Delta^{\dagger} H]+ \lambda_{\eta \Delta} (\eta^{\dagger} \eta){\rm Tr} \left[ \Delta^{\dagger}\Delta \right] + \lambda^{'}_{\eta \Delta} {\rm Tr} \left[ \eta^{\dagger} \Delta \Delta^{\dagger} \eta \right],
\end{eqnarray}

where $\tilde{H}=i \sigma_{2} H^{*}$, $\tilde{\eta}= i \sigma_{2} \eta^{*}$ and the mass parameters $\mu^{2}_{H}>0$ so that the neutral component of $H$ obtains non-vanishing VEV i.e $\langle H^{0}\rangle=v \simeq 246$ GeV. On the other hand, we consider $\mu^2_{\eta, \Delta}>0$ so that they do not take part in spontaneous symmetry breaking. While $\eta$ does not acquire a non-zero VEV at any stage keeping the $Z_2$ symmetry intact, the neutral component of scalar triplet acquires an induced VEV $\langle \Delta^{0} \rangle=v_{\Delta}$ after electroweak symmetry breaking.

The physical masses for the neutral and charged scalars can be obtained by the minimization of the scalar potential. Since $\eta$ is odd under the $Z_{2}$ symmetry, it doesn't get any VEV. In this model, we assume the RHNs are heavier than the
$\eta$ scalar such that the lightest neutral component of $\eta$ plays
the role of DM. After the EWSB, the three scalar multiplets can be written in the following form (assuming unitary gauge)
\begin{equation}
H \ = \ \begin{pmatrix} 0 \\  \frac{ v +h }{\sqrt 2} \end{pmatrix} , \qquad \eta \ = \ \begin{pmatrix} \eta^\pm\\  \frac{\eta_R+i\eta_I}{\sqrt 2} \end{pmatrix}, \qquad \Delta \ = \ \begin{pmatrix} \Delta^+/\sqrt{2} & \Delta^{++}\\  \frac{\Delta^0_R+ v_{\Delta}+ i\Delta^0_I}{\sqrt 2} & -\Delta^+/\sqrt{2}  \end{pmatrix}
\label{eq:idm}
\end{equation}

Upon minimization of the potential the masses of the charged and neutral scalars can be found out be 

\begin{eqnarray}
m_{h}^{2} & \simeq &  \lambda_{H} v^{2}, \\
m_{\Delta_{R}^{0}}^{2} & \simeq & \dfrac{\mu_{H \Delta}v^{2}}{\sqrt{2}v_{\Delta}}, \\
m_{\Delta^0_{I}}^{2} & = & \dfrac{\mu_{H \Delta}}{\sqrt{2}v_{\Delta}} \left( v^{2}+4v_{\Delta}^{2}\right), \\
m_{\Delta^{\pm}}^{2} & = & m_{\Delta^{\pm \pm}}^{2}= \left(  \dfrac{\mu_{H \Delta}}{\sqrt{2}v_{\Delta}}+ \dfrac{1}{4} \lambda_{H\Delta}^{'} \right) \left( v^{2}+4v_{\Delta}^{2}\right), \\
\label{eq:mass_diff}
m_{\eta_{R,I}}^{2} & = & \dfrac{1}{2} \left( 2 \mu_{\eta}^{2}+(\lambda_{H \eta}+\lambda_{H\eta}^{'}\pm \lambda_{H \eta}^{''})v^{2}+(\lambda_{\eta \Delta}v_{\Delta} \mp 2 \sqrt{2}\mid\mu_{\eta \Delta } \mid)v_{\Delta} \right), \\
m_{\eta^{\pm}}^{2} & = & \dfrac{1}{2} \left( 2 \mu_{\eta}^{2}+ \lambda_{H \eta}v^{2}+(\lambda_{\eta \Delta}+ \lambda_{\eta \Delta}^{'})v_{\Delta}^{2} \right).
\end{eqnarray} 

The masses for the CP even scalars are approximated considering $\mu_{H \Delta} \sim \mathcal{O}(100)$ keV. As mentioned earlier the neutrino masses are generated from both loop-level scotogenic and tree-level type-II seesaw mechanisms in this model and the resultant neutrino mass matrix can be written as 

\begin{equation}
m_{\nu}=\left( Y^{N} \right)^{T}\Lambda Y^{N}+Y^{\Delta} v_{\Delta}.
\end{equation}

Here $\Lambda$ is an effective loop-suppressed RHN mass scale and is given by \cite{Ma:2006km}
\begin{eqnarray}
\Lambda_{ii} & = & \dfrac{M_{i}}{16\pi^{2}} \left[ \dfrac{m_{\eta_{R}}^{2}}{M_{i}^{2}-m_{\eta_{R}}^{2}}\ln\left(\dfrac{M_{i}^{2}}{m_{\eta_{R}}^{2}}\right) - \dfrac{m_{\eta_{I}}^{2}}{M_{i}^{2}-m_{\eta_{I}}^{2}} \ln \left( \dfrac{M_{i}^{2}}{m_{\eta_{I}}^{2}} \right) \right].
\label{eq:lambda}
\end{eqnarray}
The two types of Yukawa couplings are related to the neutrino oscillation data by the following formulae

\begin{eqnarray}
\label{eq:YN}
Y^{N}_{i\alpha} & = & F_{I}^{1/2} \left( \Lambda^{-1/2}\mathcal{O} \hat{m_{\nu}}^{1/2} U^{\dagger}_{\rm PMNS} \right), \\ 
\label{eq:YD}
Y^{\Delta}_{\alpha \beta}  & = & F_{II}v_{\Delta}^{-1}(U_{\rm PMNS}^{*}\hat{m_{\nu}}U_{\rm PMNS}^{\dagger}), 
\end{eqnarray} 

where $\hat{m_{\nu}}= diag\left(m_{\nu_{1}},m_{\nu_{2}},m_{\nu_{3}}  \right) $ is the diagonal neutrino mass matrix and $U_{PMNS}$ is the PMNS lepton mixing matrix. Here $\mathcal{O}$ is an any arbitrary orthogonal matrix. We have used the Casas Ibarra (CI) parametrisation \cite{Casas:2001sr} to determine the Yukawa couplings. We also assume the two seesaw contributions to be equal namely $F_I \approx F_{II} = 1/2$.

The general expression for the thermally averaged cross sections for the processes are given by \cite{Gondolo:1990dk}

\begin{eqnarray}
\langle \sigma v \rangle_{i_{1}i_{2} \longrightarrow f_{1}f_{2}} & = & \dfrac{1}{2Tm_{i_{1}}^{2}m_{i_{2}}^{2} K_{2}(m_{i_{1}}/T)K_{2}(m_{i_{2}}/T)} \nonumber \\ & & \int_{s_{\rm in}}^{\infty} \int_{-1}^{1} \dfrac{1}{32 \pi} \dfrac{\mid \mathcal{M}\mid^{2}}{\sqrt{s}}p_{i_{1}i_{2}}p_{f_{1}f_{2} } K_{1}(\sqrt{s}/T) \,ds \hspace{2pt} d(\cos\theta),
\end{eqnarray}

where $T$ is the temperature, $K_{i}$ are the modified Bessel functions of order i, $\mathcal{M}$ is the amplitude for the process $i_{1}i_{2} \longrightarrow f_{1}f_{2}$, and 
\begin{eqnarray}
p_{i j} & = & \dfrac{1}{2} \sqrt{\lambda(s,m_{i}^{2},m_{j}^{2})/s},  \\ 
s_{\rm in} & = & {\rm max}[(m_{i_{1}}+m_{i_{2}})^{2},(m_{f_{1}}+m_{f_{2}})^{2}],   \\
\lambda(x,y,z) & = & x^{2}+y^{2}+z^{2}+2xy+2yz+2xz.
\end{eqnarray}
The amplitudes relevant for WIMPy leptogenesis are given below \cite{Dasgupta:2019lha}
\begin{eqnarray}
\mid \mathcal{M}(\eta \eta \longrightarrow LL) \mid^{2} & = & \dfrac{\hat{m_{\nu}^{2}}}{v_{\Delta}^{2}} \dfrac{F_{I}^{2}\mu_{\eta \Delta}^{2}s}{(s-m_{\Delta}^{2})^{2}+m_{\Delta}^{2}\Gamma_{\Delta}^{2}}+ \sum_{i} F_{II}^{2}\dfrac{\hat{m_{\nu}^{2}}}{\Lambda_{ii}^{2}} M_{i}^{2} s \left[ \dfrac{1}{t-M_{i}^{2}}+\dfrac{1}{u-M_{i}^{2}} \right]^{2} \nonumber \\ & & + \dfrac{F_{I}F_{II}\mid \mu_{\eta \Delta}\mid (s-m_{\Delta}^{2})}{(s-m_{\Delta})^{2}+m_{\Delta}^{2}\Gamma_{\Delta}^{2}} \sum_{i} \dfrac{\hat{m_{\nu}}^{2}}{\Lambda_{ii}v_{\Delta}}M_{i}s \left[ \dfrac{1}{t-M_{i}^{2}}+\dfrac{1}{u-M_{i}^{2}}\right] \\
\mid M(\eta \bar{L} \longrightarrow \eta L) \mid^{2} & = & \dfrac{\hat{m_{\nu}}^{2}}{v_{\Delta}^{2}} \dfrac{F_{I}^{2}\mu^{2}(m_{\eta}^{2}-t)}{(t-m_{\eta}^{2})^{2}}+ \sum_{i}  \dfrac{\hat{m_{\nu}^{2}}}{\Lambda_{ii}^{2}} \dfrac{F_{II}^{2}M_{i}^{2}s}{(s-M_{i}^{2})^{2}+M_{i}^{2}\Gamma_{N_{i}}^{2}}  \nonumber \\ & & \dfrac{F_{I}F_{II}\mid \mu_{\eta \Delta} \mid (m_{\eta}^{2}-t)}{(t-m_{\Delta}^{2})^{2}} \sum_{i} \dfrac{\hat{m_{\nu}^{2}}}{\Lambda_{ii}v_{\Delta} }M_{i} s \left[ \dfrac{s-M_{i}^{2}}{(s-M_{i}^{2})^{2}+M_{i}^{2}\Gamma_{N_{i}}^{2}} \right]
\end{eqnarray}

The asymmetry generated from the annihilation, $\eta \eta \longrightarrow \ell \ell$, at the amplitude level, is given by

\begin{eqnarray} \label{eq:asymmetry}
\delta & = & \mid  \mathcal{M}\mid^{2} - \mid  \bar{\mathcal{M}} \mid^{2} \nonumber \\ & & =\sum_{i}4 {\rm Im} \left[ \mu_{\eta \Delta} (Y^{N}Y^{\Delta *}(Y^{N})^{T})_{ii}  \right] \nonumber \\ & & \times \dfrac{s M_{i}m_{\Delta} \Gamma_{\Delta}}{(s-m_{\Delta}^{2})^{2}+m_{\Delta}^{2}\Gamma_{\Delta}^{2}} \left[ \dfrac{1}{t-M_{i}^{2}}+\dfrac{1}{u-M_{i}^{2}}  \right],
\end{eqnarray}

where $\Gamma_{\Delta}$ is the triplet scalar decay width. The imaginary part appearing in the CP asymmetry parameter can be parametrised as follows 
\begin{eqnarray}
{\rm Im}\left[ \mu_{\eta \Delta} (Y^{N}Y^{\Delta *}(Y^{N})^{T})_{ii} \right] & = & F_{I} F_{II} v_{\Delta}^{-1} {\rm Im} \left[ \mu_{\eta \Delta} (\Lambda^{-1/2}\mathcal{O}\hat{m_{\nu}}^{2} \mathcal{O}^{T}\Lambda^{-1/2})_{ii} \right].
\label{eq:Im}
\end{eqnarray}

We use the $\delta$ or the amplitude squared difference defined above in thermally averaged cross section to evaluate $\langle \sigma v \rangle^{\delta}_{\eta \eta \rightarrow \ell \ell} $ which enter the Boltzmann equations for $L$ asymmetry.

\section{Analytical calculation of asymmetry}
\label{sec:appen2}

The thermally averaged rate for the asymmetric part of the annihilation of $\eta$ is given by

\begin{eqnarray} \label{eq:asymmetry_therm}
\langle \sigma v\rangle^{\delta}_{\eta \eta \longrightarrow ll} & = & \dfrac{1}{2T m_{\eta}^{4} K_{2}^{2}(m_{\eta}/T)} \int_{s_{\rm in}}^{\infty} \int_{-1}^{1} \dfrac{1}{32\pi} \dfrac{\delta}{\sqrt{s}} p_{\eta \eta} p_{ll} K_{1}(\sqrt{s}/T)\, ds \, d(\cos\theta)
\end{eqnarray}

In the narrow-width approximation, the asymmetry in \eqref{eq:asymmetry} at the resonance point can be simplified to be 

\begin{eqnarray} \label{eq:asymmetry2}
\delta \simeq 4 \sum_{i} {\rm Im} \left[ \mu_{\eta \Delta} (Y^{N}Y^{\Delta^{*}} (Y^{N})^{T})  \right] sM_{i} \delta(s-m_{\Delta}^{2})  \left[ \dfrac{1}{t-M_{i}^{2}} +\dfrac{1}{u-M_{i}^{2}} \right], 
\end{eqnarray}
where we have used  $$ \dfrac{\Gamma_{\Delta}m_{\Delta}}{(s-m_{\Delta}^{2})^{2}+m_{\Delta}^{2}\Gamma_{\Delta}^{2}} \xrightarrow{\Gamma_{\Delta}/m_{\Delta} \longrightarrow 0} \pi \delta(s-m_{\Delta}^{2}). $$ 

Replacing \eqref{eq:asymmetry2} in \eqref{eq:asymmetry_therm} the $\langle \sigma v\rangle_{\eta \eta \longrightarrow ll}^{\delta}$ is calculated and is given by 

\begin{eqnarray} \label{eq:Asymmetry_ana}
\langle \sigma v\rangle_{\eta \eta \longrightarrow ll}^{\delta} & \approx & \dfrac{16\pi}{m_{\eta}^{4}}\dfrac{1}{\tilde{\mu}_{\eta \Delta}^{2} \sum_{\alpha,\beta} \mid Y^{\Delta}_{\alpha \beta} \mid^{2}} \Gamma_{\Delta \longrightarrow \eta \eta} \Gamma_{\Delta \longrightarrow ll} \sum_{i} \left[ \mu_{\eta \Delta} (Y^{N}Y^{\Delta^{*}} (Y^{N})^{T})  \right] \nonumber \\ & & \dfrac{f[m_{\Delta},m_{\eta},M_{i}] r_{N_{i}}}{\sqrt{m_{\Delta}^{2}-4m_{\eta}^{2}}} \dfrac{zK_{1}(r_{\Delta}z)}{K_{2}^{2}(z)}
\end{eqnarray}
where $\tilde{\mu}_{\eta \Delta} = \mu_{\eta \Delta}/m_{\Delta}$, $r_i = m_i/m_\eta$. We have corrected the above expression by incorporating an additional factor missing in \cite{Dasgupta:2019lha}. Here, the function $f[m_{\Delta},m_{\eta},M_{i}]$ is given by

\begin{eqnarray}
f[m_{\Delta},m_{\eta},M_{i}] & = & 2 \ln \left[\dfrac{2M_{i}^{2}+m_{\Delta}^{2}-2m_{\eta}^{2}-\sqrt{m_{\Delta}^{2}}\sqrt{m_{\Delta}^{2}-4m_{\eta}^{2}}}{2M_{i}^{2}+m_{\Delta}^{2}-2m_{\eta}^{2}+\sqrt{m_{\Delta}^{2}}\sqrt{m_{\Delta}^{2}-4m_{\eta}^{2}}} \right] \nonumber \\ & - & 2 \ln \left[\dfrac{-M_{i}^{2}-m_{\Delta}^{2}/2+m_{\eta}^{2}-\dfrac{1}{2}\sqrt{m_{\Delta}^{2}}\sqrt{m_{\Delta}^{2}-4m_{\eta}^{2}}}{-M_{i}^{2}-m_{\Delta}^{2}/2+m_{\eta}^{2}+\sqrt{m_{\Delta}^{2}}\sqrt{m_{\Delta}^{2}-4m_{\eta}^{2}}} \right].
\end{eqnarray}

\begin{figure}[h] 
\includegraphics[scale=.45]{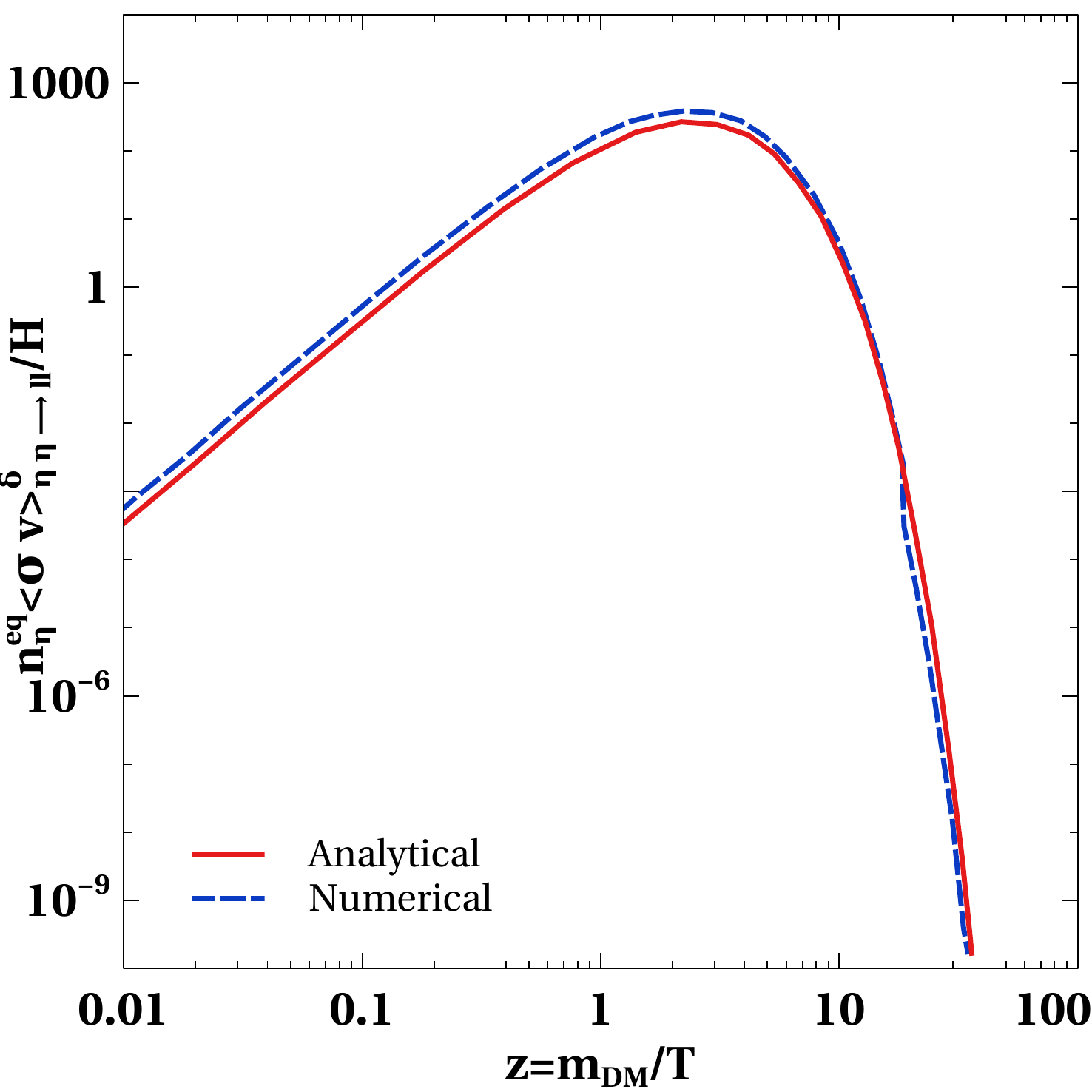}
\caption{Comparison plot showing the rates for $\langle \sigma v\rangle_{\eta \eta \longrightarrow ll}^{\delta}$ using the approximate analytical formula in Eq. \eqref{eq:Asymmetry_ana} and the exact numerical integration in Eq. \eqref{eq:asymmetry_therm}.}
\label{fig:comparison}
\end{figure}

In Fig. \ref{fig:comparison} we show the comparison of rates of $\langle \sigma v\rangle_{\eta \eta \longrightarrow ll}^{\delta}$ calculated from the the analytical expression in Eq. \eqref{eq:Asymmetry_ana} and the exact numerical integration given in Eq. \eqref{eq:asymmetry_therm}. It can be seen that there is no significant difference between the two results. Nevertheless, we have used the exact numerical integration result for $\langle \sigma v \rangle_{\eta \eta \longrightarrow ll}$ in our analysis.

\begin{figure}[h] 
\includegraphics[scale=.4]{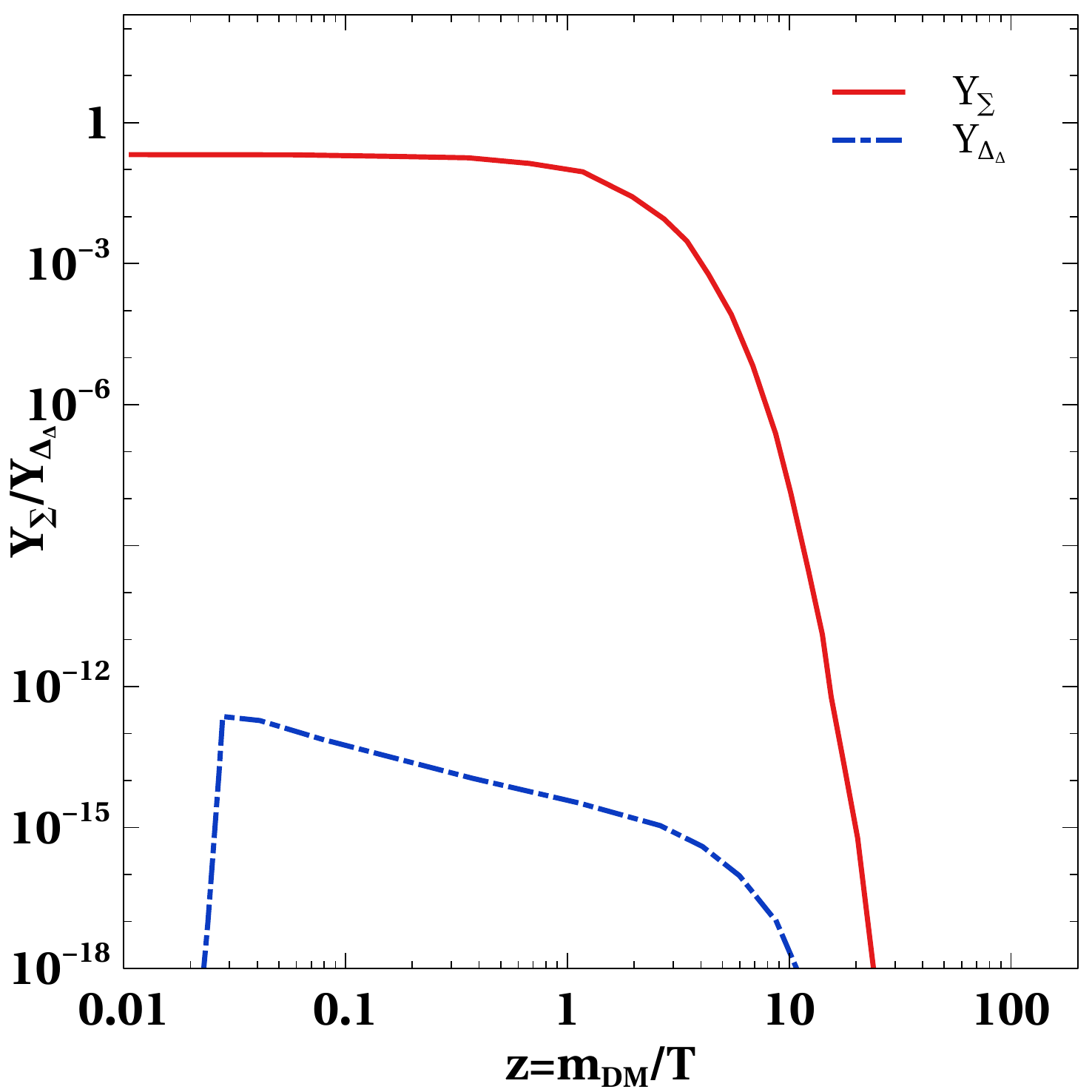}
\includegraphics[scale=.4]{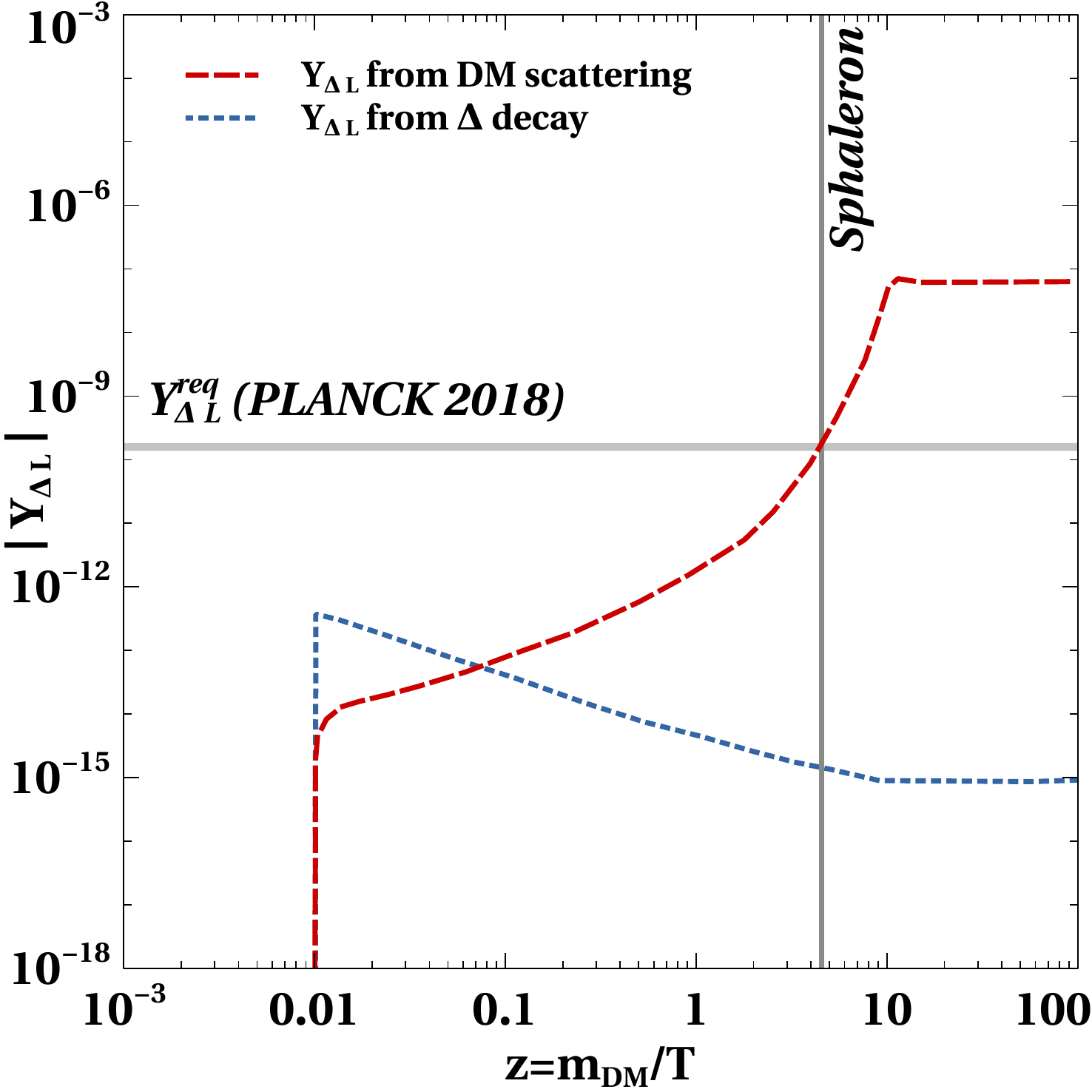}
\includegraphics[scale=.4]{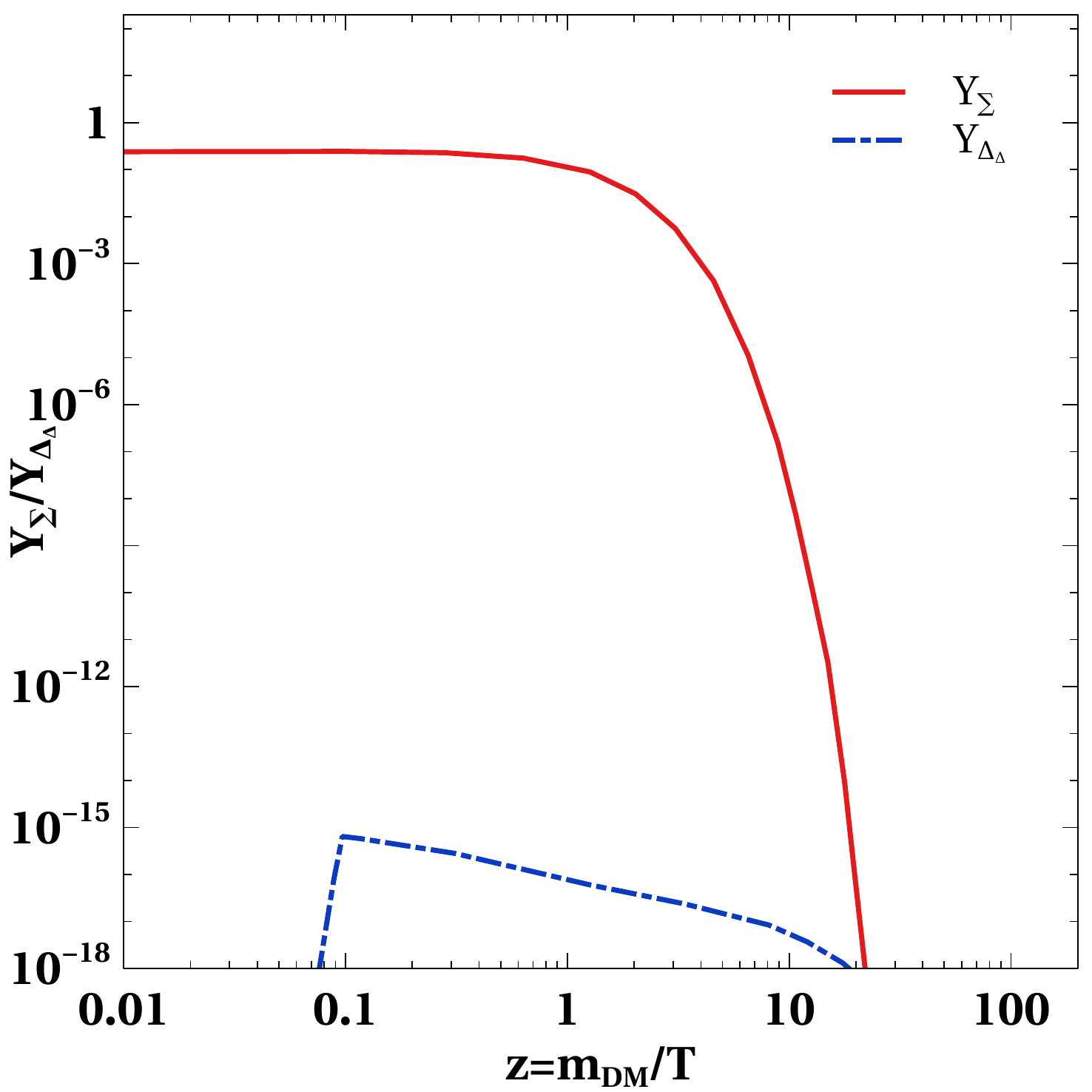}
\includegraphics[scale=.4]{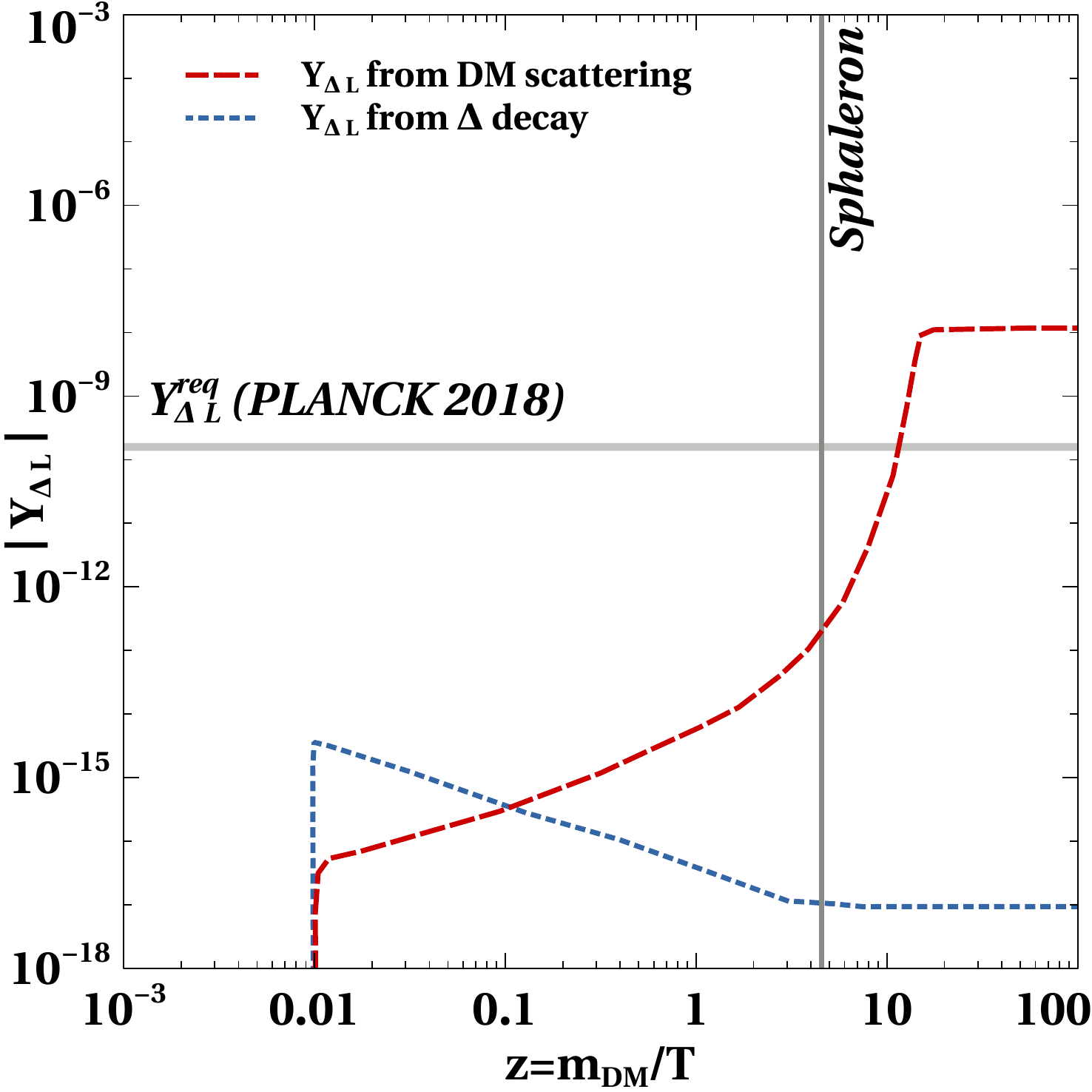}
\caption{Plot showing the evolution of the comoving number densities of $\sum$ and $\Delta$ (on the left panel) and the generated $L$ from the scattering of DM and from the decay of $\Delta$ (on the right panel plot). The important parameters are fixed at $m_{\eta_{R} }=m_{\rm DM}=600$ GeV, $m_{\Delta^{\pm}}=m_{\Delta^{\pm \pm}}=m_{\Delta^{0}}=1.2$ TeV, $M_{1}=6$ TeV, $M_{2}=6.6$ TeV, $M_{3}=7$ TeV, $\mu_{\eta\Delta}=10$i, $\lambda_{H\eta}^{''}=1\times 10^{-5}$, $v=1$ keV (for the top panel) and $v=1$ eV (for the bottom panel). }
\label{fig:deltadecay}
\end{figure}

\section{$\Delta$ decay contribution to the asymmetry}
\label{sec:appen3}

In this model there can be another source of asymmetry from the decay of the triplet to a pair of leptons.  The CP asymmetry parameter associated with this decay is given by \cite{Hambye:2003ka, Hambye:2005tk}

\begin{eqnarray}
\epsilon_{\Delta} & = & \sum_{i}\dfrac{M_{i}}{8\pi}  \dfrac{\sum_{\alpha,\beta}{\rm Im}[\mu_{\eta \Delta}(Y^N_{i\alpha}Y^N_{i\beta}(Y^{\Delta})_{\alpha \beta})]}{{\rm Tr}[(Y^{\Delta})^{\dagger}Y_{\Delta}]m_{\Delta}^{2}+\mid \mu_{H\Delta}\mid^{2}+\mid \mu_{\eta\Delta}\mid^{2}} \ln[1+\dfrac{m_{\Delta}^{2}}{M_{i}^{2}}].
\end{eqnarray}

The network of Boltzmann equations for scalar triplet leptogenesis irrespective of whether
lepton flavor effects are active or not, corresponds to a system of coupled differential
equations accounting for the temperature evolution of the triplet density $\sum=\Delta_{\Delta}+\Delta^{\dagger}$ the triplet asymmetry $\Delta_{\Delta}=\Delta-\Delta^{\dagger}$. These Boltzmann equation for leptogenesis taking the $\Delta$ decay and inverse decay into account can be written as \cite{AristizabalSierra:2014nzr}

\begin{eqnarray} \label{BE2}
\dfrac{dY_{\Sigma}}{dz} & = & - \dfrac{\gamma_{D}}{s{\bf H}z} \left( \dfrac{Y_{\Sigma}}{Y_{\Sigma}^{\rm eq}}-1 \right), \\
\dfrac{dY_{\Delta_{\Delta}}}{dz} & = & -  \dfrac{\gamma_{D}}{s{\bf H}z}\left[ \dfrac{Y_{\Delta_{\Delta}}}{Y_{\Sigma}^{\rm eq}}-\sum_{k} \left( B_{l}C_{k}^{l}-B_{H}C_{k}^{H} \right) \dfrac{Y_{\Delta_{k}}}{Y_{l}^{\rm eq}}  \right], \\
\dfrac{dY_{\Delta L}}{dz} & = & \epsilon_{\Delta} \dfrac{\gamma_{D}}{s{\bf H}z}   \left( \dfrac{Y_{\Sigma}}{Y_{\Sigma}^{\rm eq}} -1   \right) +2 \left(  \dfrac{Y_{\Delta_{\Delta}}}{Y_{\Sigma}^{\rm eq}} -\sum_{k} C_{k}^{l} \dfrac{Y_{\Delta_{k}}}{Y_{l}^{\rm eq}} \right) B_{l} \dfrac{\gamma_{D}}{s{\bf H}z}. 
\end{eqnarray}

Here $Y_{\Delta_{\Delta}}$ is the comoving number density of $\Delta_{\Delta}=\Delta-\Delta^{\dagger}$ and $Y_{\Sigma}$ is the comoving number density of $\Sigma=\Delta+\Delta^{\dagger}$. $B_{l}$ and $B_{\phi}$ are triplet decay branching ratios to lepton and scalar final states respectively. The branching ratios are defined as follows 

\begin{equation}
B_{l} = \sum_{i,j} B_{l_{ij}}= \sum_{i,j} \dfrac{m_{\Delta}}{8\pi \Gamma_{\Delta}^{\rm Tot}}\mid (Y^{\Delta})^{ij} \mid^{2},  \,
B_{H} = \dfrac{\mid \mu_{H \Delta} \mid^{2}}{8\pi m_{\Delta}\Gamma_{\Delta}^{\rm Tot}},
\end{equation}
where $\Gamma_{\Delta}^{\rm Tot}$ is the total decay width for the triplet $\Delta$ and is given by 

\begin{eqnarray}
\Gamma_{\Delta}^{\rm Tot} & = & \sum_{ij} \Gamma (\Delta \longrightarrow l_{i} l_{j})+ \Gamma (\Delta \longrightarrow HH)+ \Gamma (\Delta \longrightarrow \eta \eta) , \nonumber \\ 
    & = & \dfrac{m_{\Delta}}{8\pi} \left[ {\rm Tr}[(Y^{\Delta})^{\dagger}Y_{\Delta}]+  \dfrac{\mid \mu_{H \Delta} \mid^{2}+ \mid \mu_{\eta \Delta} \mid^{2}}{m_{\Delta}^{2}}  \right].
\end{eqnarray}

In the above Boltzmann equations, $C_{l}^{k}$ and $C_{H}^{k}$ are the conversion factor for the asymmetry in lepton sector as well as in Higgs with the fundamental asymmetries $Y_{\Delta_{\Delta}}$ and $Y_{\Delta L}$. The conversion matrices relate the asymmetries as follows 

\begin{equation}
Y_{\Delta_{l}} = -\sum_{k}C_{k}^{l}Y_{X_{k}}, \,
Y_{\Delta_{H}} = -\sum_{k}C_{k}^{H}Y_{X_{k}},
\end{equation}

where $Y_{X_{k}}$ are the elements of $Y_{X}^{T}=(Y_{\Delta_{\Delta}},Y_{\Delta L})$. The conversion matrices for the unflavored leptogenesis are \cite{AristizabalSierra:2014nzr}

\begin{equation}
C^{l} =
\begin{pmatrix}
0 & 1/2
\end{pmatrix}, \,
C^{H} =
\begin{pmatrix}
3 & 1/2
\end{pmatrix} 
\end{equation}
Here we have simplified the situation by considering the chemical potential of the $\eta$ field to be zero such that $C^{\eta}$ and hence the corresponding $B_{\phi} = \dfrac{\mid \mu_{\eta \Delta}\mid^{2}}{8\pi m_{\Delta} \Gamma_{\Delta}^{\rm Tot}}$ does not appear. For detailed analysis of scalar triplet leptogenesis in this model, including flavour effects, please refer to \cite{Datta:2021gyi}.

In Fig. \ref{fig:deltadecay} we have shown the evolution of the comoving number densities of $\sum$, $\Delta$ and the $L$ asymmetry generated from the annihilation of WIMP DM $\eta$ and from the decay of $\Delta$. While plotting we have taken both $\Delta$ and $\bar{\Delta}$ to be in equilibrium and therefore $Y_{\Delta_{\Delta}}^{initial}=0$. From Fig. \ref{fig:deltadecay}, it can be seen that the contribution to the $L$ asymmetry coming from the decay of $\Delta$ is much less compared to that coming from the scattering of DM. Therefore in this work we have neglected the $L$ asymmetry coming from the $\Delta$ decay. This is expected as the scale of $\Delta$ is higher compared to $\eta_R$ or WIMP DM. Similarly, one can have leptogenesis from decay of $Z_2$-odd RHNs as well, as discussed in earlier works \cite{Hambye:2009pw, Racker:2013lua, Clarke:2015hta, Hugle:2018qbw, Borah:2018rca, Mahanta:2019gfe, Mahanta:2019sfo, Kashiwase:2012xd, Kashiwase:2013uy, JyotiDas:2021shi}. However, the scale of RHNs is even higher than that of $\Delta$ and such high scale generation of asymmetries will be sub-dominant in final asymmetry dominantly generated by lepton number violating WIMP annihilations at lower scales.

\section{Calculation of the Sphaleron factor}
\label{sec:appen4}
Here we show the derivation of the sphaleron conversion factor in our model, following the standard approach outlined in \cite{Harvey:1990qw}. For a relativistic particle $X$ with spin $s$ and degrees of freedom (dof) $g_X$, the relation between the asymmetry of particle over antiparticle and the particle's chemical potential is given by 
\begin{eqnarray}
Y_{\Delta_{X}} & = & \dfrac{T^{2}}{6s}g_{X}\mu_{X}     \hspace{20pt} \textrm{for fermions},  \\
Y_{\Delta_{X}} & = & \dfrac{T^{2}}{6s}2g_{X}\mu_{X}  \hspace{14pt}     \textrm{for boson}.
\end{eqnarray}
In principle, there are as many as chemical potentials (an asymmetry) as the number of particles in the plasma. This number is reduced due to the constraints imposed by a set of chemical equilibrium conditions and some conservation laws in the early universe, as outlined below.

\begin{enumerate}
\item Chemical potentials for all the gauge bosons vanishes $\mu_{W^{i}}=\mu_{\mathcal{B}}=\mu_{g}=0$. Therefore the electroweak and color multiplets have the same chemical potentials. 
\item Regardless of the temperature of the universe the electric charge must be conserved and this leads to the following constraint, 
\begin{eqnarray} \label{eq:Echarge}
Q  =  \sum_{i}^{N_{f}} \left( \mu_{Q_{i}}+2\mu_{u_{i}}-\mu_{d_{i}}-\mu_{l_{i}}-\mu_{e_{i}}  \right)+\sum_{i}^{m} 2 \mu_{\phi}+\sum_{i}^{n} 6 \mu_{\Delta} & = & 0.
\end{eqnarray} 
Here $\mu_{Q_{i}}$, $\mu_{u_{i}}$, $\mu_{d_{i}}$, $\mu_{l_{i}}$, $\mu_{e_{i}}$ are the chemical potential for the left handed quark doublets, right handed up type quarks, right handed down type quarks, left handed lepton doublets and right handed charged leptons respectively. $N_{f}$ is the number of Fermion generations present in the model. $\mu_{\phi}$ and $\mu_{\Delta}$ are the chemical potentials for the the doublets and triplet scalars respectively. $m$ and $n$ are the number of scalar doublet and scalar triplet generations in the model.   

\item Non-perturbative electroweak sphaleron and QCD instanton processes, while in thermal equilibrium, imposes the following constraints
\begin{eqnarray}
& \sum_{i}^{N_{f}}\left( 3\mu_{Q_{i}}+\mu_{l_{i}} \right) = 0 
\Rightarrow3\mu_{Q}+\mu_{l} = 0, \nonumber \\
& \sum_{i}^{N_{f}}\left( 2\mu_{Q_{i}}-\mu_{d_{i}}-\mu_{u_{i}} \right) = 0.
\end{eqnarray}
\item The Yukawa interactions, while in equilibrium, bring the following constraints 

\begin{eqnarray} 
\mu_{u_{i}}-\mu_{Q_{i}}-\mu_{\phi} & = & 0, \hspace{20pt}  \textrm{$\phi^{0} \longleftrightarrow \bar{u_{L}}+u_{R}$}  \\
\mu_{d_{i}}-\mu_{Q_{i}}+\mu_{\phi} & = & 0, \hspace{20pt} \textrm{$\phi^{0} \longleftrightarrow \bar{d_{R}}+d_{L}$}    \\
\mu_{e_{i}}-\mu_{l_{i}}+\mu_{\phi} & = & 0, \hspace{20pt} \textrm{$\phi^{0}\longleftrightarrow e_{iL}+\bar{e_{iR}}$}   \\
\mu_{\Delta}-2\mu_{l_{i}} & = & 0, \hspace{20pt} \textrm{$\Delta \longleftrightarrow 2l_{iL}$} 
\end{eqnarray}
Since we are assuming equilibrium among different generations, the generation index $i$ in subscript can be dropped from the above equations. Replacing $\mu_{u_{i}}$, $\mu_{d_{i}}$, $\mu_{e_{i}}$ and $\mu_{\Delta}$ in Eq. \eqref{eq:Echarge}, $\mu_{\phi}$ can be expressed in terms of $\mu_{Q_i} \equiv \mu_{Q}$ as follows
\begin{eqnarray}
\mu_{\phi} & = & \dfrac{36n-8N_{f}}{2m+4N_{f}}\mu_{Q}
\end{eqnarray}
Similarly, we can express other chemical potentials in terms of $\mu_Q$. 

Now, the baryon number ($B$) in terms of the chemical potentials is given by
\begin{eqnarray} \label{eq:Bnumber}
B & = & \sum_{i}^{N_{f}} \left( 2\mu_{Q_{i}}+\mu_{u_{i}}+\mu_{d_{i}} \right)  = 4N_{f}\mu_{Q}.
\end{eqnarray}
Similarly, the lepton number ($L$) is given by
\begin{eqnarray} \label{eq:Lnumber}
L & = & \sum_{i}^{N_{f}} \left( 2\mu_{l_{i}}+\mu_{e_{i}}  \right) = 3N_{f}\mu_{l}-N_{f}\mu_{\phi}=-9N_{f}\mu_{Q}-\dfrac{N_{f}(36n-8N_{f})}{4N_{f}+2m}\mu_{Q}.
\end{eqnarray}
From Eq. \eqref{eq:Bnumber}, \eqref{eq:Lnumber}, we can now find
\begin{eqnarray}
B & = & \dfrac{8N_{f}+4m}{22N_{f}+13m+18n} (B-L),  \\
B & = & -\dfrac{8N_{f}+4m}{14N_{f}+9m+18n} L,
\end{eqnarray}
which for our model with $m=2, n=1$ gives $\Delta B = -\frac{16}{39} \Delta L$.

\end{enumerate}

\providecommand{\href}[2]{#2}\begingroup\raggedright\endgroup

\end{document}